\documentclass[prd,aps]{revtex4}
\usepackage{amssymb,graphicx}

\newcommand{\be}{\begin{equation}}
\newcommand{\ee}{\end{equation}}
\newcommand{\bea}{\begin{eqnarray}}
\newcommand{\eea}{\end{eqnarray}}
\newcommand{\beay}{\begin{eqnarray*}}
\newcommand{\eeay}{\end{eqnarray*}}

\begin{document}
\renewcommand{\arraystretch}{2.}
\setlength{\tabcolsep}{3mm}
\title{Multi-block simulations in general relativity:
  high order discretizations, numerical stability, and applications}
\author{Luis Lehner$^1$, Oscar Reula$^2$, and Manuel Tiglio$^{1,3,4}$}
\affiliation{$1$ Department of Physics and Astronomy, Louisiana State
University, Baton Rouge, LA 70803-4001\\
$2$ FaMAF, Universidad Nacional de C{\'o}rdoba, Ciudad Universitaria, 5000 C{\'o}rdoba, Argentina \\
$3$ Center for Computation and Technology, 302 Johnston Hall, Louisiana State
University, Baton Rouge, LA 70803-4001 \\
$4$ Center for Radiophysics and Space Research, Cornell University, Ithaca, NY 14853}

\begin{abstract}
The need to smoothly cover a computational domain of interest generically requires 
the adoption of several grids. To
solve the problem of interest under this grid-structure one must ensure
the suitable transfer of information among the different grids involved.
In this work we discuss a technique that allows one to construct
finite difference schemes of arbitrary high order which are guaranteed to satisfy linear numerical
and strict stability. The technique relies on the use of difference operators
satisfying summation by parts and {\it penalty techniques} to transfer
information between the grids. This allows the derivation of 
semidiscrete energy estimates for problems admitting such estimates at the
continuum. We analyze several aspects of this technique when used in
conjuction with high order schemes and illustrate 
its use in one, two and three dimensional numerical relativity model
problems with non-trivial topologies, including truly spherical black hole excision. 
\end{abstract}
\maketitle

\section{Introduction}
\label{sec:Introduction}
Many systems of interest have a non-trivial natural topology that  
a single cubical computational domain cannot accommodate in a smooth manner.
Examples of these topologies in three-dimensional settings are  
$S^2\times R$ (or a subset of it), encountered when dealing with spacetimes
with smooth 
outer boundary and inner boundaries where required to excise 
singularities,  and $S^2\times S^1$ or $S^3$ topologies, commonly found in
cosmological problems.

The need to treat these scenarios naturally leads one to consider multiple coordinate patches
in order to cover the region of integration. These, in turn, translate into having to adopt multiple grids
at the implementation level. Each one of these represents a region of discrete space, 
{\it a patch}, which might come equipped with a discrete Cartesian 
coordinate system or discrete charts; that is, invertible maps from discrete space to regions of $Z^3$
(i.e. the integers labeling coordinate grid points in each direction).

Covering a spacetime by charts is commonly done at the continuum level when considering
the differential geometry of the spacetime (see for instance
\cite{wald}). These charts are usually thought of as defining a map
of a portion of the spacetime into a subset of $R^3$, and the combination of charts (which usually overlap in some regions)
covers the whole spacetime. Points belonging to an overlapping region are considered as belonging
to any of the involved charts. Here a well defined coordinate transformation
between the charts is naturally
defined by the combination of the maps and their inverse in between the spacetime and the charts.

At the discrete level one can in principle adopt an analog of the above construction. 
However, it is often the case that in the overlapping region a grid point in one of the charts
does not have a corresponding one in the other. Consequently,  the coordinate transformation is not defined.
This presents a problem at the practical level as communication between patches must take place.
This issue is commonly solved in two different ways: (I) By introducing further points via interpolation
where needed, (II) By considering patches that only {\it abut}, i.e. do not
overlap.

In the first case -- commonly referred to as {\it overlapping-grids approach} -- non-existent points in a given grid within the overlapping region
are defined where needed by appropriate interpolations. Although this can be done
in a straightforward manner with a relatively simple multiple-grid structure, the drawback 
of this approach is the introduction of a new
ingredient -- the interpolation -- which does not have a counterpart at the continuum.
This complicates the assessment of stability of even simple evolution problems as the details of the interpolation
itself are intertwined with any attempt in this direction in an involved manner. As a consequence, 
there exist few stability proofs for such evolution schemes and have so far been restricted to one-dimensional settings. Notwithstanding
this point a number of implementations in Numerical Relativity, 
 where the possible truncation-error driven inconsistencies
at the interfaces are dealt with by introducing a certain amount of
dissipation or filtering, make use of this approach with good results (see for instance,
\cite{overlapping1,overlapping2,overlapping3,overlapping4}). 

In the second case -- commonly referred to as {\it multi-block approach} -- grids are defined in a way such that there is no overlap and
only grid points at boundaries are common to different grids. This requirement translates into
having to define the multiple grids with greater care than in the previous option.
 This extra effort, however,
has as one pay off that schemes preserving important continuum properties can
be constructed. In particular, this allows the construction of stability analyses
which are similar to those of a single grid. More explicitly, following Abarbanel, Carpenter,
Nordstrom, and Gottlieb \cite{penalties0,penalties1,penalties2} one can construct
schemes of arbitrary high order for which semi-discrete energy estimates are straightforward to derive in a
general way. 
The availability of stability results for this second approach makes it a very
attractive option in involved problems --like those typically found when evolving
 Einstein equations-- where schemes eliminating
spurious sources of instabilities provide a strong starting point 
for a stable implementation of the problem.

In this paper we discuss and analyze the use of this multi-block approach 
in the context of Numerical Relativity. At the core of the technique to treat
outer and patch interfaces is the addition of suitable {\it penalty terms} to the
evolution equations  \cite{penalties0,penalties1,penalties2}. 
In the case of hyperbolic systems these terms penalize the possible {\it mismatches} 
between the different values the characteristic fields take
at the interface between several patches.

Not only does this method provide a consistent way to communicate information between the
different patches but, 
more importantly, does so in a way which allows for the derivation of energy estimates at the semi-discrete level.
Consequently, numerical stability can be ensured for a large set of problems.
These estimates can be obtained with difference operators of any accuracy order, provided
they satisfy the summation by parts (SBP) property and the penalty terms are constructed appropriately.

In this work we discuss this technique in a context relevant to numerical relativity,
analyze its properties and illustrate it in specific examples. In particular we show  results for
the case of the $S^2\times R$ topology used in black hole excision techniques.

This work is organized as follows. Section  II includes a description of 
the numerical analysis needed to attain stability in the presence of multiple grids 
and summarize how the penalty method of Refs. 
\cite{penalties0,penalties1,penalties2} allows for achieving this goal.

In Section  III we study some aspects of Strand's \cite{strand} high order operators satisfying SBP
 with respect to diagonal norms, when combined with the penalty technique. We find that 
in some cases, typically used operators that minimize the bandwidth have a very
 large spectral radius, with corresponding 
limitations in the Courant-Friedrich-Levy (CFL) factor when used in evolution equations. We 
therefore construct operators that minimize 
the spectral radius instead.  Additionally, we examine the behavior of the convergence rate and 
the propagation behavior that different modes have when employing different higher order operators. 

In Section  IV we present and analyze different tests relevant to numerical
relativity employing derivative operators of different order of accuracy and the
penalty technique to deal with multiple grids. These tests cover from linearized Einstein equations
(in effectively one-dimensional scenarios) to propagation of three-dimensional fields in curved backgrounds.

We defer to appendices the discussion of several issues. Appendix A presents details of
the higher derivative operators and diagonal norms which we employ in this work. Appendix B discusses
our construction of high order dissipative operators which are negative definite with
respect to the corresponding SBP scalar product. Last, Appendix C lists some useful properties
that finite difference derivative operator satisfy, which help in our
construction of dissipative operators.

\section{Interface treatment for symmetric hyperbolic problems in multiple blocks} \label{theory}
As  mentioned, we are interested in setting up a computational
domain which consists of several grids which just abut. 
This domain provides the basic arena on which symmetric hyperbolic
systems are to be numerically implemented. The basic strategy is
to discretize the equations at each individual grid or block, treating
boundary points in a suitable way. Boundary points at each
grid either represent true boundary ones from a global perspective or
lie at the interface between grids. In the latter case, since these points are 
common to more than one grid the solution at them can be regarded
as multi-valued. As we show below, this issue can be dealt with consistently
and stably, ensuring that any possible mismatch converges to zero with resolution. 

At the core of the technique is the appropriate communication of 
these possibly different values of the solution at the interfaces.
Intuitively, since we are dealing with symmetric
hyperbolic systems, a natural approach would be to communicate the
characteristic variables from one domain to the other one. However, this is
not known to be numerically stable. There exists nonetheless a technique based on this strategy
which does guarantee numerical stability \cite{penalties0,penalties1,penalties2}.  This relies in 
adding {\it penalty terms} to the evolution equations of characteristic fields which penalize: 
a) in the interface case the mismatch  between the different values each characteristic field takes at
the interface of several grids; b) in the outer boundary case the difference
between each incoming characteristic field and the boundary conditions one wishes to impose to it.

These penalty terms are constructed so as to guarantee the stability of the
whole composite 
grid if it can be guaranteed at each individual grid through the energy method. To this end, the use of schemes
with difference operators satisfying SBP are employed. Hence, on each single grid there
exists a family of natural semidiscrete energies, defined by both a symmetrizer of the
continuum equations {\em and} a discrete scalar product with respect to which
SBP holds \footnote{Symmetrizers are not unique, already at the
continuum level. In the case in which
there is a preferred one there is, similarly, a single, preferred
semidiscrete energy on each grid. This is the case, for example, when one can
derive a sharp energy estimate at the continuum, which gives rise to the
construction of a strictly stable scheme,  see \cite{strict} for a detailed
discussion in the context of Numerical Relativity.}.  
One can then define an energy for the whole domain by simply adding the different energies of each grid. 
The use of operators satisfying SBP allows one to get an energy estimate, up to 
outer boundary and interface terms left after SBP. The penalties 
allow to control their contribution, thus obtaining an 
estimate for the global grid.  
This is achieved if the contribution to the time derivative of the energy due
to the  interface and outer boundary terms (in
the latter case when, say, homogenous maximally dissipative outer boundary
conditions are imposed) left after SBP is non-positive. 
When these terms are exactly zero the penalty treatment of Refs.
\cite{penalties0,penalties1,penalties2} 
is, in a precise sense, 
``energy non-dissipative''. On the other hand, if these terms are negative 
the scheme is numerically stable but at
fixed resolution a damping of the energy (with respect to the growth one would
obtain in the absence of an interface) in time arises. 
This damping is proportional to either the mismatch of a given characteristic variable at each
interface or its failure to  satisfy an outer
boundary condition. As we describe below, these interface and boundary terms
left after SBP are controlled precisely
by the mentioned penalties, each of which depends on: {\it the possible
  mismatch; the characteristic speeds, the corresponding SBP scalar product at the interface, 
the resolution at each intervening grid, and a free parameter which regulates
the strength of the penalties}.

Next, we explicitly describe how this penalty technique allows one to derive semidiscrete
energy estimates. We first discuss in detail the one-dimensional
example of an advection equation on a domain with a single interface. The more
general case of systems of equations in several dimensions follows essentially
the same principles, applying the $1$d treatment to each characteristic
field. We illustrate this by discussing a general
constant-coefficient system in a given two-dimensional setting. From this, the
generalization to the three-dimensional general case is straightforward, and we
therefore only highlight its salient features. 

\subsection{A one-dimensional example}
Consider a computational domain represented by a discrete grid consisting
of points $i= i_{min}\ldots i_{max}$ and gridspacing $h$ covering $x \in
[a,b]$.  A $1$d difference
operator $D$ on such a domain is 
said to satisfy SBP with respect to a positive definite scalar product (defined by its coefficients 
$\sigma_{ij}$)
\begin{equation}
\langle u,v \rangle = h\sum_{i,j} u_i v_j \sigma_{ij} \label{prod} \, ,
\end{equation}
if the property 
$$
\langle u,Dv \rangle + \langle v,Du \rangle = \left(uv\right)|_a^b 
$$
holds for all gridfunctions $u,v$. The scalar product/norm is said to be {\it
 diagonal} if $\sigma_{ij}= \sigma_{ii} \delta_{i,j}$\footnote{Here $\delta_{i,j}$ is 
the Kronecker delta ($\delta_{i,j}=1$ if $i=j$ and zero otherwise)}. 
One advantage of $1$d difference operators satisfying SBP wrt 
diagonal norms is that SBP is guaranteed to hold in several dimensions if the $1$d operator is 
used on each direction 
(which is not known to hold in the non-diagonal case in general)
\cite{olsson}. Even in $1$d, in the variable coefficients and non-diagonal case the commutator
between $D$ and the principal part might not be bounded for all resolutions
(something that is {\it generically} \footnote{That is, unless the scheme can be
written in strictly stable form.}  needed for an energy estimate to hold) \cite{svard}. Another advantage is that 
the operators are, for a given order in the interior, simpler in their expressions. 
The disadvantage is that their order 
at and close to boundaries is half that one in the interior, while in the
non-diagonal case the operators loose only one order with respect to the
interior \cite{sbp_order,strand}.  
Throughout this paper we will mostly restrict our treatment to the
use of diagonal norms. \\

As an example of how to impose interface or outer boundary conditions through
penalty terms, we concentrate next on the advection equation for $u$ propagating
with speed $\Lambda$,
\be
\dot{u} = \Lambda \partial_x u   \;. \label{adv} 
\ee

\subsubsection{A domain with an interface}
Consider the interval $(-\infty,\infty)$ with appropriate fall-off conditions at
infinity. We consider two grids: a {\it left} one covering $(-\infty,0]$, and 
a {\it right} one covering $[0,\infty)$. We refer to the gridfunction $u$ on each grid by
$u^l$ and $u^r$ (corresponding to the left and right grids, respectively).
Both of these
gridfunctions have a point defined at the  $x=0$ interface and they need not
coincide there, 
except at the initial time. Therefore, the numerical solution will in principle be multivalued at $x=0$,
though, as we shall see, the penalty technique is designed to keep this difference small. 

The problem is discretized using on the right and left grids, 
respectively, with gridspacings $h^l,h^r$ --not necessarily equal-- and 
difference operators $D^l,D^r$ satisfying SBP with respect to scalar products
given by the weights $\sigma^l, \sigma^r$ {\it at their individual grids}. 
That is, these scalar products are defined through 
$$
\langle u^l,v^l \rangle =  h^l\sum_{i,j=-\infty}^{0} \sigma^l_{ij} u^l_i v^l_j  \;\;\; , \;\;\;
\langle u^r,v^r \rangle =  h^r\sum_{i,j=0}^{\infty} \sigma^l_{ij}  u^r_i v^r_j 
$$

The semidiscrete equations are written as
\begin{eqnarray}
\dot{u}^l_i &=& \Lambda D^l u^l_i + \frac{\delta_{i,0} S^l}{h^l\sigma^l_{00}}(u^r_0-u^l_0) \label{advl} \, ,\\
\dot{u}^r_i &=& \Lambda D^r u^r_i + \frac{\delta_{i,0} S^r}{h^r\sigma^r_{00}}(u^l_0-u^r_0) \label{advr} \, .
\end{eqnarray}
Notice in the above equations the second term on each right
hand side, which constitutes the penalty added to the problem. They are defined by the possible mismatch,
the grid-spacing, the inner product employed and the free parameters $\{S^l,S^r\}$, which will
be determined by requiring an energy estimate to hold.

We define a natural energy for the whole domain, which is the sum of the energies 
for each grid (for this simple example with a trivial symmetrizer), 
\[
E := \langle u^l,u^l \rangle + \langle u^r,u^r \rangle \, .
\]
Taking a time derivative of this energy, using the semidiscrete equations 
(\ref{advl},\ref{advr}) and the SBP property one gets
\begin{equation}
E_t = (\Lambda - 2S^l) (u^l_0)^2 + (- \Lambda - 2S^r) (u^r_0)^2 + 2(S^l+S^r )
u^l_0 u^r_0  \label{term} \, .
\end{equation}
In order to get an energy estimate the above interface term (i.e., 
the right-hand side of Eq. (\ref{term})) must be non-positive for all 
$u^l_0,u^r_0$. It is straightforward to check that this is equivalent to the three following
conditions holding: 
\begin{eqnarray}
\Lambda - 2S_l & \leq & 0 \\
- \Lambda - 2S_r & \leq & 0 \\
(\Lambda + S_r - S_l)^2 &\leq & 0
\end{eqnarray}
From there, it is clear that we need $\Lambda + S_r - S_l = 0$. And with this condition the
other two become $S_l + S_r \geq 0$. There are three possibilities:

\begin{itemize}

\item{\em Positive $\Lambda$:} we can take 
\be
S_l = \Lambda + \delta, \;\;\; S_r = \delta, \;\;\; \mbox{ with } \delta \geq
- \frac{\Lambda}{2} \label{pos_lam}
\ee
The time derivative of the energy with this choice becomes
$$
E_t = -(u_0^l-u^r_0)^2(\Lambda + 2 \delta) \leq 0
$$
\item {\em Negative $\Lambda$:} similarly, we can take 
\be
S_r = -\Lambda + \delta, \;\;\; S_l = \delta, \;\;\; \mbox{ with } \delta \geq
 \frac{\Lambda}{2} \label{neg_lam}
\ee
The corresponding time derivative of the energy with this choice becomes
$$
E_t = (u_0^l-u^r_0)^2(\Lambda - 2 \delta) \leq 0
$$
\item {\em Vanishing $\Lambda$}: this can be seen as the limiting case of any of the above two, 
and we can take
\be
S_l = S_r= \delta,\;\;\; \mbox{ with } \delta \geq0 \; . \label{zero_lam}
\ee
Hence,
$$
E_t = -(u_0^l-u^r_0)^2 2 \delta \leq 0
$$
\end{itemize}
The coefficients $S_l$ and $S_r$ need not be equal, but the following symmetry is important, under the 
change $\Lambda \to -\Lambda$ we should have $S_l \to S_r$ and vice-versa, since it transforms incoming
modes to outgoing ones. This is clearly satisfied by the choice above.\\

Summarizing, there is a freedom in the penalty factors of Eqs.(\ref{advl},\ref{advr}), 
encoded in the parameter $\delta$, which has to satisfy $\delta \ge
-|\Lambda|/2$. If $\delta = -|\Lambda/2|$ there is no interface term in
the estimate (that
is, exact energy conservation in the above model), while if  If
$\delta > -|\Lambda/2|$ there is a negative definite interface term in
the estimate (which represents a damping in the energy proportional to the mismatch). 

As we will see below, one proceeds similarly in the more general case of
systems of equations in several dimensions. The penalty terms are applied to
the evolution equation of 
each characteristic mode, with factors given by
Eqs.(\ref{pos_lam},\ref{neg_lam},\ref{zero_lam}) (where $\Lambda$ in the
general case is the corresponding characteristic speed).

\subsubsection{A domain with an outer boundary}

The penalty method also allows to treat outer boundaries in a similar way. As an example,
consider again the advective equation Eq.(\ref{adv}), but now on the domain
$(-\infty, 0]$, 
with gridpoints $i=-\infty \ldots 0$. Assume $\Lambda > 0$; boundary
conditions therefore need to be given at $x=0$; say $u(x=0,t)=g$. 
The semidiscrete equations are written as 
\begin{eqnarray}
\dot{u}_i &=& \Lambda D u_i + \frac{\delta_{i,0}T}{h\sigma_{00}} (g - u_0) \label{adv_outer} \, . \\
\end{eqnarray}
Defining the energy to be
$$
E = \langle u,u \rangle, 
$$
its time derivative is 
\begin{eqnarray}
\dot{E} &=& (\Lambda - 2T)u_0^2 + 2 g u_0 T \label{simpleestimate1} \, ,\\
        &\leq& (\Lambda - T ) u_0^2 +  T g^2 \label{simpleestimate2} \, .
\end{eqnarray}
As in the interface case with positive speed [c.f. Eq.(\ref{pos_lam})], 
we can therefore take $T=\Lambda + \delta$. For the homogeneous case, $g=0$,
the equality (\ref{simpleestimate1}) holds and we have
$$
\dot{E} = (\Lambda - 2T)u_0^2 \, ,
$$
indicating that for $\delta \geq -\Lambda/2$ the energy will not increase.
For the non-homogeneous case ($g\neq 0$) the inequality (\ref{simpleestimate2}) yields
$$
\dot{E} \leq \Lambda g^2 + \delta (g^2 - u_0^2) \, ;
$$
Note that for $\delta=0$ one trivially recovers the continuum estimate. For other values of
$\delta$ the consistency with the continuum estimate follows from the observation that
$u_0$ converges to $g$ to the $s+1$-th order if the SBP derivative operator has accuracy $s$ at
the boundary point~\cite{olivierthanks}. This implies that both the numerical implementation
and the corresponding energy estimate are consistent with those defined at the continuum level.

At this point we find it important to remark the following.
Notice that just having an energy inequality is not enough, as one further
needs to ensure consistency of the discrete equations with respect to the continuum ones.
In the case of the penalty approach this not straightforward as 
the penalty term diverges when the grid size decreases unless
$u$ converges to $g$ sufficiently fast (as mentioned above, $u$ does converge
to $g$ fast enough if $u$ is an incoming mode). In fact, the penalty term can be viewed at the continuum as approximating
the original equation and boundary conditions through the introduction of a suitable 
delta function at the boundary. 
The argument of the delta function must be consistent with the underlying problem.
For instance, if one were to put a penalty term on a boundary where the mode
is outgoing --and thus the value
of the function there is determined by the evolution itself-- the inconsistency
would manifest itself through a lack of convergence. In the above example 
this would be the case if we take $\Lambda < 0$
but insist in putting a penalty term at $x=0$. Note that in such situation the
energy inequality would still hold if 
$T \geq \Lambda/2$. This would imply that the numerical solution is still bounded in the $L^2$ norm,
 but no more than that; indeed, numerical experiments show that a high frequency solution 
traveling in the incoming direction (that is, with velocity opposite to that
one at the continuum, see Section  \ref{groupspeed}), whose amplitude depends on the size of
the penalty term, is generated.
Naturally, if $T < \Lambda/2$ 
the energy inequality is violated and the solution 
blows up exponentially with a rate increasing with the highest frequency that can be accommodated by
the grid being employed.

\subsection{The two-dimensional case}
\begin{figure}[ht]
\begin{center}
\includegraphics*[height=4cm]{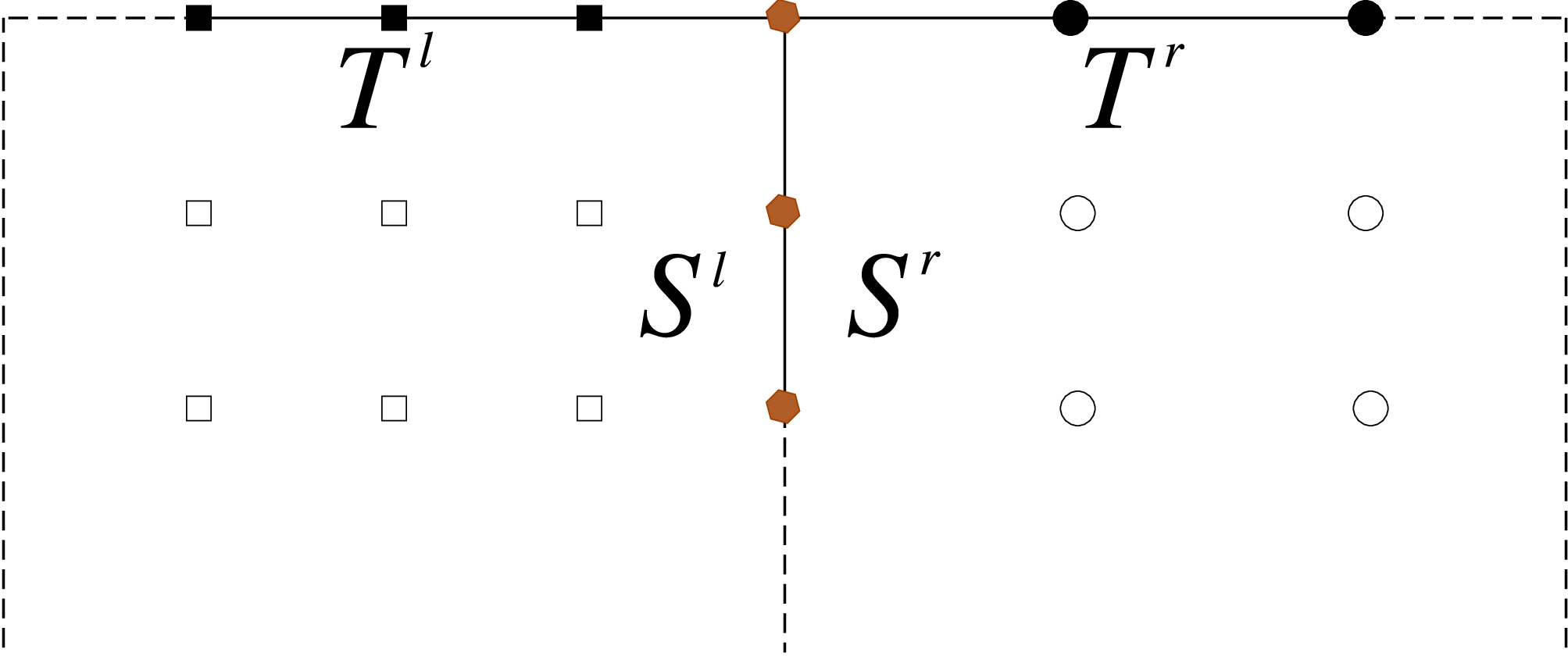}
\caption{Example of a multi-block domain in two dimensions. Only the spacing in the
vertical direction need be the same in both grids so as to ensure boundary points --represented by grey hexagons-- coincide.}.
\label{2d} 
\end{center}
\end{figure}

Consider now the system of equations
\[
\dot{u} = A^{\mu} \partial_{\mu} u = A^x\partial_x u + A^y \partial_y u  \, ,
\]
where $u$ is a vector-valued function, $A^x,A^y$ are symmetric and, for simplicity, constant coefficient
matrices. The domain is composed of two grids: a {\it left} one covering $
(-\infty,0]\times (
-\infty,0]$ and
 a {\it right} one covering 
$[0,\infty)\times(-\infty,0]$, with an interface at $x=0$ and an
outer boundary at $y=0$ (see Figure \ref{2d}). At $y=0$ we impose homogeneous maximally dissipative boundary
conditions by setting to zero the incoming characteristic fields. 

We assume that the scalar products associated with the different $1$d difference
operators are diagonal. This, as already mentioned, ensures certain properties that guarantee an energy estimate. 

We denote the gridspacings, difference operators and
associated scalar products corresponding to the $x,y$ directions by $h_x,h_y$,
$D_x,D_y$, and $\sigma_{(x)},\sigma_{(y)}$, respectively. These quantities
need not coincide on the different subdomains, except for the gridspacing in
the transversal direction at an interface (so that 
gridpoints belonging to different grids align with each other). 
To all quantities we add an $l$ or $r$ supraindex 
to denote quantities belonging to the left or right
domain. Thus, the only condition we require is $h_y^l =
h_y^r =: h_y$. 

On each subdomain the $2$d scalar product is defined as the product of the
scalar product on each direction,
\begin{equation}
\langle u,v \rangle = h_xh_y\sum_{i,j} (u_{ij}, v_{ij}) \sigma_{(x)i} \sigma_{(y)j} \label{prod2d}
\end{equation}
where $(u,v)$ is the pointwise Euclidean scalar product of two vectors.

As in $1$d, the total energy is defined as the sum of the energies on each subdomain
(in this case with a trivial symmetrizer, as the equations are already in
symmetric form):
$$
E = E^l + E^r \, ,
$$
where 
\begin{eqnarray}
E^l &=& h^l_xh_y\sum_{i\le 0}\sum_{j\le 0} (u^l_{ij}, u^l_{ij})
\sigma^l_{(x)i} \sigma^l_{(y)j} \label{el} \, , \\
E^r &=& h^r_xh_y\sum_{i\ge 0}\sum_{j\le 0} (u^r_{ij}, u^r_{ij})
\sigma^r_{(x)i} \sigma^r_{(y)j} \label{er} \, .
\end{eqnarray}

The evolution equations are a combination of Eqs.(\ref{advl},\ref{advr}) and Eq.(\ref{adv_outer}),
\begin{eqnarray}
\dot{u}_{ij}^l &=& A^{\mu} D^l_{\mu} u^l_{ij} + 
\frac{\delta_{i,0}S^l}{h_x^l\sigma^l_{(x)i=0}}(u^r_{0j} - u^l_{0j})  -
\frac{T^l}{h_y^l\sigma^l_{(y)j=0}} u^l_{i0} \label{udot_l} \\
\dot{u}_{ij}^r &=& A^{\mu} D^r_{\mu} u^r_{ij} + 
\frac{\delta_{i,0}S^r}{h_x^r\sigma^r_{(x)i=0}}(u^l_{0j} - u^r_{0j})  -
\frac{T^r}{h_y^r\sigma^r_{(y)j=0}} u^r_{i0} \label{udot_r}
\end{eqnarray}
In the above expressions, $S^l,S^r,T^l,T^r$ are operators (as opposed to
scalars), since we are dealing with a system of equations. The first two
correspond to penalty terms added to handle grid interfaces while
the latter two for imposing outer boundary conditions.
The goal of these
operators is to transform to characteristic variables and apply to the
evolution equation of each characteristic mode suitable penalty terms, as in 
Eqs.(\ref{advl},\ref{advr},\ref{adv_outer}). 

Taking a time derivative of the energies defined in Eq.(\ref{el},\ref{er}),
using the evolution equations (\ref{udot_l},\ref{udot_r}),  and
employing the SBP property along each direction, one gets
\begin{eqnarray}
\dot{E}^l&=&  h_y\sum_{j\le0} \sigma^l_{(y)j}\left[
  (u^l_{0,j},(A^x-2S^l)u^l_{0,j}) + 2(u^l_{0,j},S^lu^r_{0,j}) \right] +  
h_x^l\sum_{i\le0}\sigma_{(x)i}^l(u^l_{i,0},(A^y-2T^l)u^l_{i,0}) \\
\dot{E}^r&=&  h_y\sum_{j\le0} \sigma^r_{(y)j}\left[
  (u^r_{0,j},(-A^x-2S^r)u^r_{0,j}) + 2(u^r_{0,j},S^ru^l_{0,j}) \right] +
h_x^r\sum_{i\ge0}\sigma_{(x)i}^r(u^r_{i,0},(A^y-2T^r)u^r_{i,0}) 
\end{eqnarray}
where we have assumed that $S$ and $T$ are hermitian matrices. 

In order to control the interface terms in $\dot{E}^l+\dot{E}^r$ we can take
\[
S^l = 
\frac{1}{\sigma^l_{(y)j}} \left[ (\Lambda_a^+ +\delta^+_a) P_+^a + \delta^-_a
  P_-^a  + \delta^0 P_0 \right] \, ,
\]
\[
S^r = 
 \frac{1}{\sigma^r_{(y)j}} \left[ (-\Lambda_a^- +\delta^-_a) P_-^a + \delta^+_a
  P_+^a  + \delta^0 P_0 \right] \, ;
\]
where a sum over the index $a$ is assumed,
and $\{ P_+^a,P_-^a,P_0 \}$ are projectors to the sub-spaces of eigenvectors of $A^x$ with
eigenvalues $\{ \Lambda_a^+,\Lambda_a^-,\Lambda^0 \}$ respectively. With this choice
$\dot{E}^l+\dot{E}^r$ becomes 
\begin{eqnarray}
\dot{E}^l+\dot{E}^r &=& h_y\sum_{j\leq 0} \left[  (\Lambda^-_a -
2\delta^-_a)||u_-^{a,l}-u_-^{a,r}||^2 
-  (\Lambda^+_a +
2\delta^+_a)||u_+^{a,l}-u_+^{a,r}||^2 
- 2\delta^0 ||u_0^{l}-u_0^{r}||^2 \right] \\
& +& h_x^l\sum_{i\le0}\sigma_{(x)i}^l(u^l_{i,0},(A^y-2P^l)u^l_{i,0}) + 
h_x^r\sum_{i\ge0}\sigma_{(x)i}^r(u^r_{i,0},(A^y-2P^r)u^r_{i,0})  \, .
\end{eqnarray}
Clearly, in order to obtain an estimate, the following conditions must be satisfied, 
$$
\Lambda^-_a - 2\delta^-_a \leq 0 , \;\;\; \Lambda^+_a +
2\delta^+_a \geq 0, \;\;\; \delta^0 \geq0 \, ;
$$
which is analogous to the one-dimensional case,
Eqs.(\ref{pos_lam},\ref{neg_lam},\ref{zero_lam}). Similarly, 
the outer boundary terms in $\dot{E}^l+\dot{E}^r$ (i.e. the sums over $i$) can
be controlled on each domain separately. We need 
\[
u^r(A^y - 2T^r)u^r \leq 0 \, ,
\]

\[
u^l(A^y - 2T^l)u^l \leq 0 \, .
\]
We can therefore take, as in the one-dimensional case, 
\[
P^r = P^l = 
 (\Lambda_a^+ +\delta_a^+)P^+_a \ ,
\]
where now $P_+^a$ are projectors to the spaces of eigenvectors of $A^y$ of
eigenvalues $\Lambda_a^+$. With these
choices the final expression for the time derivative of the energy is
\begin{eqnarray}
\dot{E}^l+\dot{E}^r &=& h_y\sum_{j\leq 0} \left[  (\Lambda^-_a -
2\delta^-_a)||u_-^{a,l}-u_-^{a,r}||^2 
-  (\Lambda^+_a +
2\delta^+_a)||u_+^{a,l}-u_+^{a,r}||^2 
- 2\delta^0 ||u_0^{l}-u_0^{r}||^2 \right] \\
& +& h_x^l\sum_{i\ge0}\sigma_{(x)i}^l (-\Lambda^-_a -
2\delta^-_a)||u_-^{a,l}||^2 + h_x^r\sum_{i\ge0}\sigma_{(x)i}^r   (-\Lambda^-_a -
2\delta^-_a)||u_-^{a,r}||^2 
\end{eqnarray}
and, again, the possible ranges for the different $\delta$'s are as in the $1$d
case, Eqs.(\ref{pos_lam},\ref{neg_lam},\ref{zero_lam}).

Notice that nothing special has to be done at a corner, as each direction is
treated and controlled independently. 

\subsection{The general case}

The general case follows the same rules. Namely, 
 we must add penalty terms on the characteristic modes corresponding to each
of the boundary matrices separately and accordingly. 

For example, what to do at the vertices of three patches meeting in the
cubed-sphere case discussed later in this paper? 
As we will see, there we have three meshes with coordinates (at a constant
radius) $(a^1,b^1)$,  $(a^2,b^2)$, and $(a^3,b^3)$
arranged in a clockwise distribution according to the indices, and
intersecting at a point. In that case, 
the contribution to the energy (without the penalty terms added to the
evolution equations) is proportional to

\[
(u^1_{0N},(A^{a^1}+A^{b^1})u^1_{0N}) + (u^2_{00},(A^{a^2}+A^{b^2})u^2_{00}) 
+ (u^3_{0N},(A^{a^3}+A^{b^3})u^3_{0N}) 
\]

Since the interfaces are aligned to the grids we know that the normals coincide on both
sides, therefore we have:

\[
A^{a^1} = A^{a^3} \;\;\;\; A^{b^1} = A^{b^2} \;\;\;\; A^{b^3} = A^{a^2}
\]

So we include penalty terms on each side, including the end-points
of the grids in each direction, (which constitute vertices and edges). Note that the
characteristic modes at these points are computed with the normal with respect to the side that
contains this direction. Consequently, points at edges/vertices of a
(topologically) cubical grid will have
two/three penalty terms.

\section{High order difference operators with diagonal norms}
\label{HO}

In this section we analyze some aspects of Strand's $1$d difference 
operators satisfying SBP with respect to diagonal metrics, when used in
conjuction with the penalty technique to construct high order schemes for handling domains
with interfaces. 

In particular, we discuss operators with accuracy of 
order two, four, six and eight at interior points. The requirement
that these operators satisfy the SBP property with respect to diagonal norms implies that their respective
accuracy order at and close to boundaries is one, two, three and four, respectively. 
We will therefore 
refer to these operators as $D_{2-1},D_{4-2},D_{6-3}$, and $D_{8-4}$. 
Some of these operators are not unique, as the accuracy order and SBP requirements still 
leave in some cases additional freedom in their construction. Indeed, while the
 first two operators ($D_{2-1},D_{4-2}$) are unique, the  
$D_{6-3}$ one comprises a mono-parametric family, and  $D_{8-4}$ a three-parametric one. This
freedom can be exploited for several purposes. For instance, to minimize
the operator's bandwidth or its spectral radius. While the former produces operators
which are more compact, the latter can have a significant impact on the CFL
limit when dealing with evolution equations. Indeed, for the $D_{8-4}$ case, minimizing its bandwidth 
leads to a considerably larger spectral radius (though this does
not happen in the $D_{6-3}$ case) which, in turns, requires one to employ a rather small
CFL factor for the fully discrete scheme to be stable.

To analyze this in each case, we numerically solve and discuss the eigenvalues of the amplification 
matrix of the advection equation with speed one, $u_t = u'$, under periodic boundary conditions.
The periodicity is imposed  through an interface with penalty terms and hence the scheme does depend on
the penalty parameter $\delta$ and so will the discrete eigenvalues obtained. As discussed in Section  \ref{theory},
in the case in which $\delta = -1/2$, SBP holds across
the interface, and the energy for this model is strictly conserved. In
other words, the amplification matrix is anti-symmetric and the eigenvalues are purely
imaginary (see Section  \ref{theory}). On the other hand, if $\delta > -1/2$ 
there is a negative definite interface term left after
SBP, and a negative real component in the eigenvalues must
appear in the spectrum of the amplification matrix (see Section  \ref{theory}). 

We additionally discuss the global convergence factor for these operators, and recall a
feature associated with the mode with highest possible group speed at a given
number of gridpoints. Namely, that it
travels in the ``wrong'' direction, and that the absolute value of its speed
increases considerably with the order of the operator. 

Appendix \ref{appendix_der} lists, for completeness, some typos in
Ref. \cite{strand} in some of the coefficients for these high order
operators. 

\subsection{Spectrum}
In the following we discuss the range of discrete eigenvalues obtained for the different
derivative operators and their dependence on $\delta$. We pay particular attention on
the impact different values of $\delta$ and the chosen derivative operator
have on the CFL limit. 

\subsubsection{Second order in the interior, first order at 
  boundaries ($D_{2-1}$ scheme)}

\begin{figure}[ht]
\begin{center}
\includegraphics*[height=7cm]{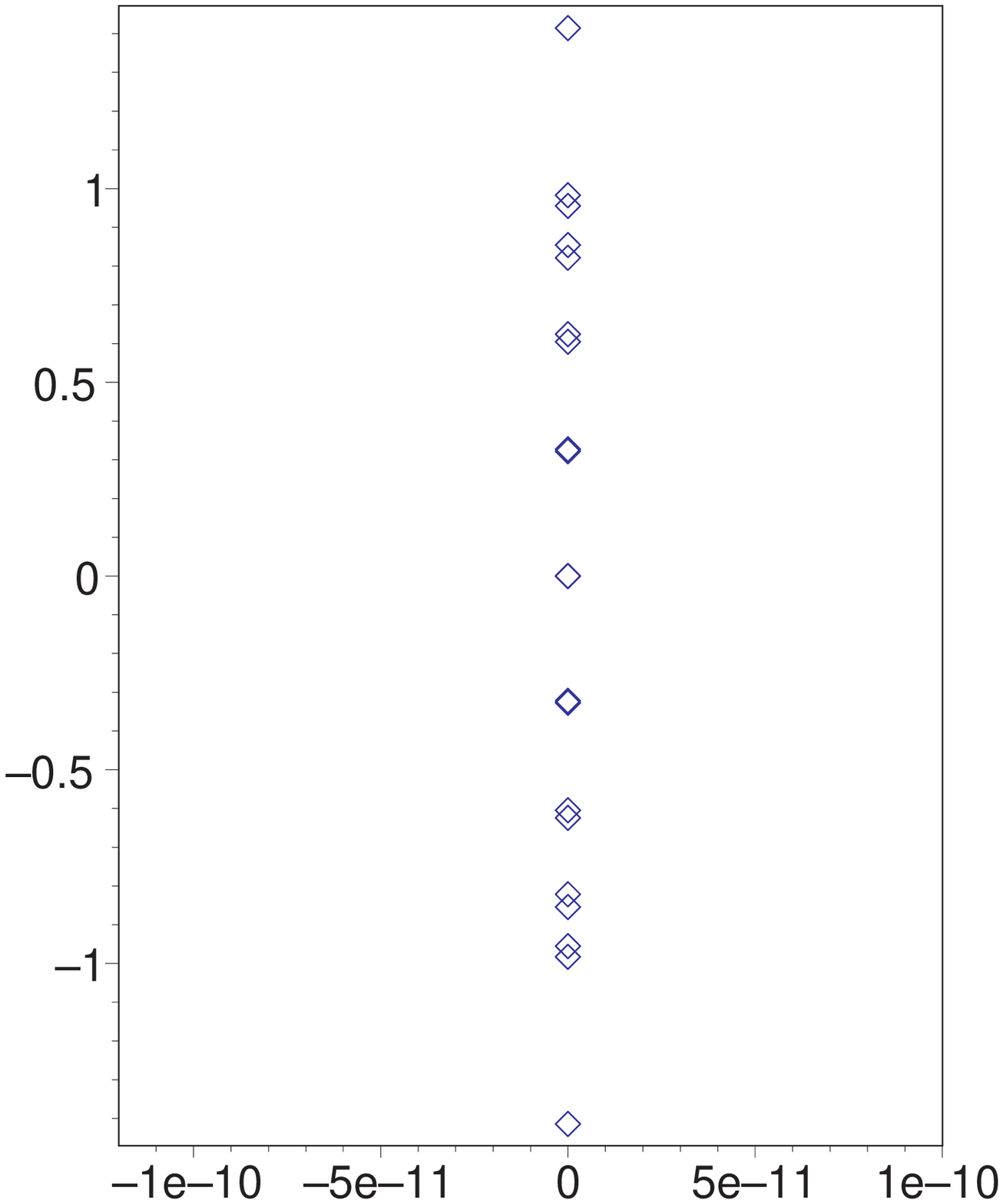}
\includegraphics*[height=7cm]{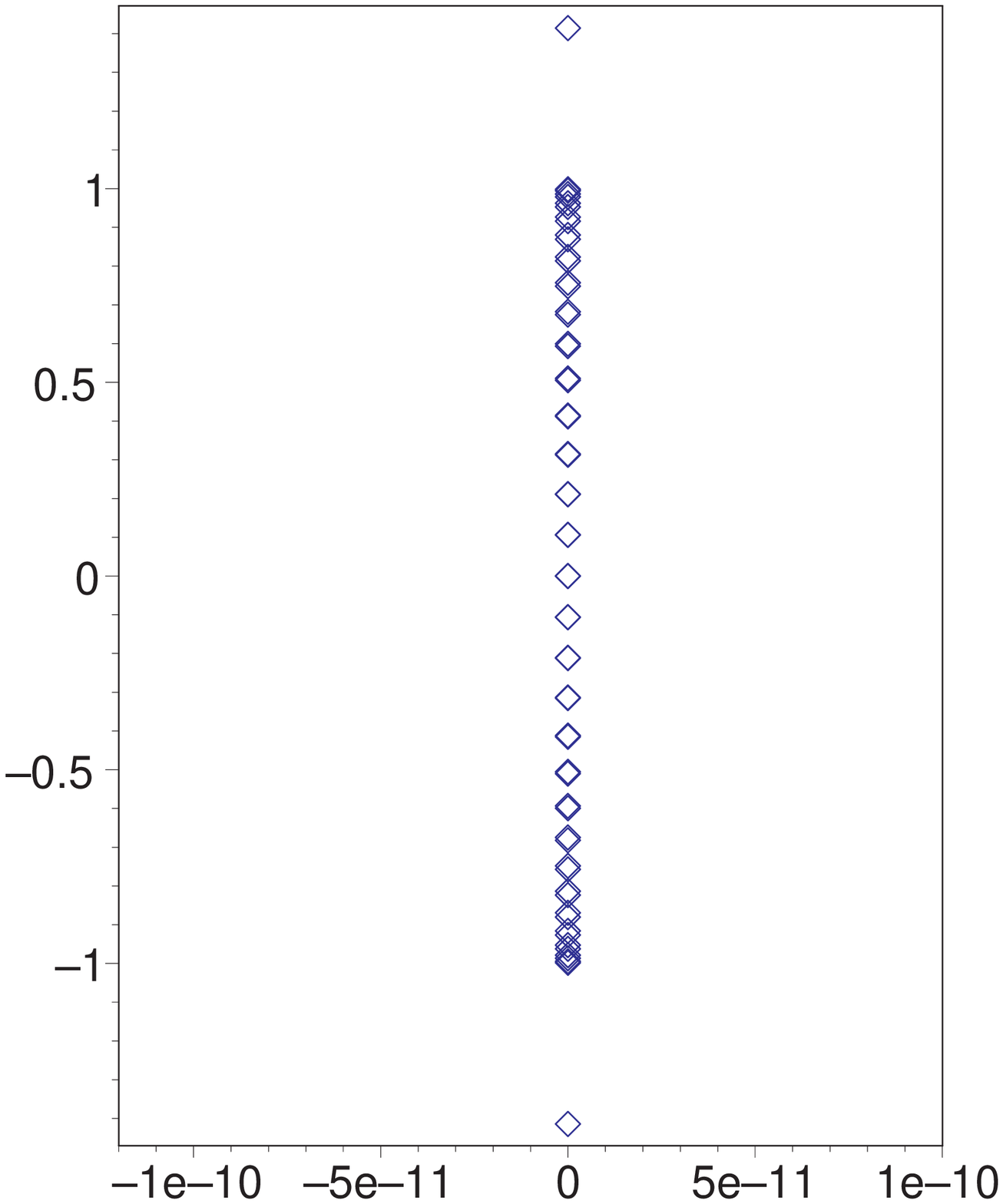}
\includegraphics*[height=7cm]{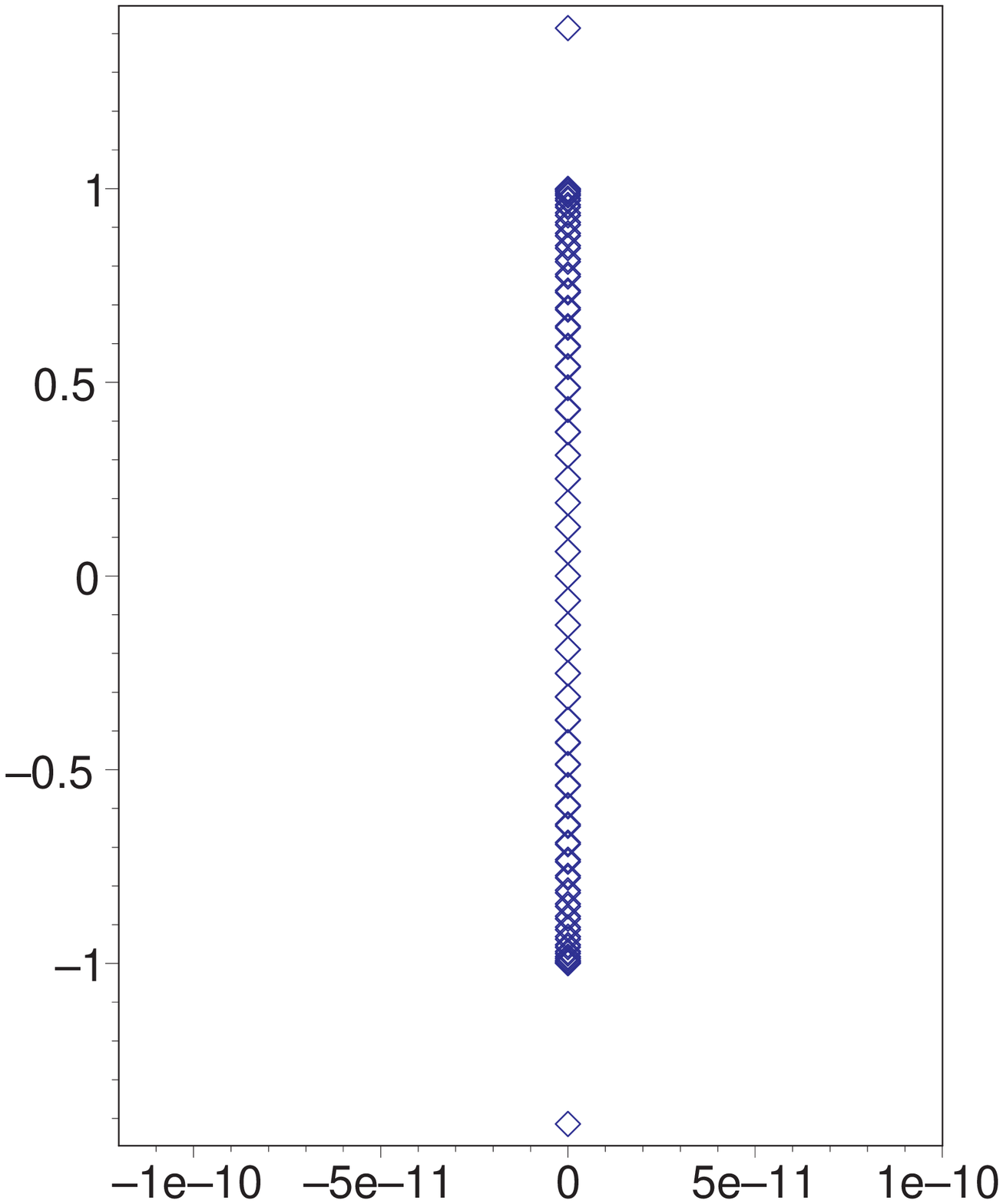}
\caption{Numerically obtained eigenvalues (in the complex plane) corresponding to the $D_{2-1}$ 
operator for $\delta =  -1/2$ (purely imaginary case). From left to right the plots
illustrate the results obtained with a grid containing $20, 60, 100$ points respectively.
It is clear from the figures that these correspond, indeed, to a purely imaginary case.}
\label{eigenvalues2} 
\end{center}
\end{figure}

Figure \ref{eigenvalues2} shows the eigenvalues obtained using $20,60,100$ gridpoints 
and penalty term $\delta = -1/2$ (that is, the purely
imaginary spectrum case, see Section  \ref{theory}) for the $D_{2-1}$ case. 
The maximum and minimum values are, approximately, 
$\pm 1.414$, and they seem to be related to the operator near the boundary, as
their absolute value does not seem to increase with the number of points 
(instead, the region between the maximum 
and minimum is filled out). As discussed below, in the higher order cases the
maximum eigenvalues also seem to be related to the operator near the boundary.  

\begin{figure}[ht]
\begin{center}
\includegraphics*[height=7cm]{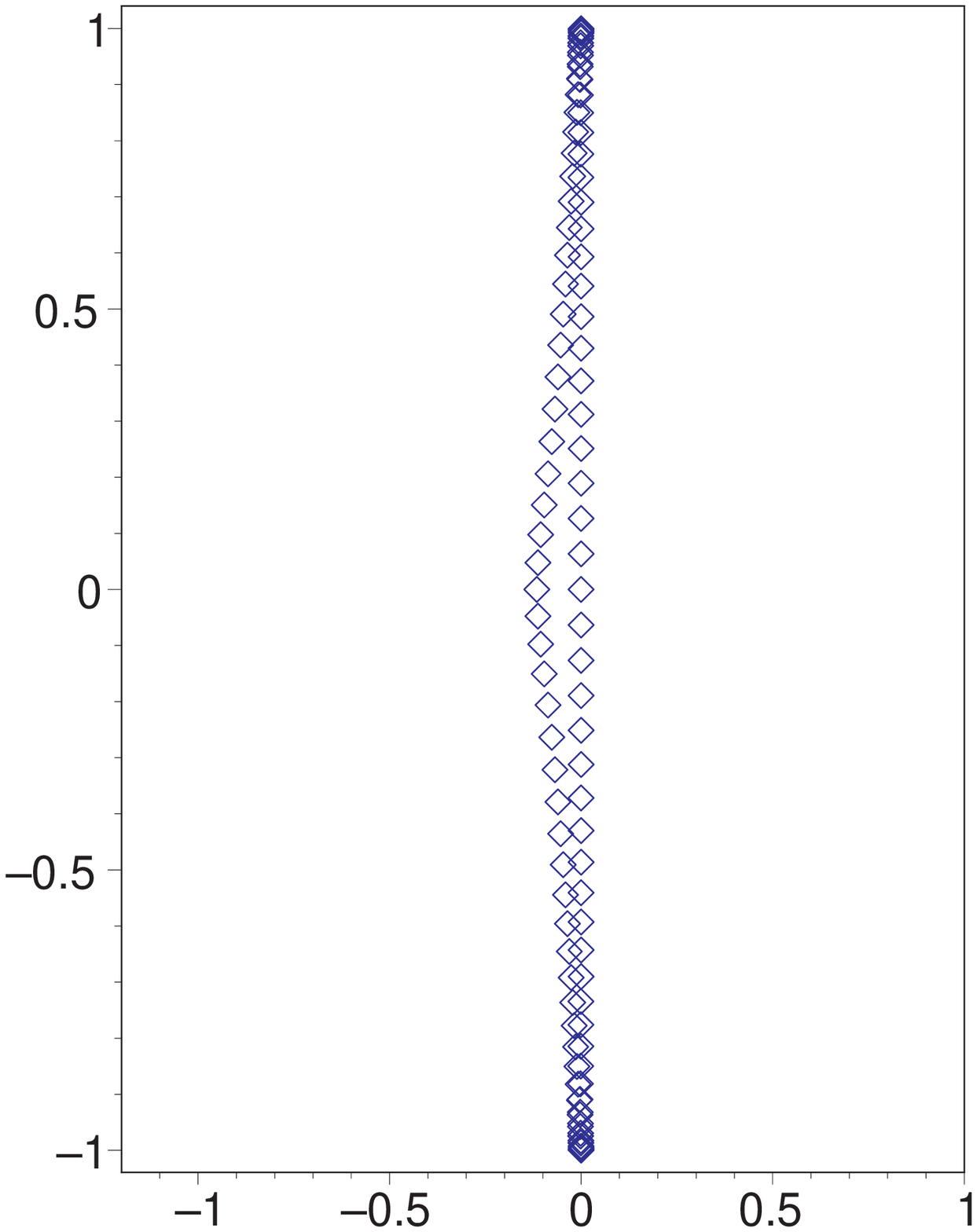}
\includegraphics*[height=7cm]{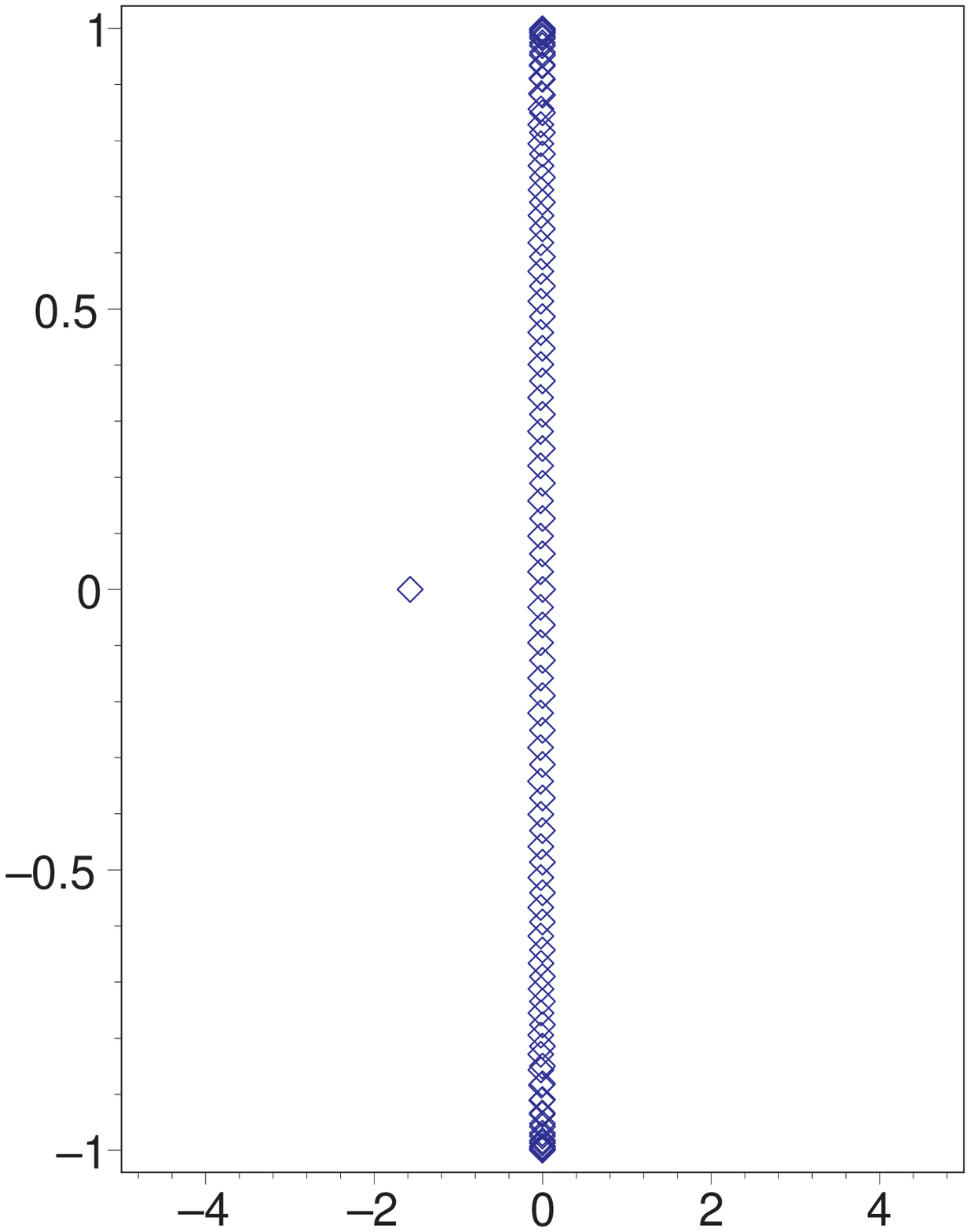}
\includegraphics*[height=7cm]{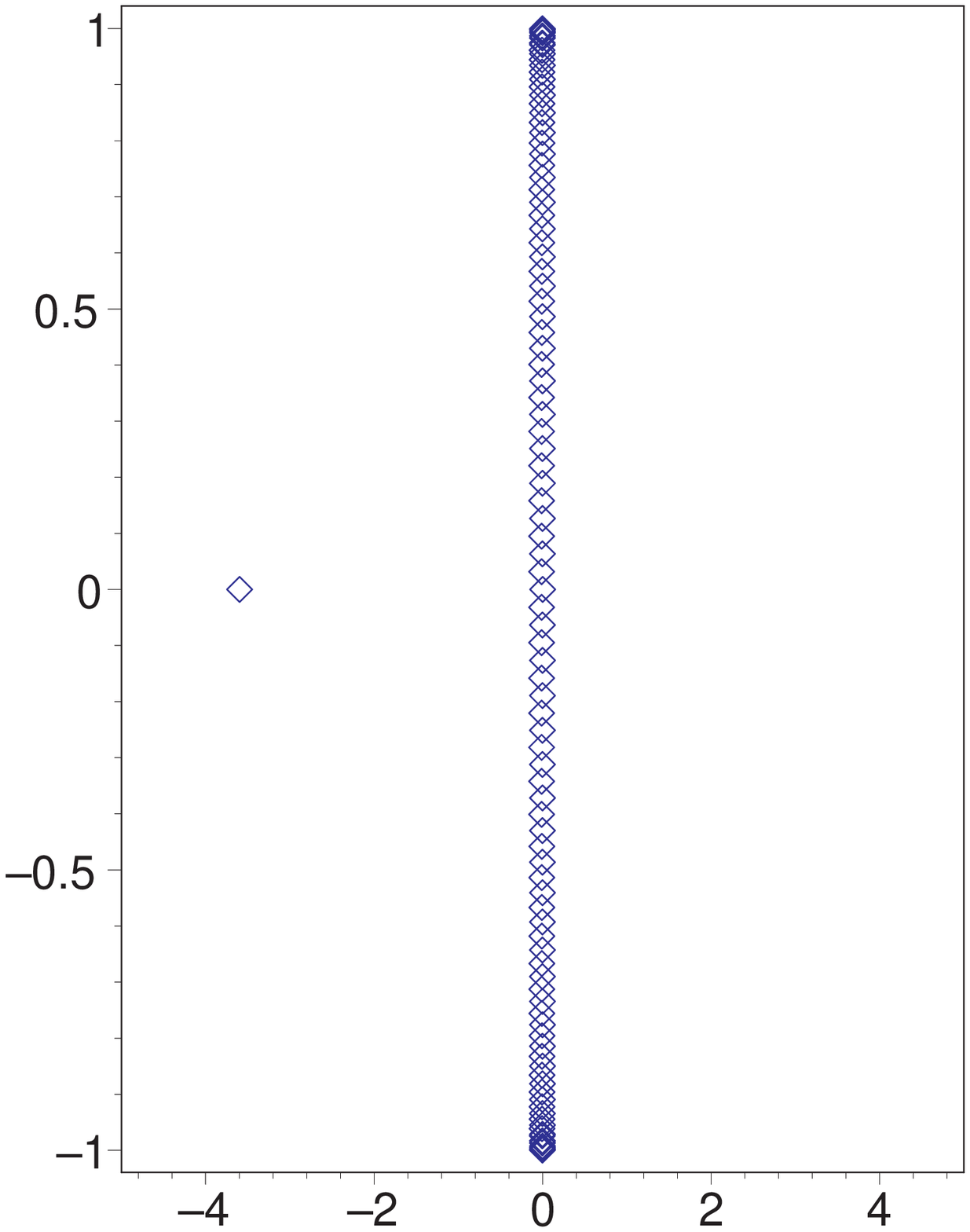}
\caption{Eigenvalues corresponding to the $D_{2-1}$ operator, obtained with a
grid containing $100$ points. From left to right the plots illustrate the
behavior for $\delta = 0, 1/10, 1/2$ respectively. As $\delta$ becomes larger,
a larger (in magnitude) negative eigenvalue on the real axis is observed (notice the
left-most diamond at $y\simeq -1.6,-3.5$ on the middle and right plots, respectively).}
\label{eigenvalues2delta} 
\end{center}
\end{figure}

Figure \ref{eigenvalues2delta}, in turn, shows the
eigenvalues computed with $100$ points and $\delta=0,1/10, 1/2$. A
negative real part appears, as it should (based on the energy
calculation),  and the maximum in the imaginary axis
 slightly decreases (to approximately $0.999$, not varying much among these
three values). However, the maximum absolute value 
in the negative real axis grows quite fast with $\delta$. For example, for
$\delta=1/10$ such maximum already dominates over the maximum in the imaginary
axis. The higher order operators analyzed below behave similarly.   

\subsubsection{Fourth order in the interior, second order at and close to boundaries ($D_{4-2}$ scheme)}

\begin{figure}[ht]
\begin{center}
\includegraphics*[height=7cm]{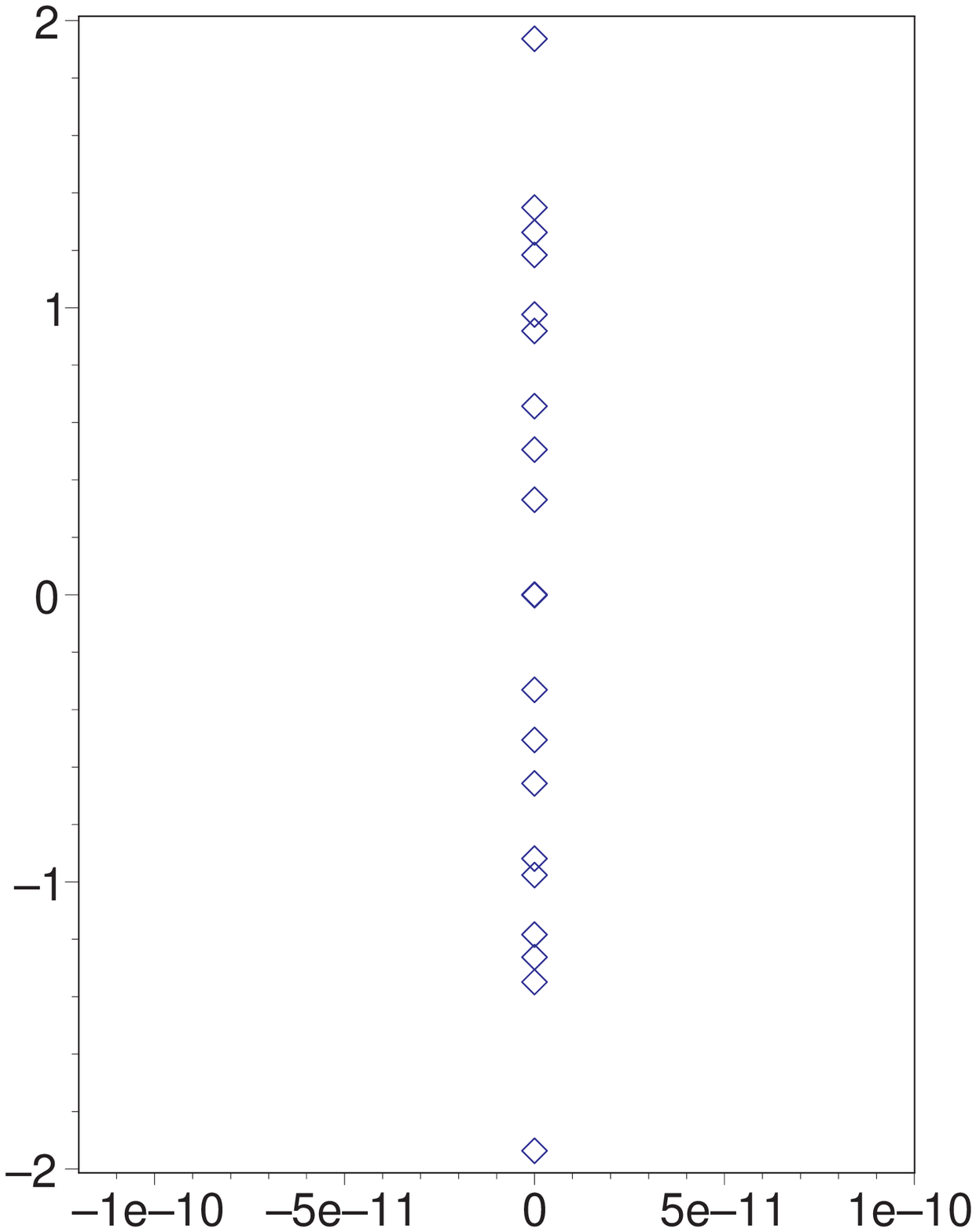}
\includegraphics*[height=7cm]{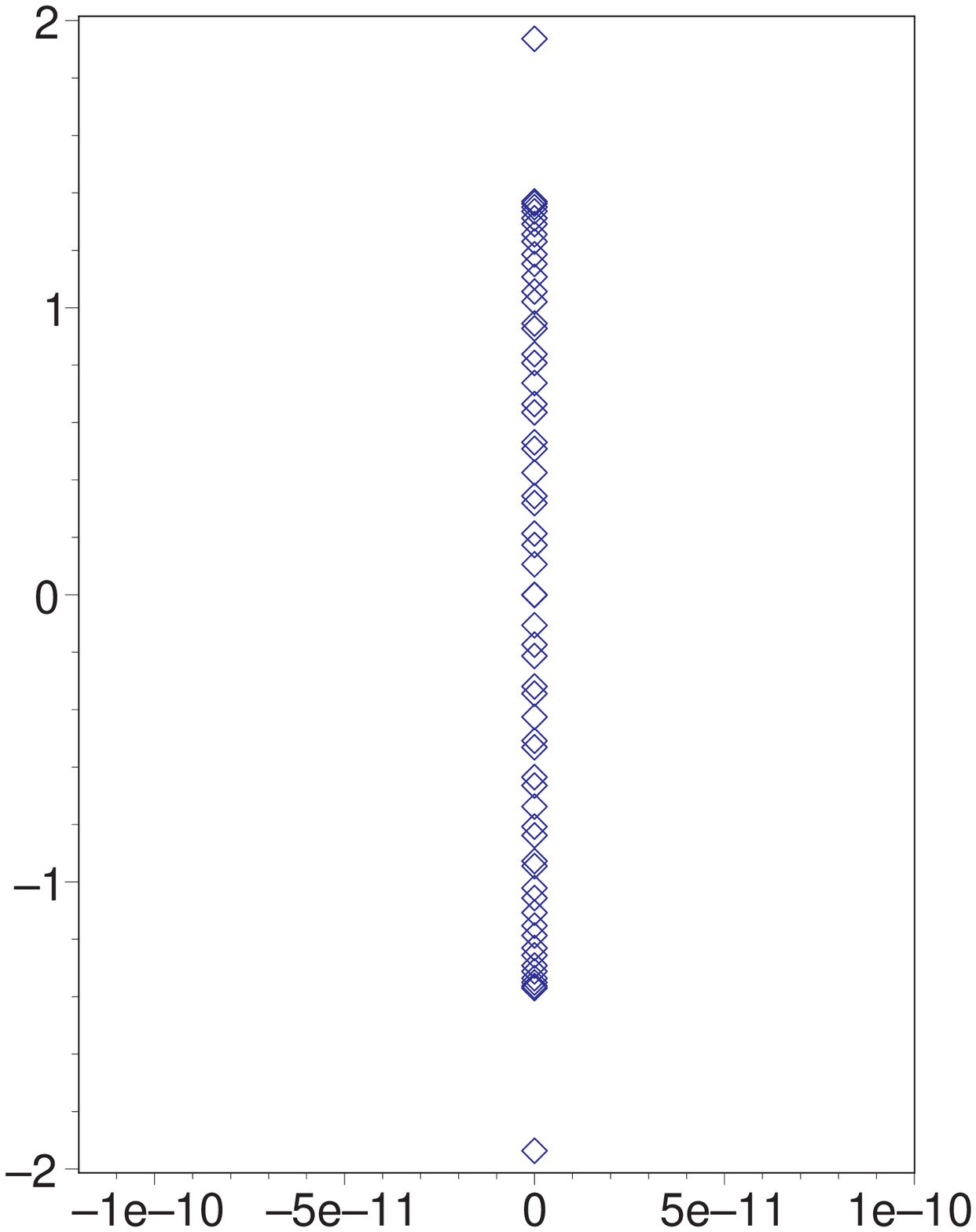}
\includegraphics*[height=7cm]{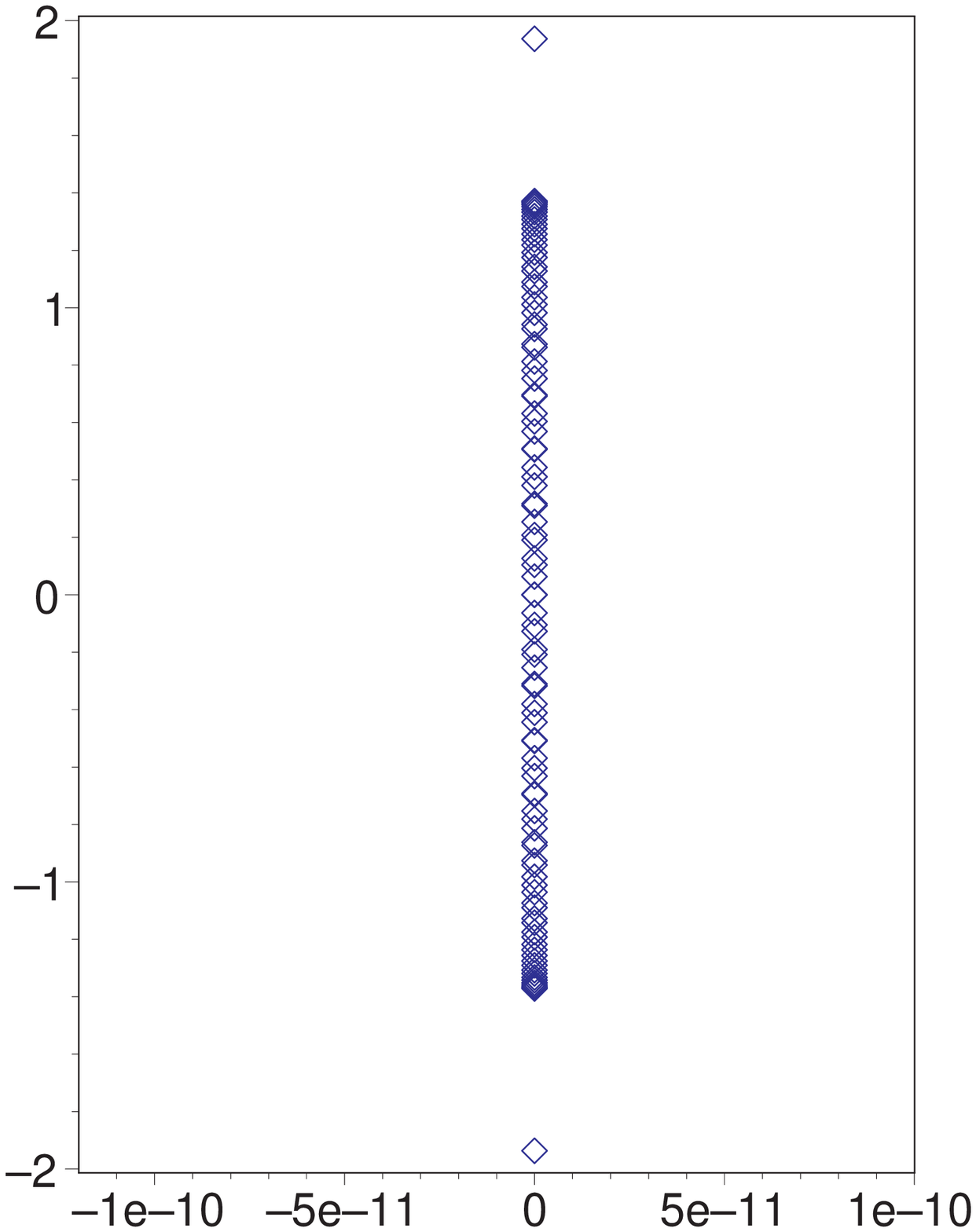}
\caption{
Numerically obtained eigenvalues corresponding to the $D_{4-2}$ 
operator for $\delta =  -1/2$ (purely imaginary case). From left to right the plots
illustrate the results obtained with a grid containing $20, 60, 100$ points respectively.
It is clear from the figures that these correspond, as in
Fig.\ref{eigenvalues2}, to a purely imaginary case.}
\label{eigenvalues4} 
\end{center}
\end{figure}

Figure \ref{eigenvalues4} shows the equivalent of Figure \ref{eigenvalues2},
but now for the $D_{4-2}$ case. The maximum is slightly
larger than the corresponding 
one for the $D_{2-1}$ case: approximately  $1.936$. 

\begin{figure}[ht]
\begin{center}
\includegraphics*[height=7cm]{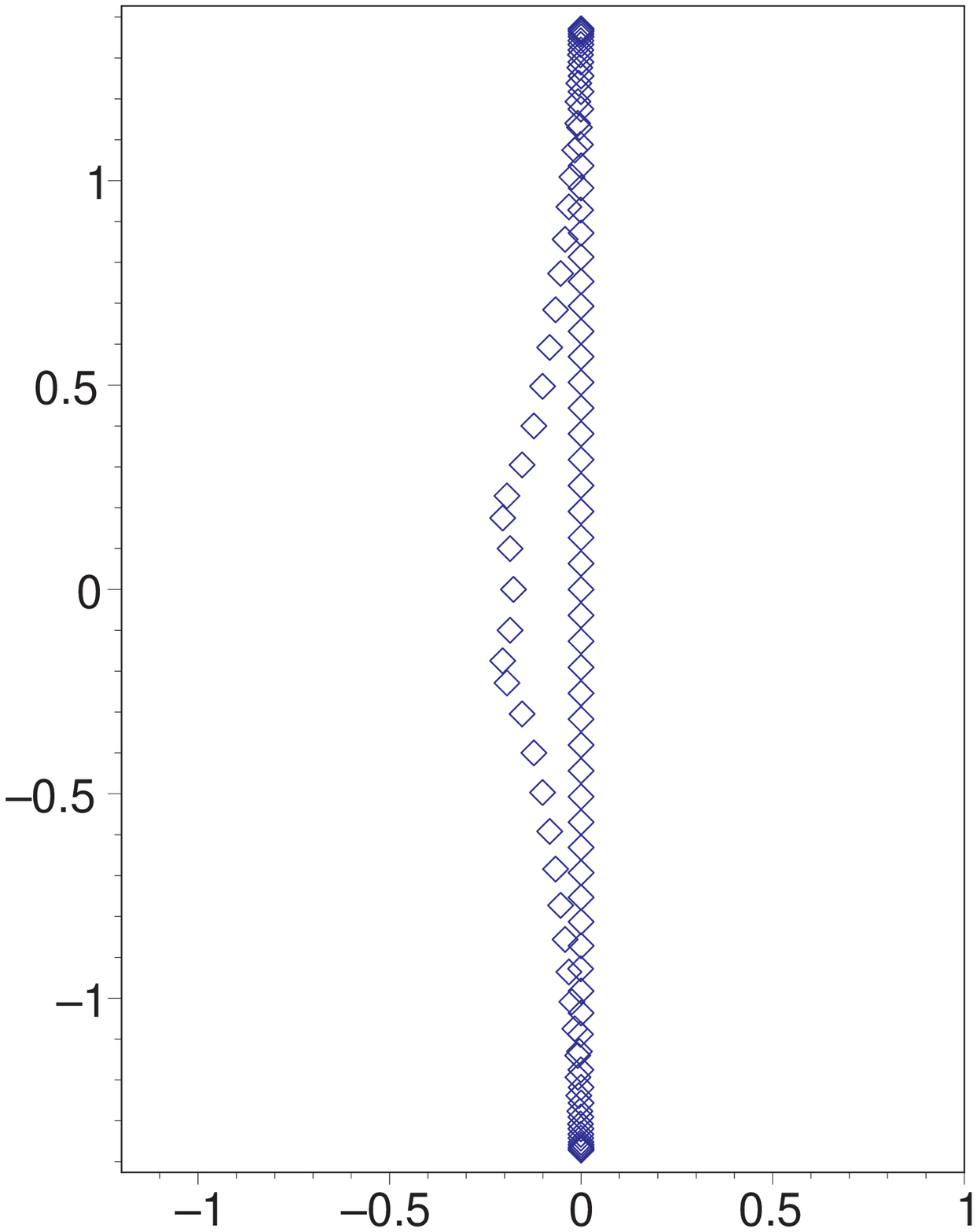}
\includegraphics*[height=7cm]{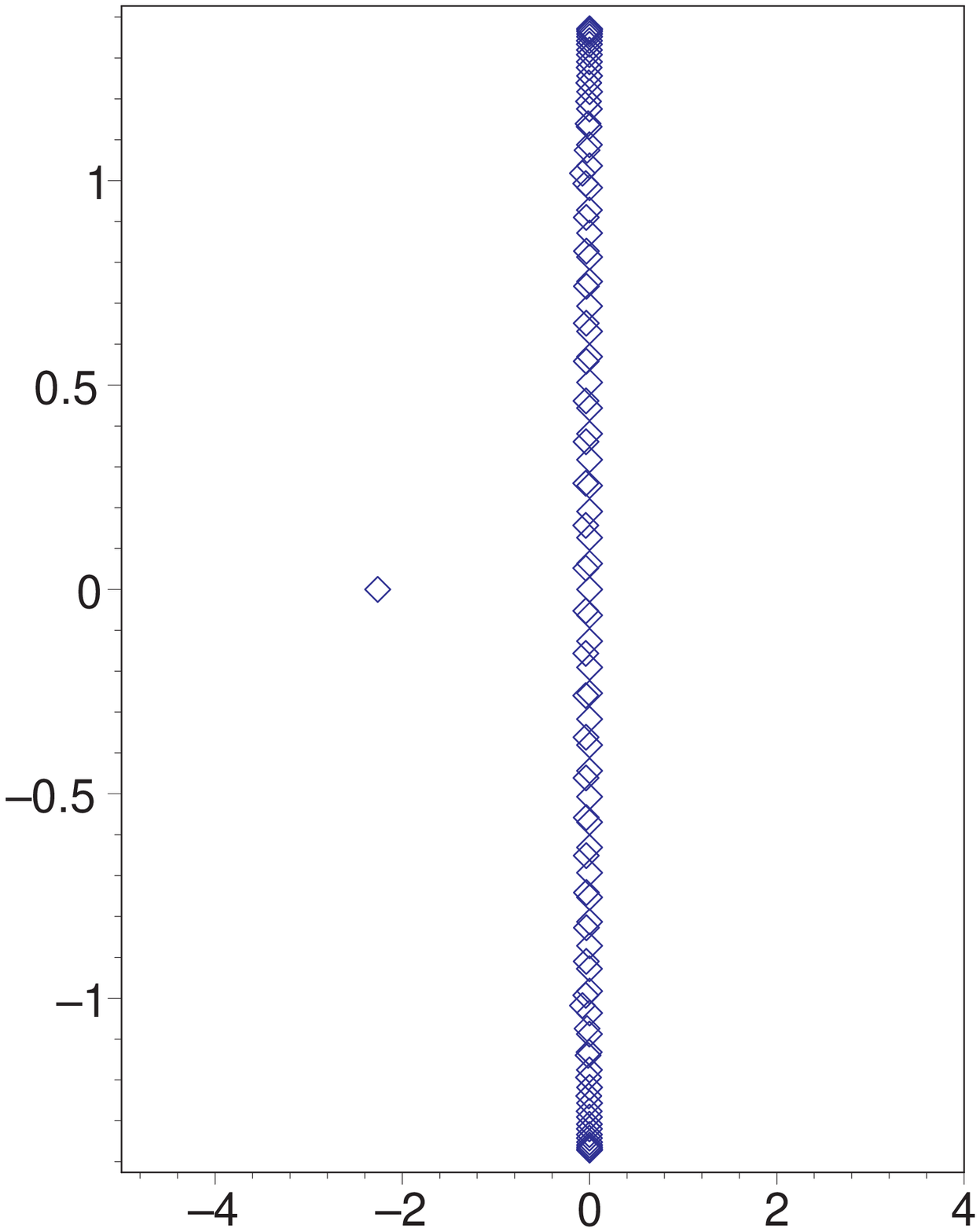}
\includegraphics*[height=7cm]{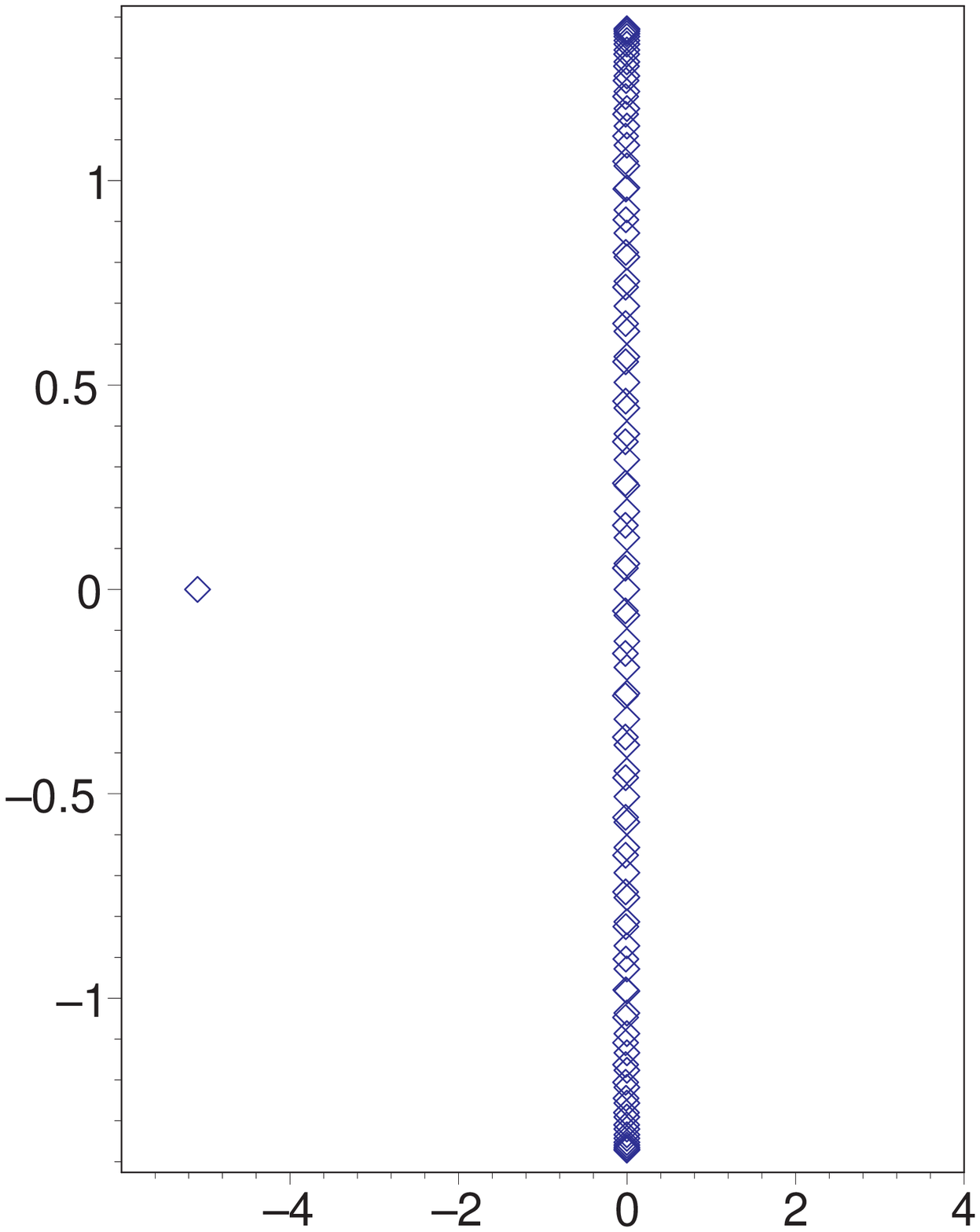}
\caption{
Eigenvalues corresponding to the $D_{4-2}$ operator, obtained with a
grid containing $100$ points. From left to right the plots illustrate the
behavior for $\delta = 0, 1/10, 1/2$ respectively. As $\delta$ becomes larger,
a larger (in magnitude) negative eigenvalue on the real axis is observed (notice the
left-most diamond at $y\simeq -2.5,-5$ on the middle and right plots, respectively).}
\label{eigenvalues4delta} 
\end{center}
\end{figure}

Figure \ref{eigenvalues4delta}, in turn, shows the equivalent of Figure 
\ref{eigenvalues2delta} for the current case. As before, a 
negative real part appears and the maximum in the imaginary axis
slightly decreases (in this case to roughly $1.371$, not changing much among these
three values of $\delta$). 

\subsubsection{Sixth order in the interior, third order at and close to
  boundaries, minimum bandwidth case ($D_{6-3}$ scheme)}

\begin{figure}[ht]
\begin{center}
\includegraphics*[height=7cm]{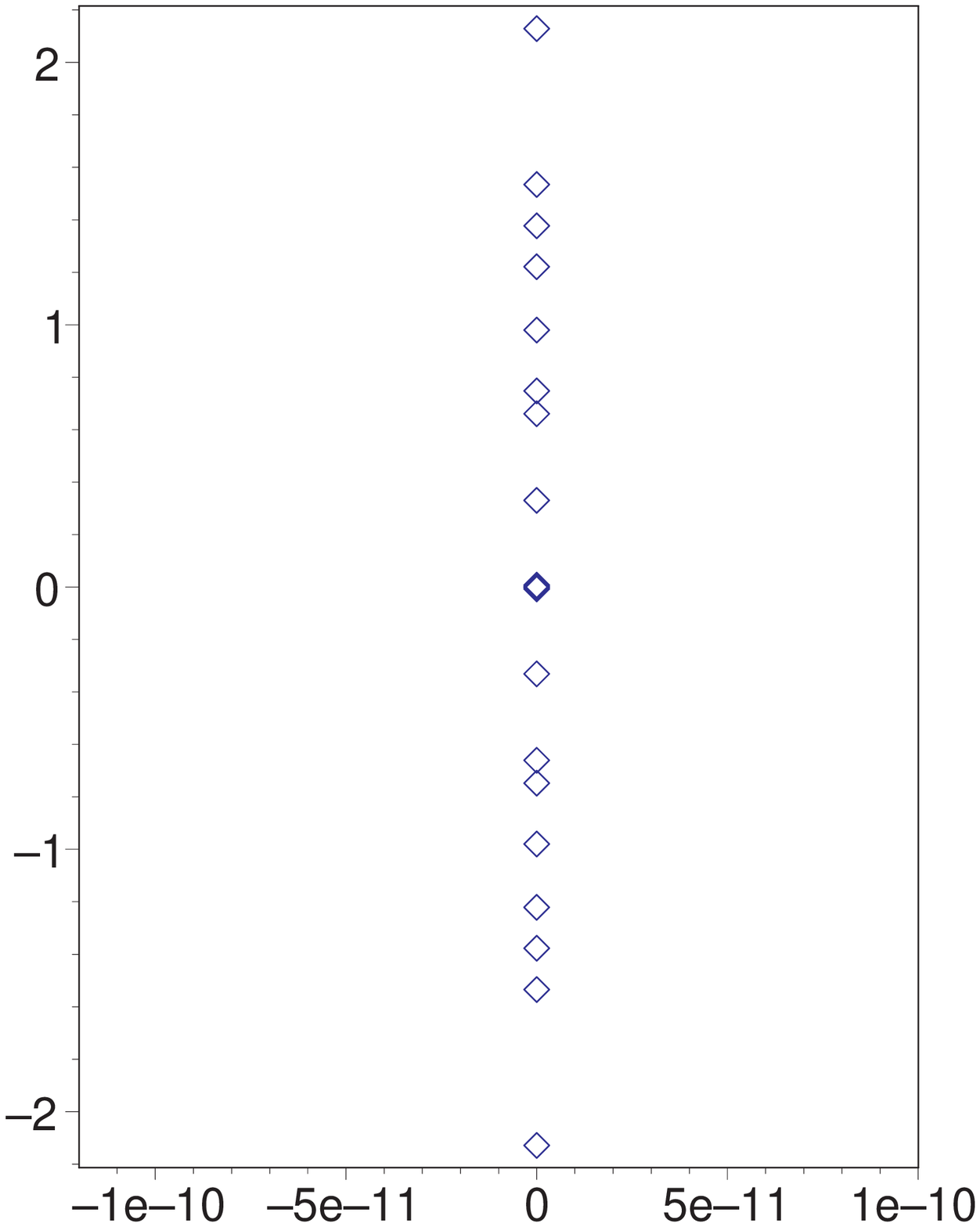}
\includegraphics*[height=7cm]{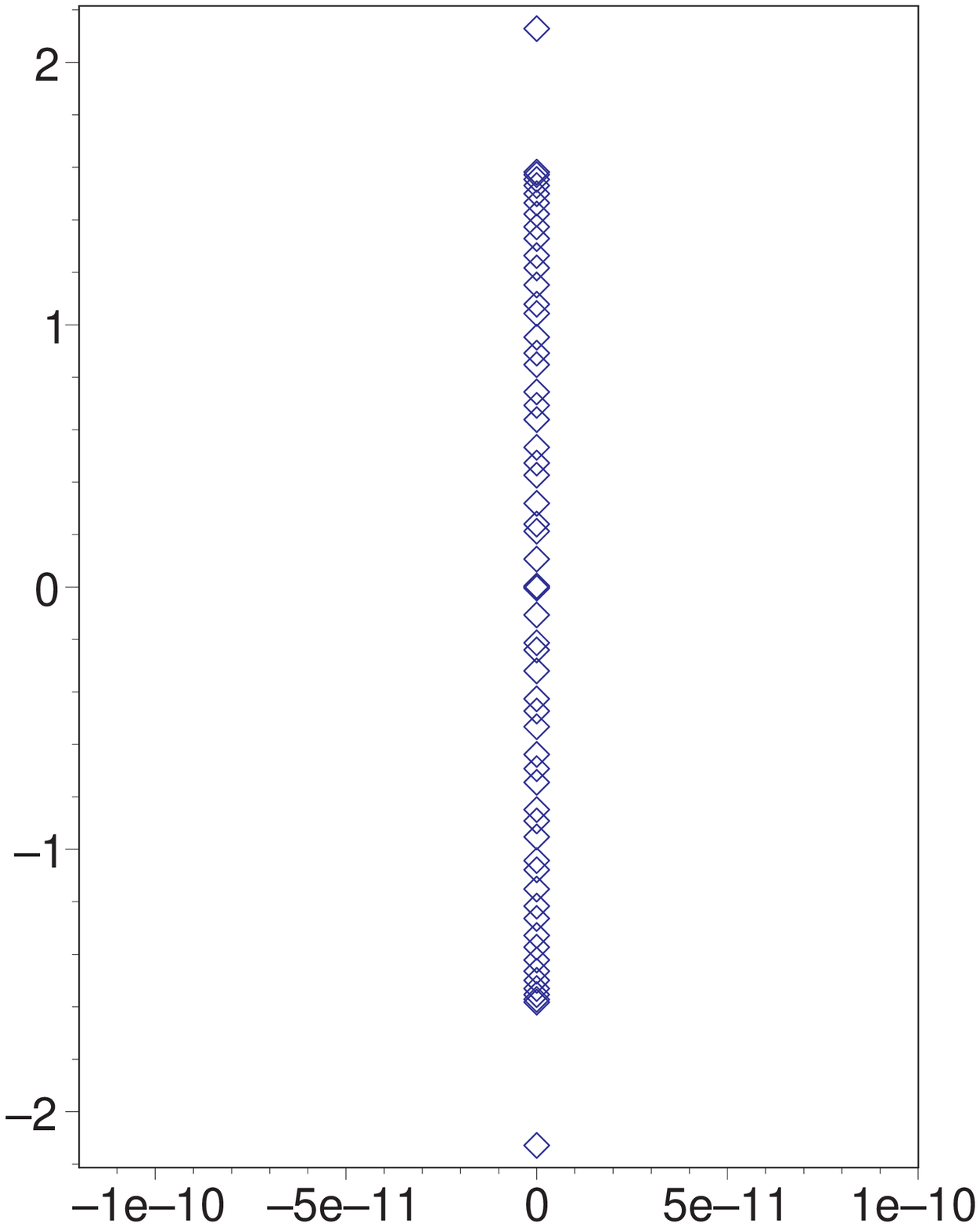}
\includegraphics*[height=7cm]{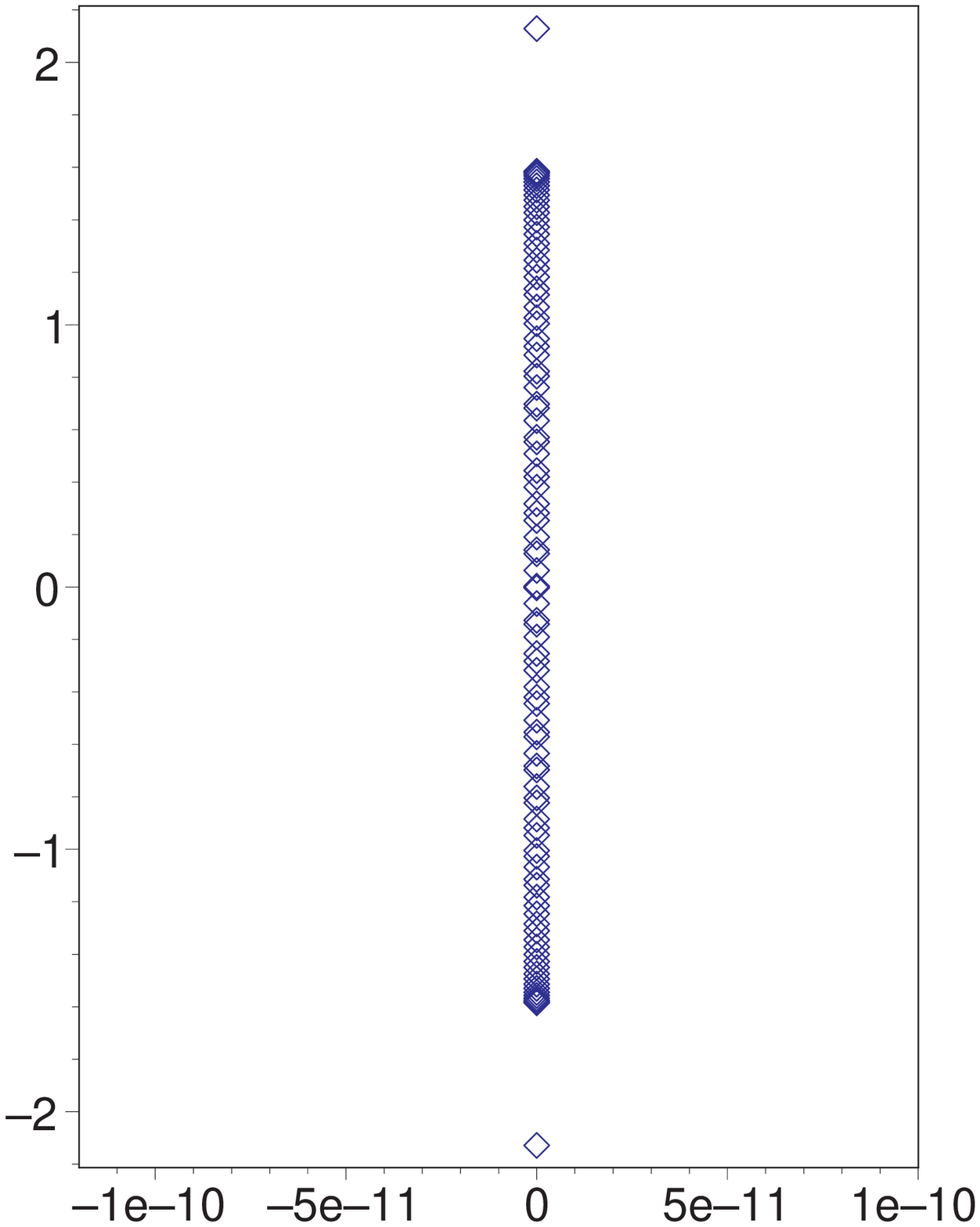}
\caption{
Numerically obtained eigenvalues corresponding to the minimum bandwidth $D_{6-3}$ 
operator for $\delta =  -1/2$ (purely imaginary case). From left to right the plots
illustrate the results obtained with a grid containing $20, 60, 100$ points respectively.
As in Figs.(\ref{eigenvalues2},\ref{eigenvalues4}), these correspond to a purely imaginary case.}
\label{eigenvalues6} 
\end{center}
\end{figure}

\begin{figure}[ht]
\begin{center}
\includegraphics*[height=7cm]{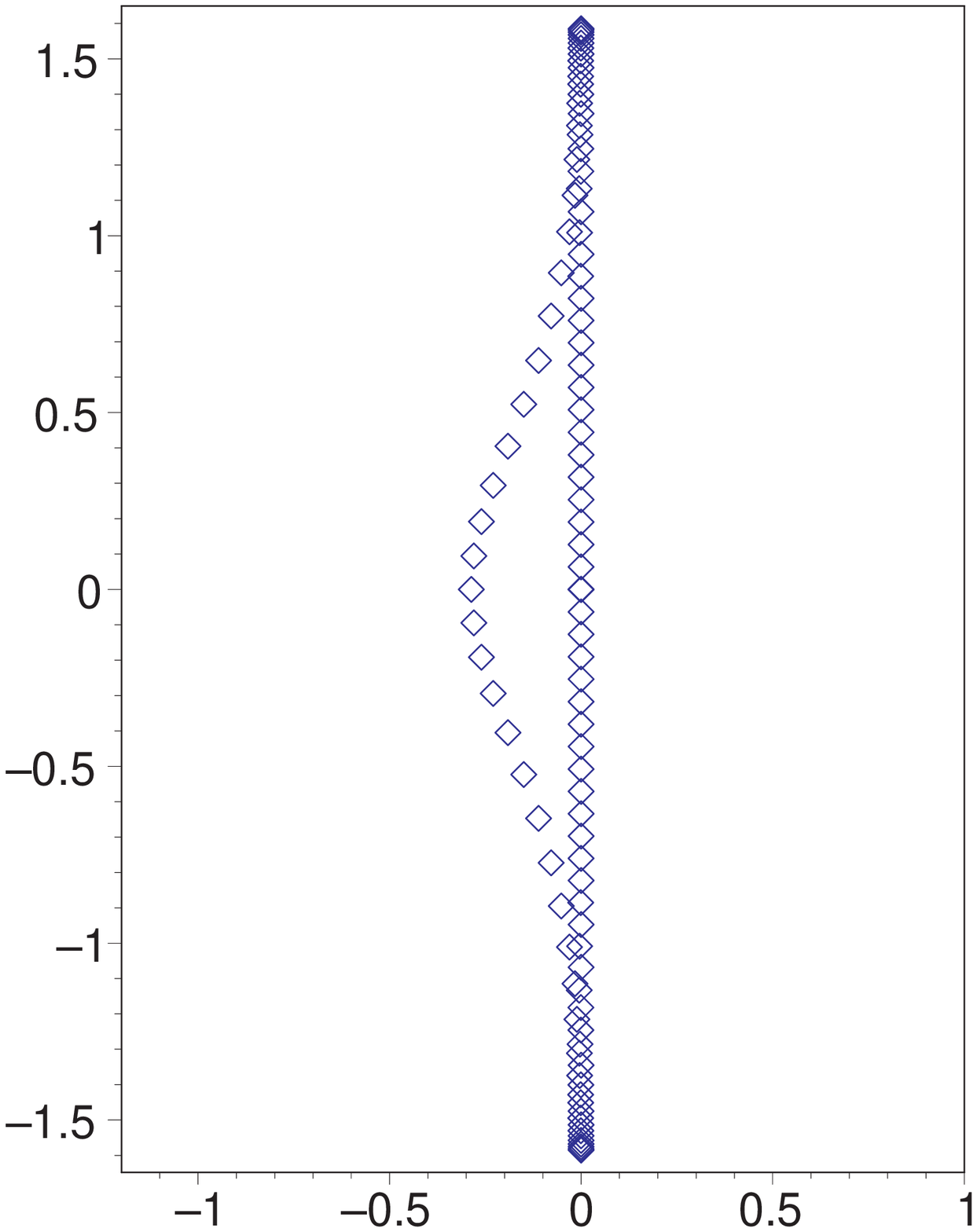}
\includegraphics*[height=7cm]{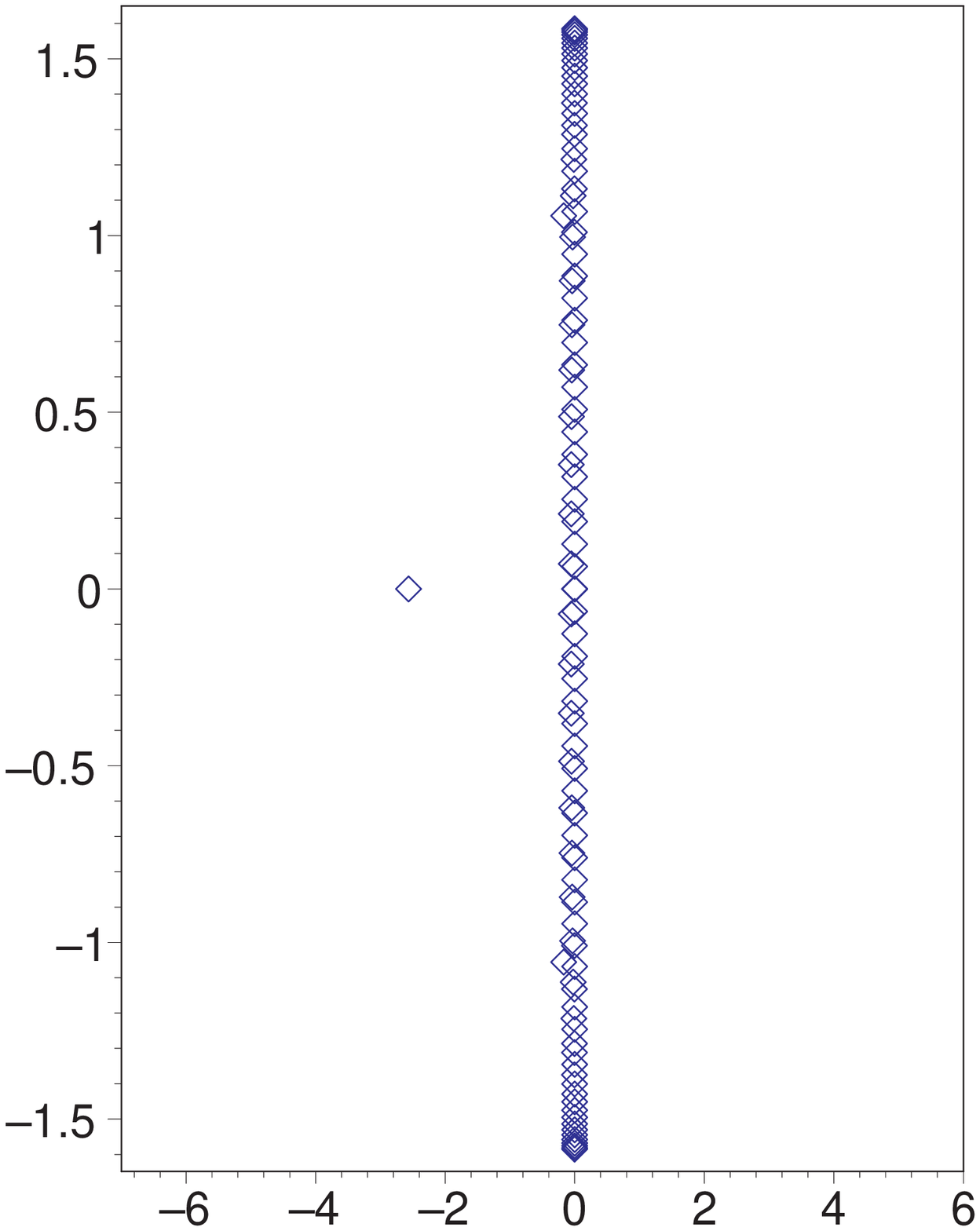}
\includegraphics*[height=7cm]{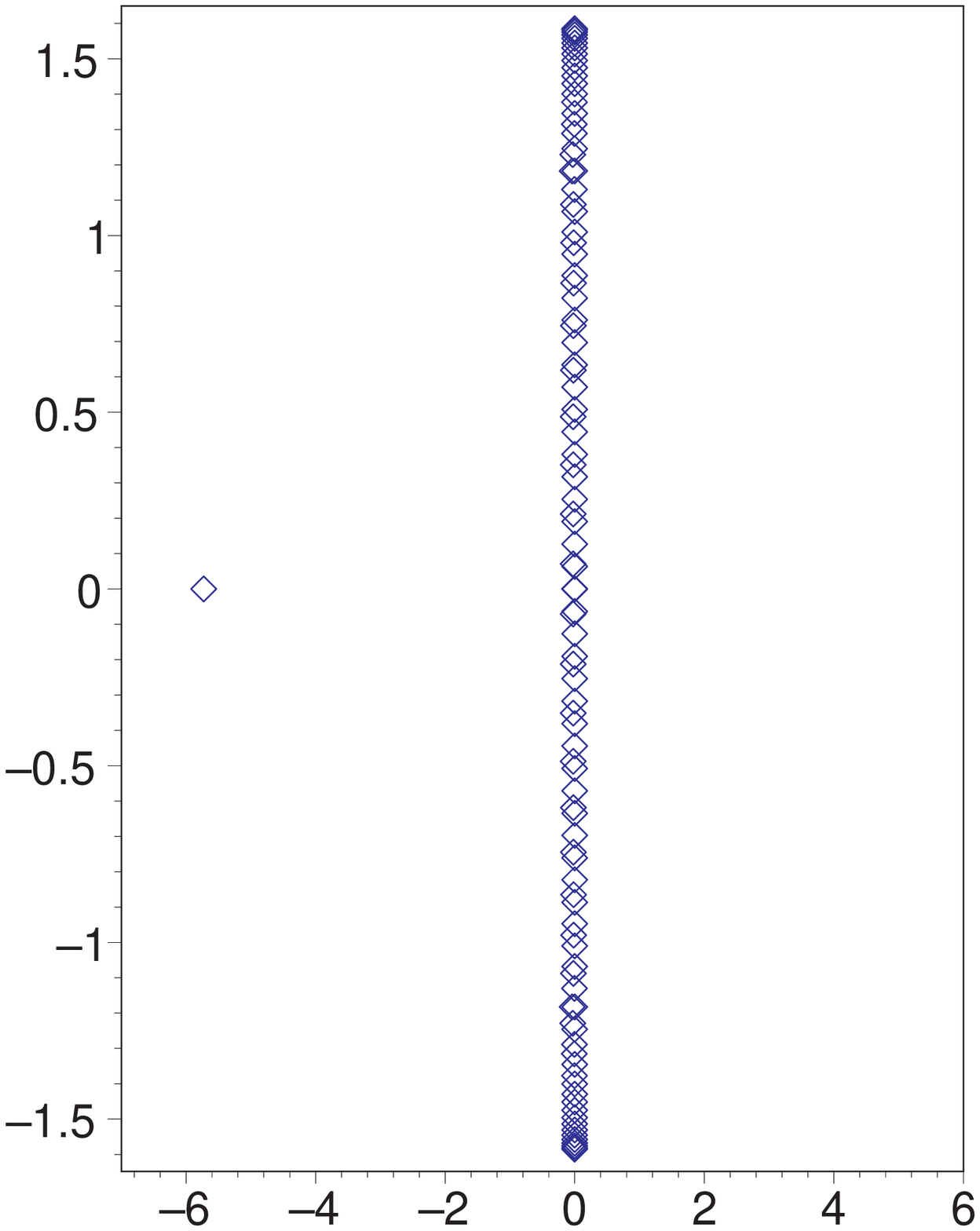}
\caption{Eigenvalues corresponding to the $D_{6-3}$ operator obtained with a
grid containing $100$ points. From left to right the plots illustrate the
behavior for $\delta = 0, 1/10, 1/2$ respectively. As $\delta$ becomes larger,
a larger (in magnitude) negative eigenvalue on the real axis is observed (notice the
left-most diamond at $y\simeq -2.5,-6$ on the middle and right plots, respectively).}
\label{eigenvalues6delta} 
\end{center}
\end{figure}

Figure \ref{eigenvalues6} illustrates the equivalent of Figures
(\ref{eigenvalues2},\ref{eigenvalues4}) for the
$D_{6-3}$ case. The maximum is
roughly $2.129$, slightly larger than those of the previous two cases.
The behavior for larger values of $\delta$ is similar to that one found in
the previous two cases, as seen in Figure \ref{eigenvalues6delta}. 
The maximum in the imaginary axis again decreases slightly compared to the
$\delta=-1/2$ case (to roughly $1.585$) and does not change much among these
three values of $\delta$. 

\subsubsection{Eight order in the interior, fourth order at and close to boundaries ($D_{8-4}$ scheme)}
The $D_{8-4}$ operator has three free parameters, denoted as $x_1,x_2,x_3$ both in
Ref.\cite{strand} and here. As mentioned, these parameters can be 
freely chosen to satisfy a given criteria. For instance, they can be fixed so as to minimize
the width of the derivative operator or yield as small a spectral radius as possible.
As we discuss next, these options can yield operators with significantly different 
stability requirements as dictated by the CFL condition.


\textbf{Minimum bandwidth operator}. 
The minimum bandwidth case corresponds to the choice (see \cite{strand})
\begin{equation}
x_1= \frac{1714837}{4354560}, \;\;\;
x_2= -\frac{1022551}{30481920}, \;\;\;
x_3= \frac{6445687}{8709120}.  \label{mbw}
\end{equation}

Figure \ref{eigenvalues6} shows for this minimum $D_{8-4}$ bandwidth case 
the eigenvalues for $20,60,100$ points,
for $\delta = -1/2$ (purely imaginary case). While for the previous operators we have seen 
that the maximum eigenvalue increases slightly with the order of the operator, in this case 
the increase is quite large: the maximum is roughly $16.04$. This translates into a 
CFL limit for this operator being almost 
an order of magnitude smaller than the limits for the previous operators. Additionally, variation
of $\delta$ does not significantly affect this behavior, as shown in Figure \ref{eigenvalues8delta}. 
That figure shows the
eigenvalues computed with $100$ points, and $\delta=0,1/10, 1/2$. The
qualitative behavior when increasing $\delta$ is similar to that one of the
previous cases. A negative real part appears in the spectrum, and the maximum in the imaginary axis
slightly decreases, to roughly $16.02$, not varying much among these
three illustrative values of $\delta$.
\begin{figure}[ht]
\begin{center}
\includegraphics*[height=7cm]{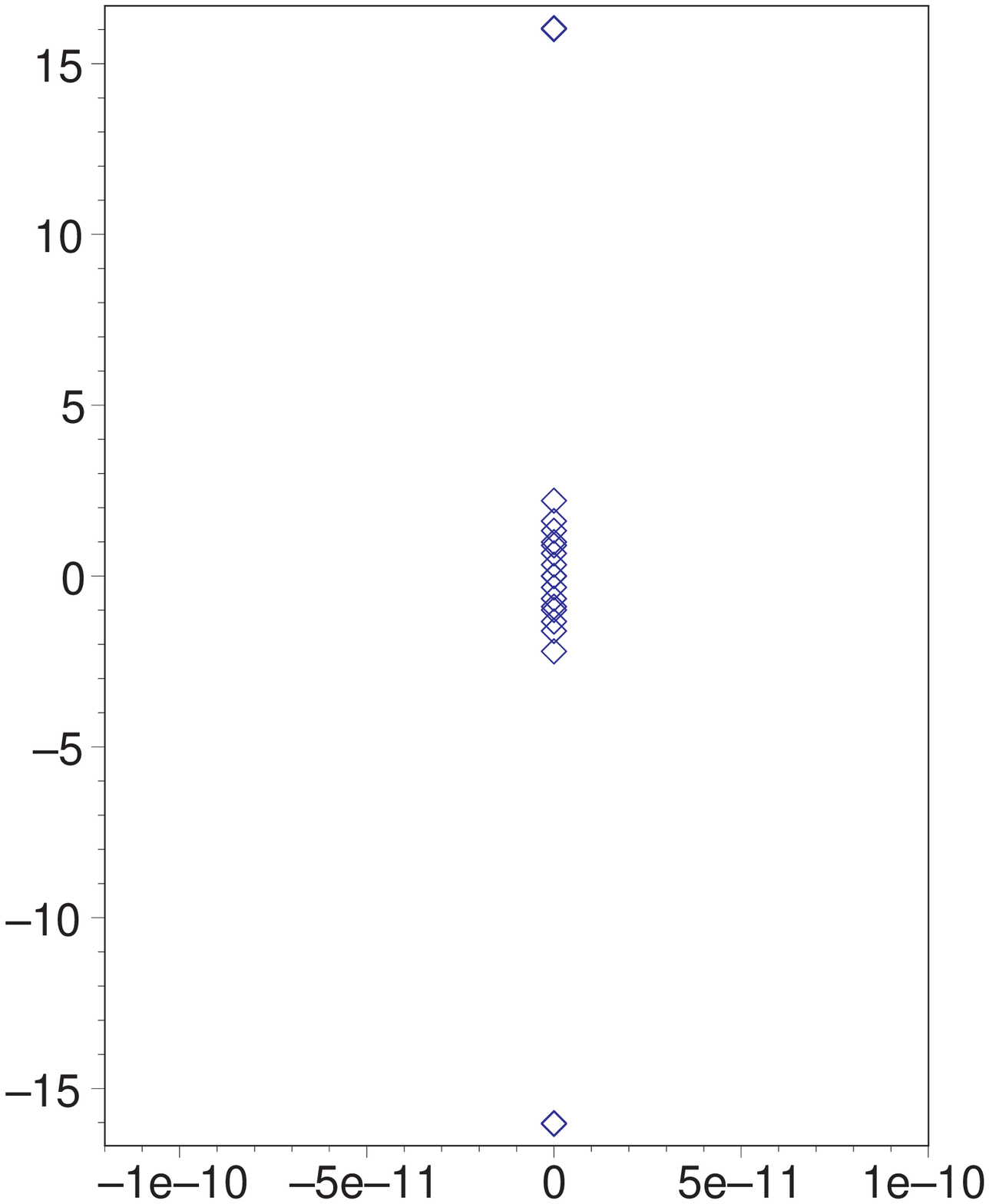}
\includegraphics*[height=7cm]{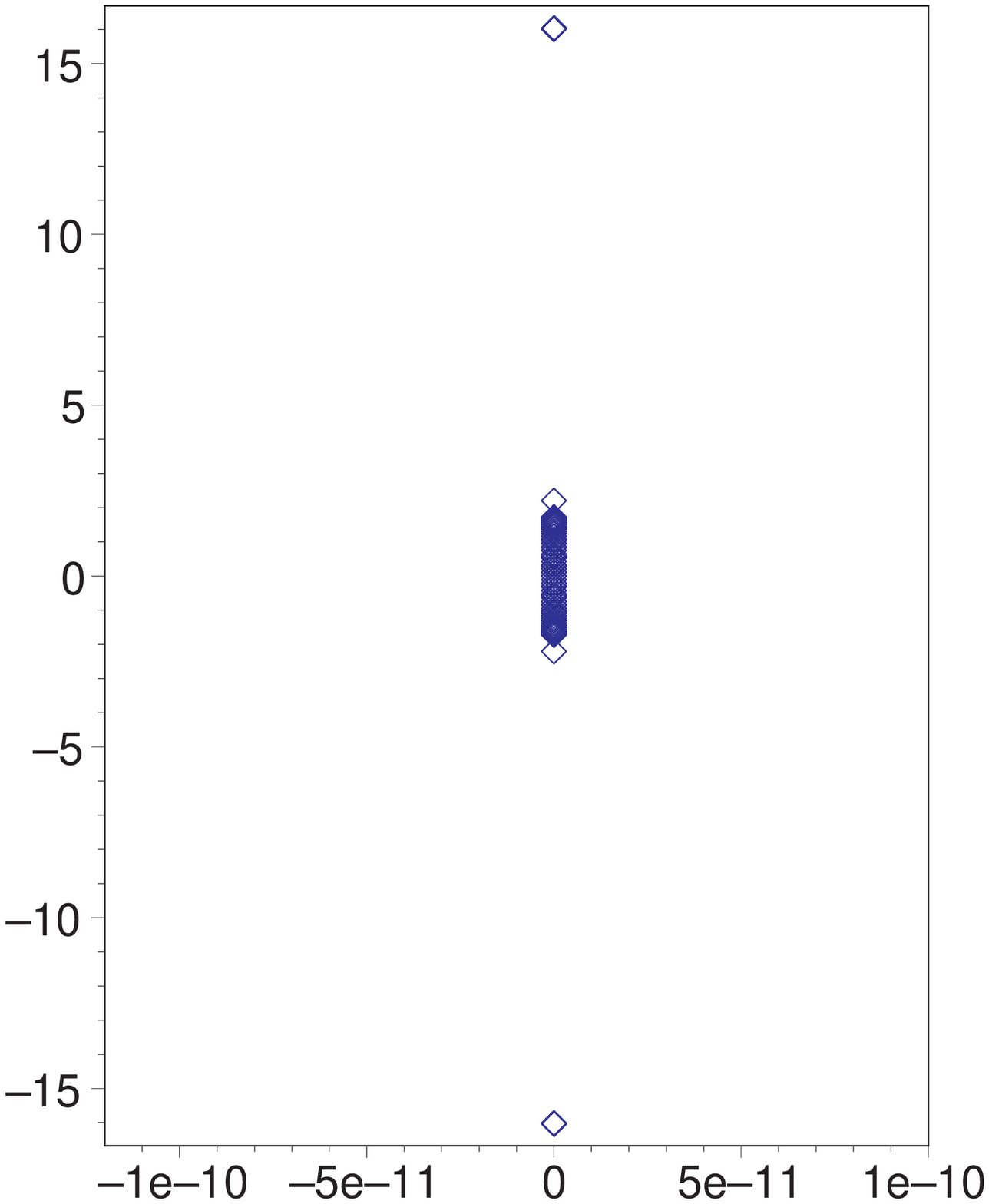}
\includegraphics*[height=7cm]{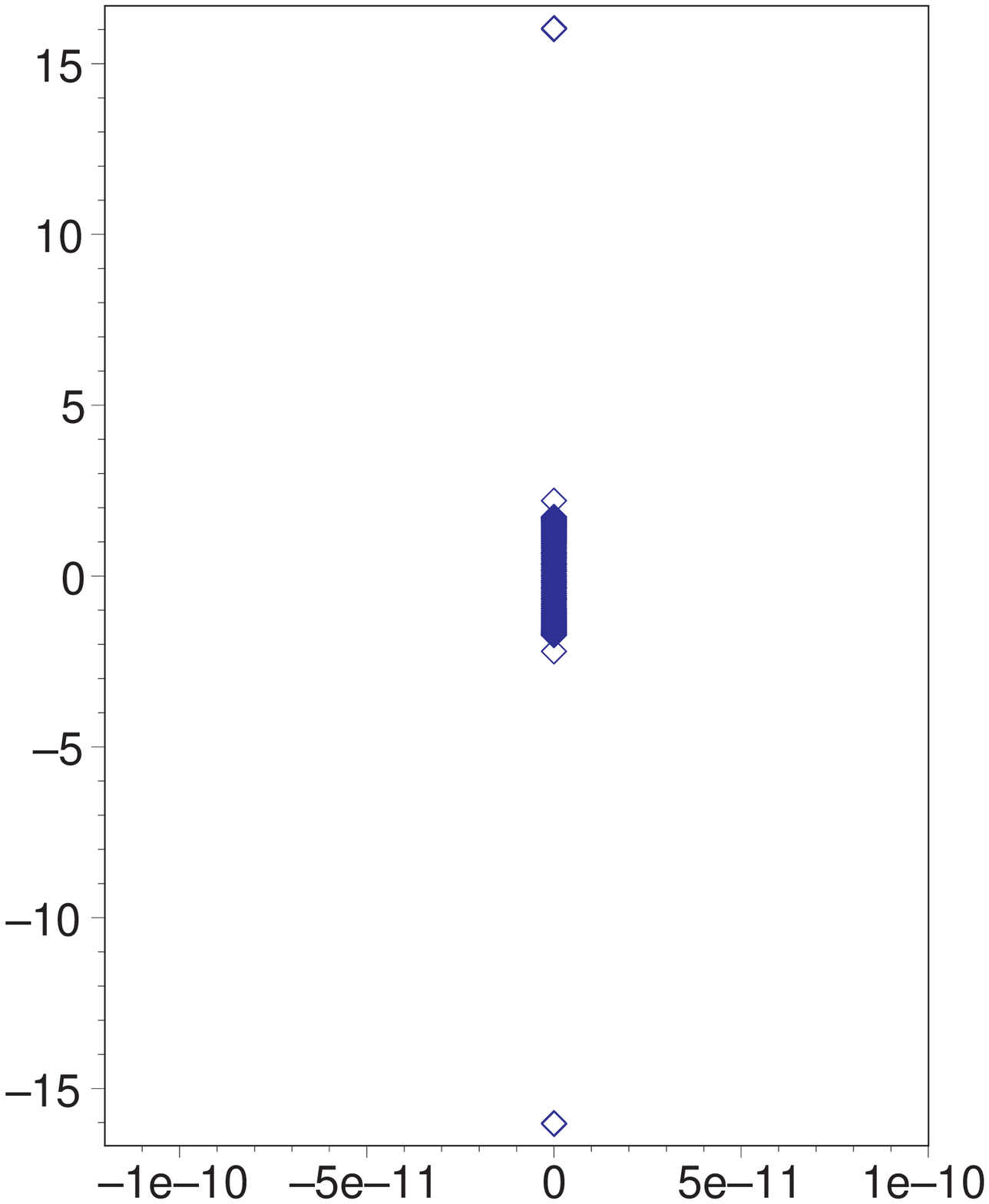}
\caption{Numerically obtained eigenvalues corresponding to the minimum bandwidth $D_{8-4}$ 
operator for $\delta =  -1/2$ (purely imaginary case). From left to right the plots
illustrate the results obtained with a grid containing $20, 60, 100$ points, respectively.
Although purely imaginary, the maximum (absolute) value in the vertical axis is approximately $16$.}
\label{eigenvalues8} 
\end{center}
\end{figure}
Such a large spectral radius for this operator motivates the search for
another one, with a more convenient radius at the expense of 
not having the minimum possible bandwidth.

\begin{figure}[ht]
\begin{center}
\includegraphics*[height=7cm]{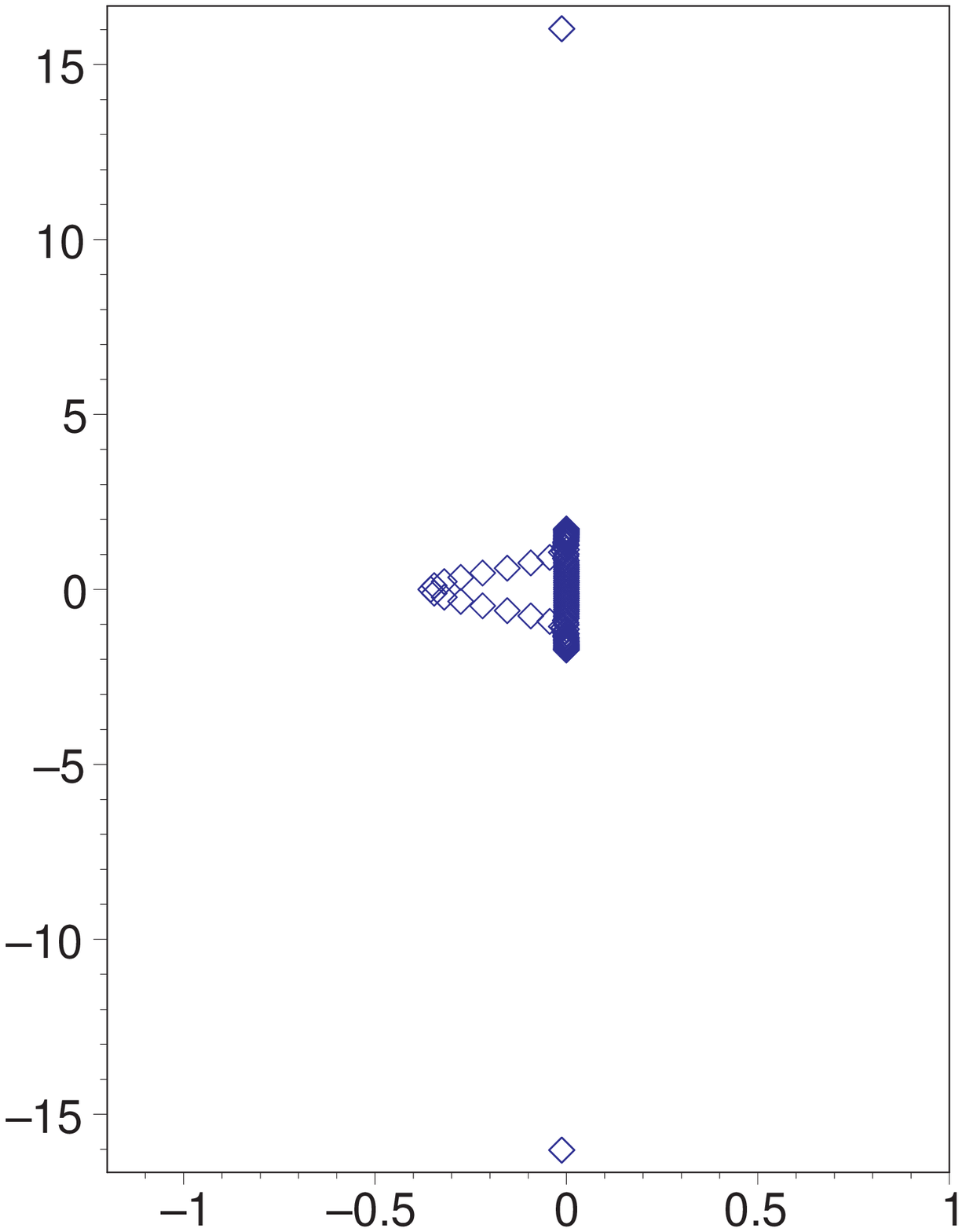}
\includegraphics*[height=7cm]{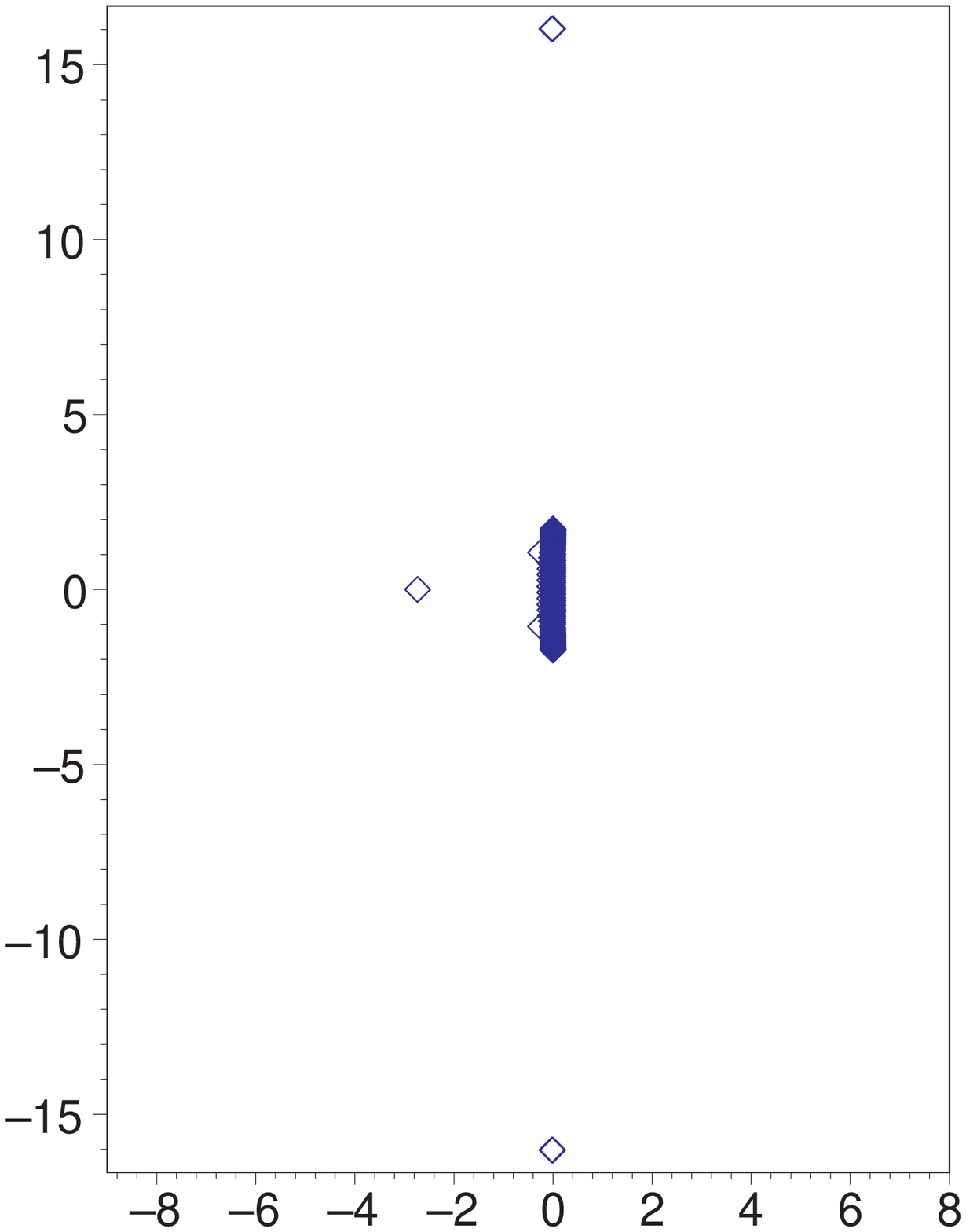}
\includegraphics*[height=7cm]{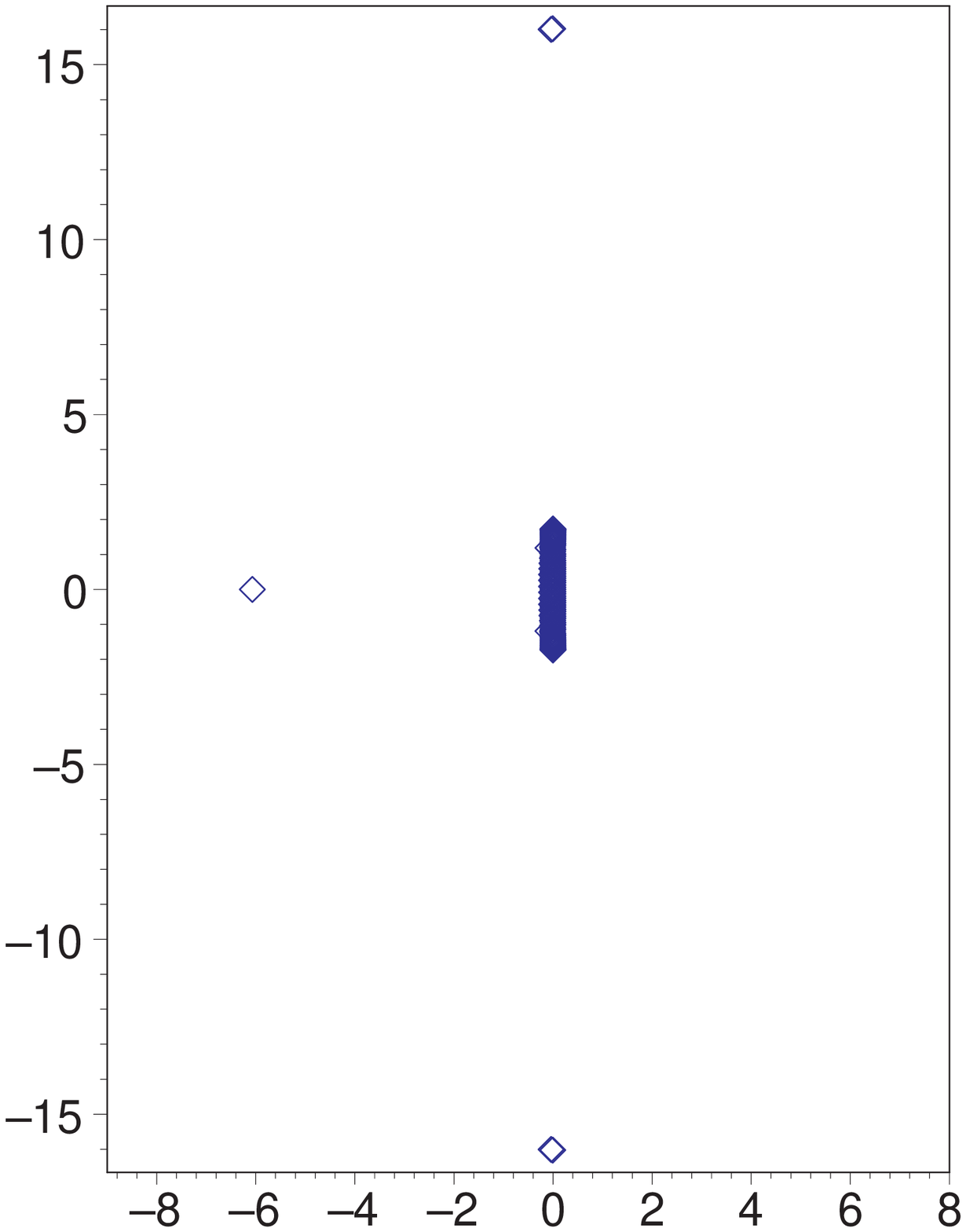}
\caption{Eigenvalues for the minimum bandwidth $D_{8-4}$ operator with $\delta = 0,
  1/10, 1/2$ (from left to right) and $100$ points.
Values of $\delta$ larger than $-1/2$ introduce a negative real part in the
spectrum but they have little effect on the maximum absolute value, which remains at approximately $16$.}
\label{eigenvalues8delta} 
\end{center}
\end{figure}

\begin{figure}[ht]
\begin{center}
\includegraphics*[height=7cm]{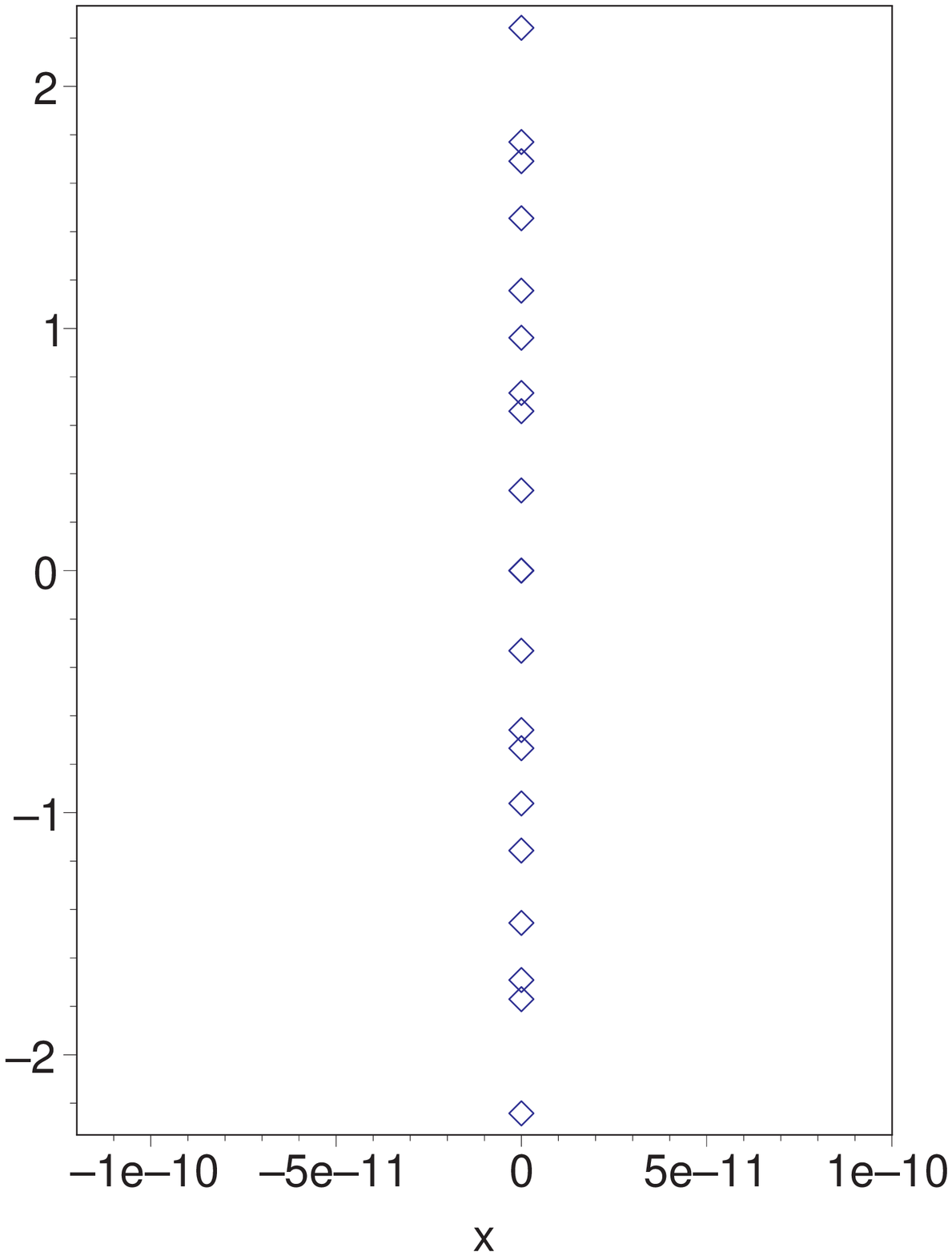}
\includegraphics*[height=7cm]{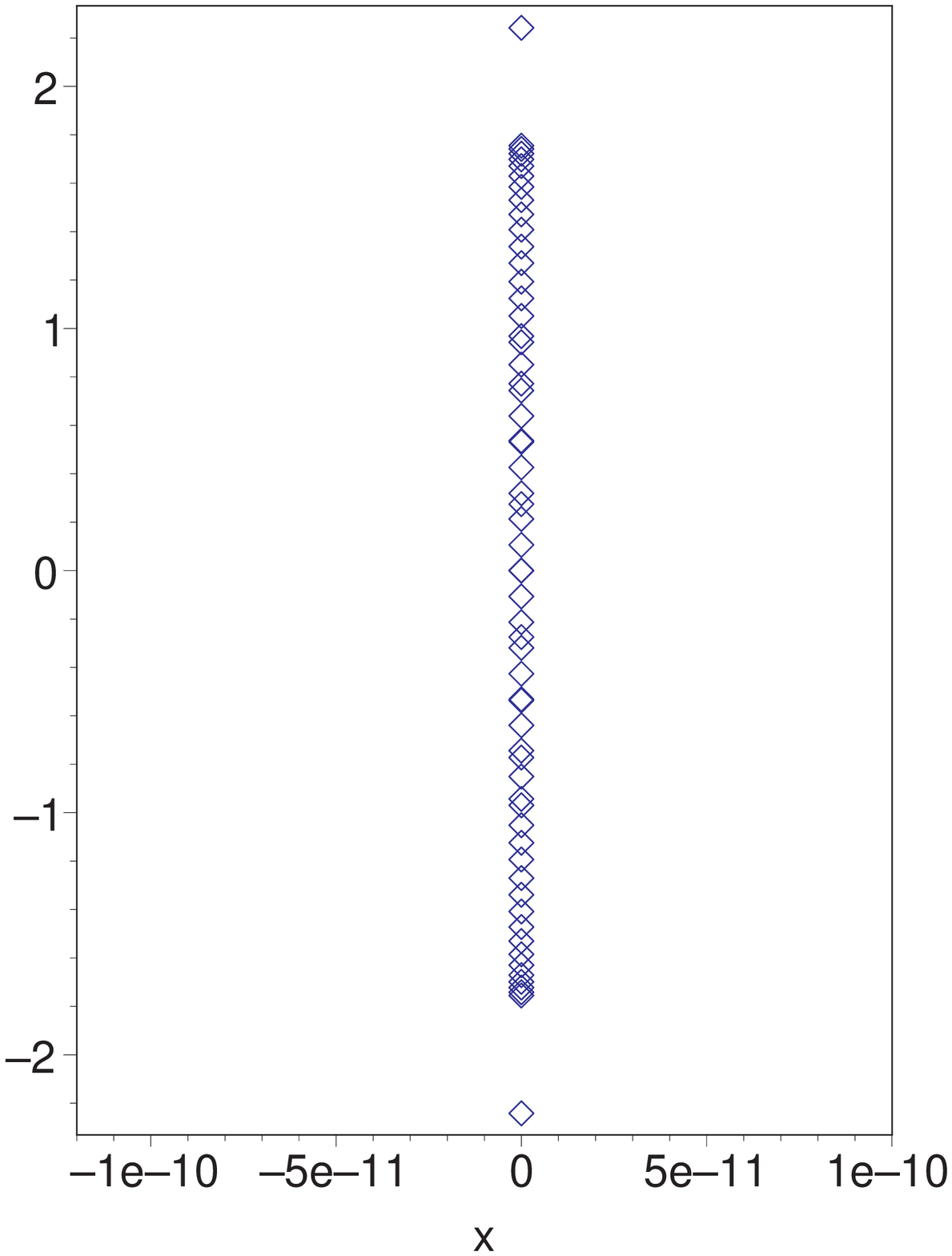}
\includegraphics*[height=7cm]{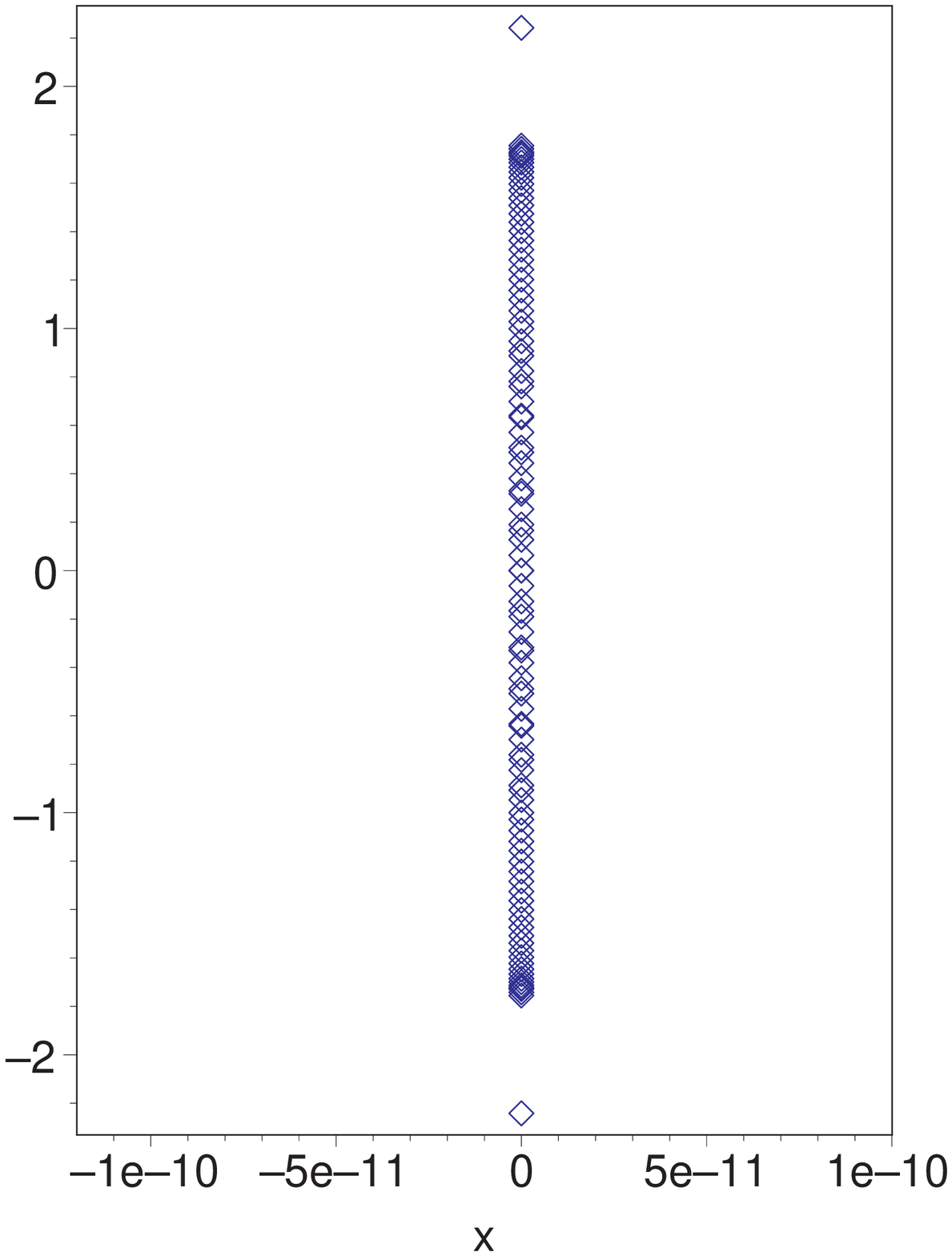}
\caption{Eigenvalues for the optimized $D_{8-4}$ operator with $\delta =
  -1/2$ (purely imaginary case), and $20, 60, 100$ points (from left to right).
Clearly, this modified operator has a much smaller spectral radius, compared to
the minimum bandwidth one. }
\label{eigenvalues8a} 
\end{center}
\end{figure}

\begin{figure}[ht]
\begin{center}
\includegraphics*[height=7cm]{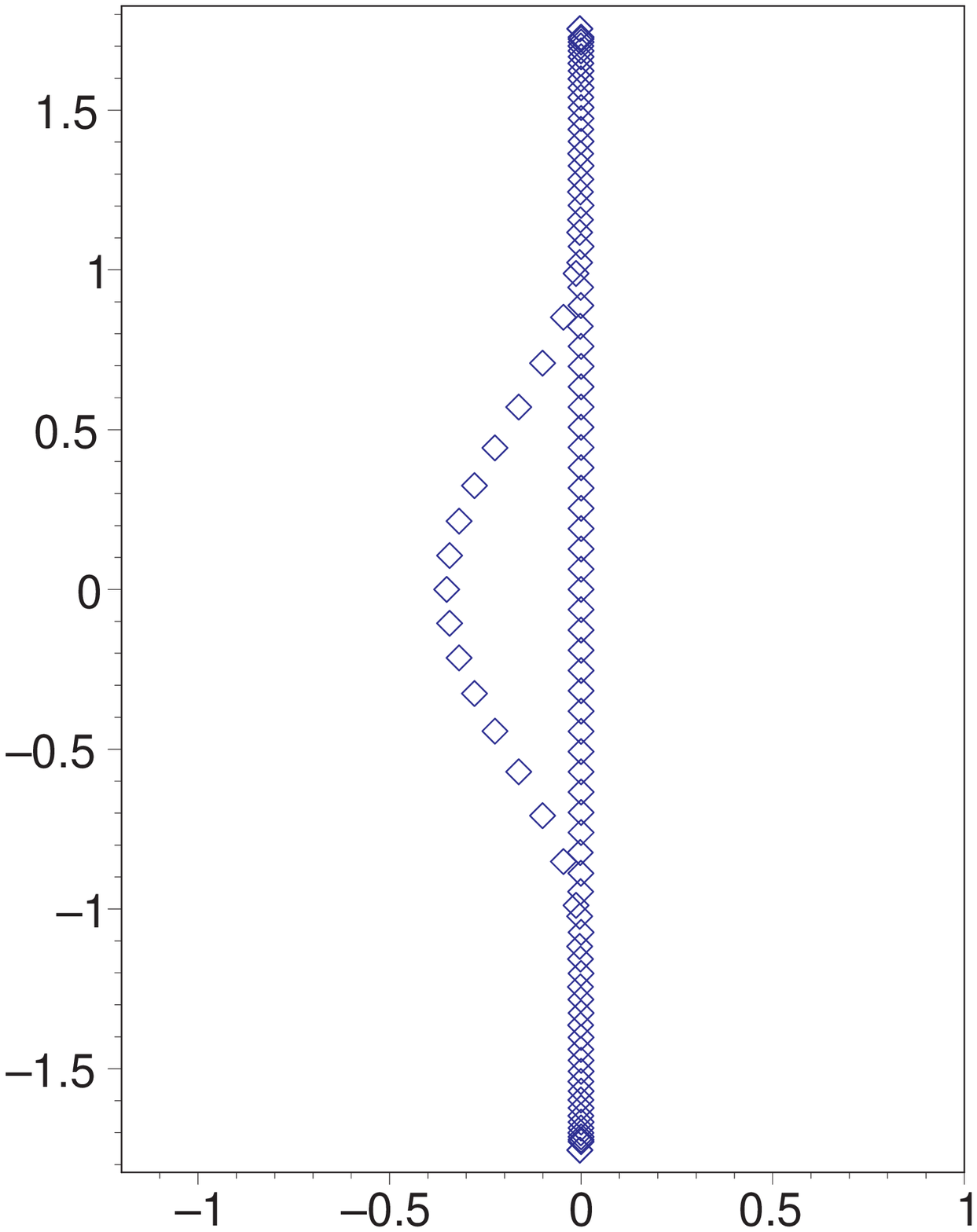}
\includegraphics*[height=7cm]{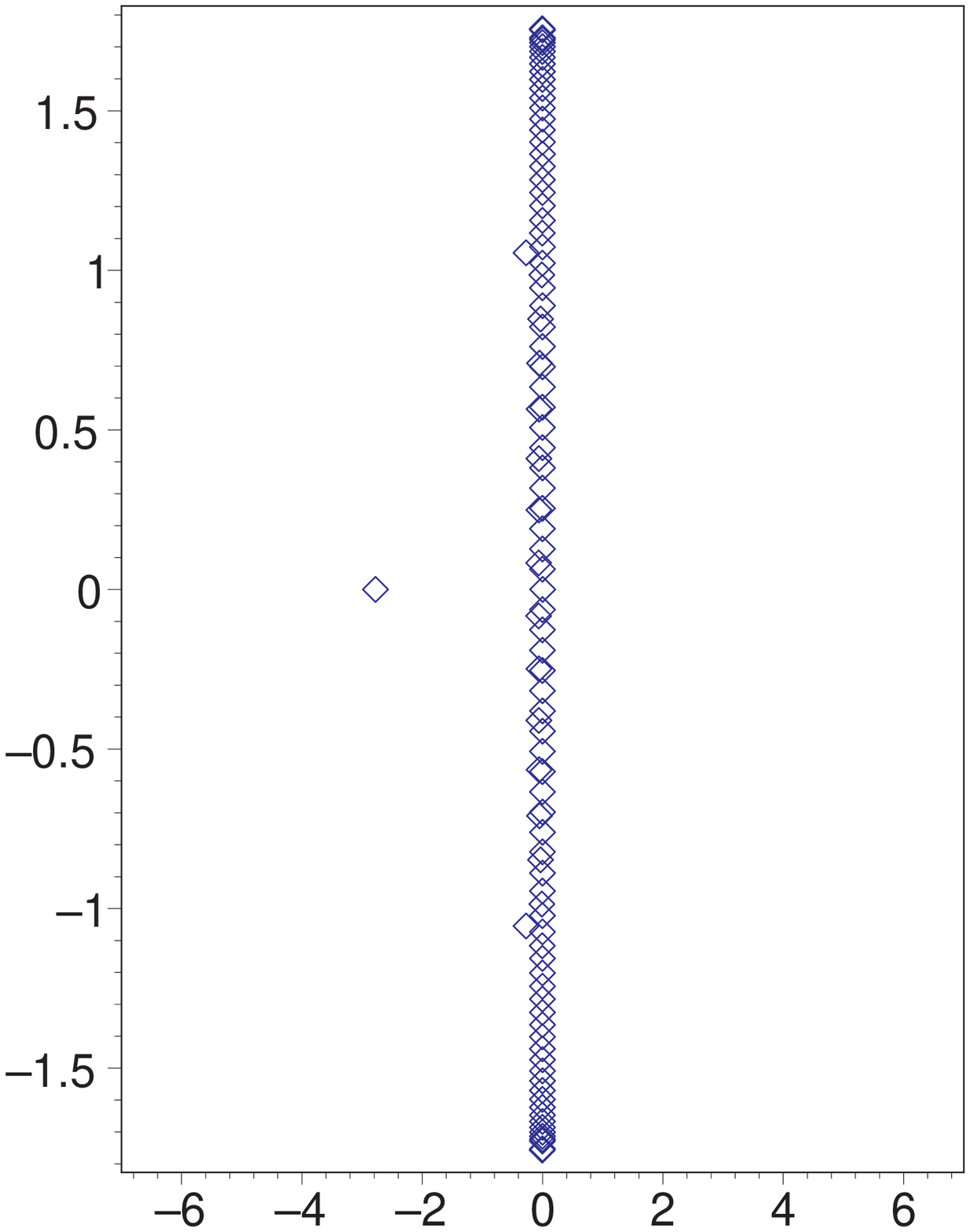}
\includegraphics*[height=7cm]{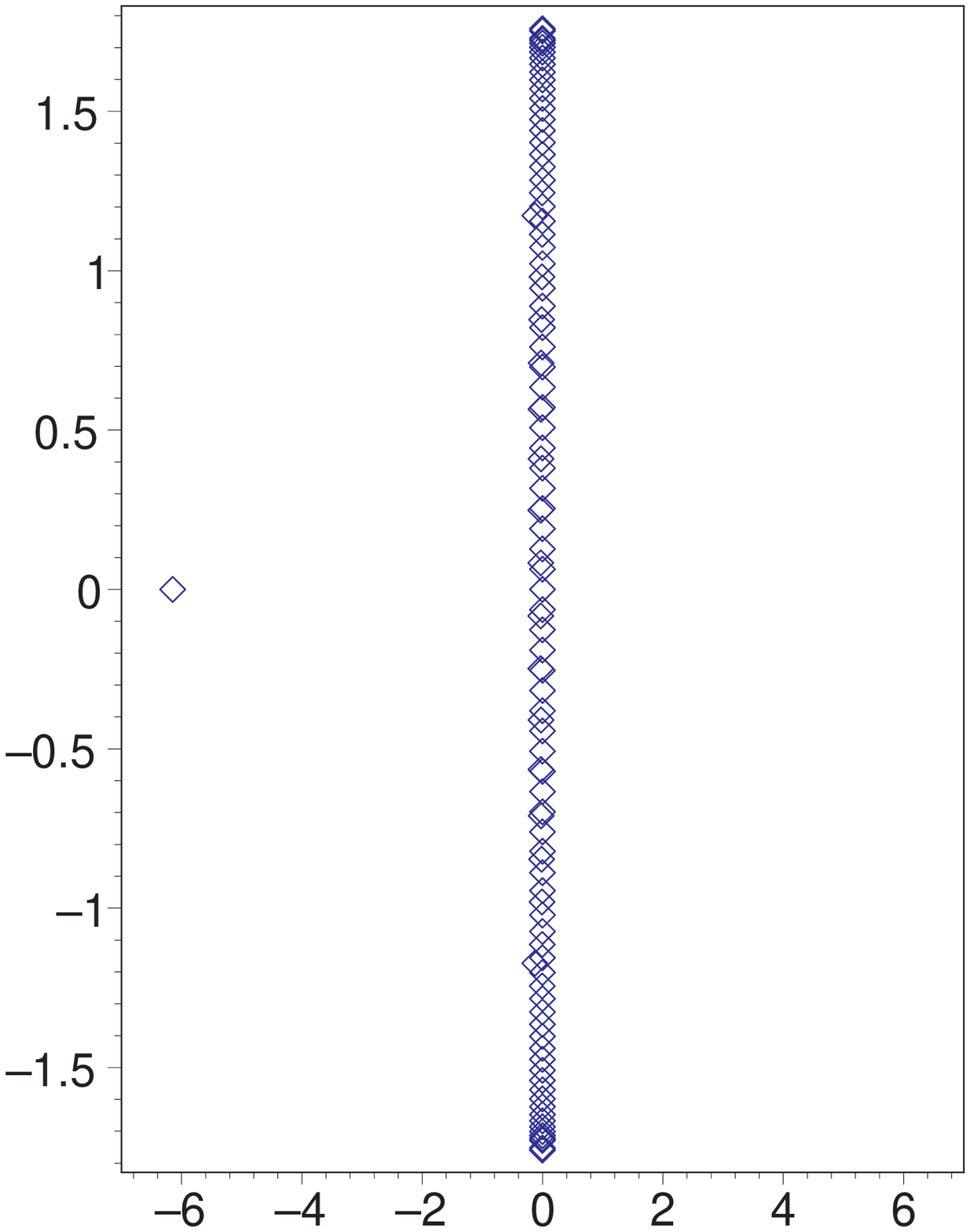}
\caption{Eigenvalues for the optimized $D_{8-4}$ operator, with $100$
  points and $\delta =0,1/10, 1/2$ (from left to right). }
\label{eigenvalues8adelta} 
\end{center}
\end{figure}

\textbf{Optimized operator}. We here construct an ``optimized $D_{8-4}$
operator'' (which we shall use from here on in the $D_{8-4}$ case) in 
the sense that it  has a spectral radius considerably smaller than that one
defined by Eq.(\ref{mbw}). More precisely, through a numerical 
search in the three-parameter space we have found that the following values 
\begin{equation}
x_1 = 0.541  , \;\;\; x_2= -0.0675, \;\;\; x_3 = 0.748 \label{good_pars}\, ,
\end{equation}
yield an operator whose maximum absolute eigenvalue in the purely imaginary case ($\delta=-1/2$) is
\begin{equation}
\lambda_{max} = 2.242 \label{good_lam} \, .
\end{equation}
This maximum eigenvalue appears to be quite sensitive on these parameters. For
example, truncating the above values to two significant digits, 
$$
x_1 = 0.54  , \;\;\; x_2= -0.067, \;\;\; x_3 = 0.75 \;, 
$$
gives $\lambda _{max}=2.698$
and truncating even more, to just one digit,  
$$
x_1 = 0.5  , \;\;\; x_2= -0.07, \;\;\; x_3 = 0.7 \;, 
$$
gives the large value $\lambda _{max} = 71.76$.
On the other hand, refining in the parameter search the values in Eq.(\ref{good_pars}) in one more digit
did not change the maximum of Eq.(\ref{good_lam}) in its four digits here shown. The
eigenvalues for $\delta = - 1/2$ for this optimized $D_{8-4}$ operator, given by the
parameters of Eq.(\ref{good_pars}), are shown in Figure \ref{eigenvalues8a},
while 
Figure \ref{eigenvalues8adelta} shows them for 
$\delta=0,1/10,1/2$ and $100$ points. As before, a negative real
component appears and the maximum in the imaginary
axis decreases (to around $1.754$). 

While completing this work we became aware of similar work by Svard, Mattson
and Nordstrom \cite{smn}, who
construct an optimized operator with different parameters by minimizing the
spectral radius of the derivative itself (rather than that of the
amplification matrix of a toy problem with an interface, as in our case), 
obtaining $x_1=0.649,x_2=-0.104,x_3=0.755$.  When using these parameters in our toy problem with
an interface, the resulting spectral radius (for twenty gridpoints) is
$\lambda_{max}=2.241259$, while for the parameters we chose
[cf. Eq.(\ref{good_pars})] is $\lambda_{max}=2.241612$ \cite{pdiener}.

\subsection{Global convergence rate}

\begin{figure}[ht]
\begin{center}
\includegraphics*[height=8cm]{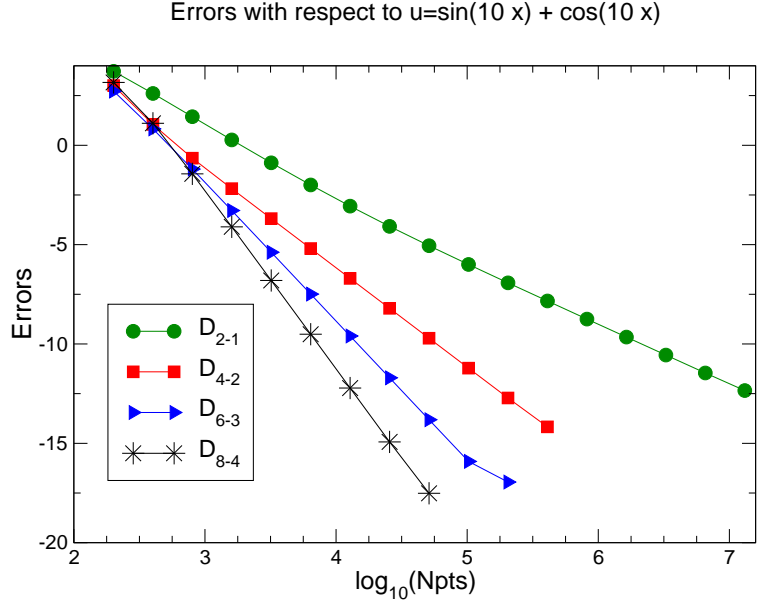} 
\caption{$L_2$ norms of the errors obtained when taking the discrete
  derivative of $\sin(10x) + \cos(10x)$ and comparing it with the analytical
  answer, using  $D_{2-1},D_{4-2},D_{6-3},D_{8-4}$ 
  operators, versus the number of gridpoints. }
\label{conv_high}
\end{center}
\end{figure}

\begin{figure}[ht]
\begin{center}
\includegraphics*[height=8cm]{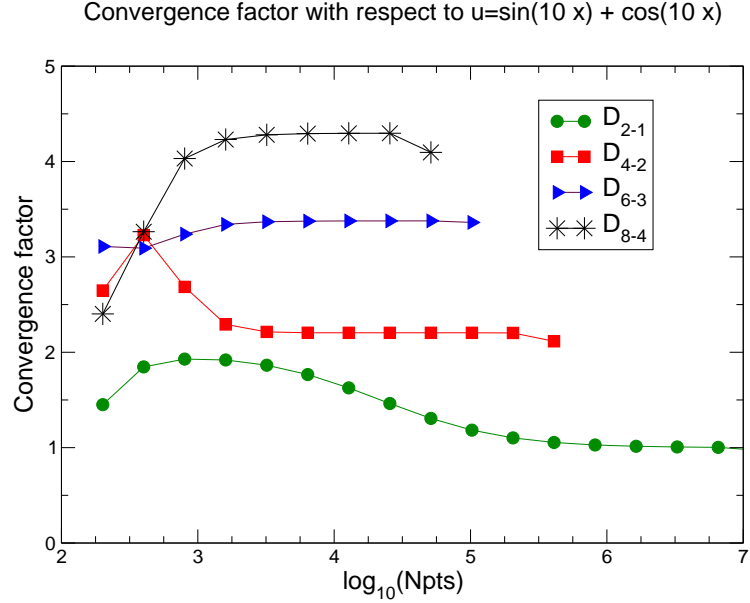} 
\caption{Convergence factors for the curves of Figure \ref{conv_high}.
As in this case,  the convergence in general will be dictated
by the lower order the derivative operators have at and close to the boundaries. Note that 
the lines corresponding to the different operators terminate at sequentially fewer points.
This is due to the corresponding errors reaching round-off levels, after which the convergence factor
calculation ceases to have a sensible meaning.}
\label{conv_high2}
\end{center}
\end{figure}

In general, the global (say, in an $L_2$ norm) convergence factor for these
operators will be dominated by the lower order at and close to
boundaries. However, it is sometimes found that 
roundoff values for the error in such a global norm are  
reached before this happens, 
and the convergence factor is different from the one expected from the boundary terms. 
The precise value is found to actually depend on the function being differentiated and whether one
reaches round-off level. To illustrate the expected behavior in a generic case, we consider the
function $\sin(10x) + \cos(10x)$ in the domain $x\in[0,2\pi]$. 
Figure \ref{conv_high}  shows the error (with respect to the exact solution) when computing 
the discrete derivative 
versus the number of gridpoints, for the difference operators
$D_{2-1},D_{4-2},D_{6-3}$, and  $D_{8-4}$. 
The errors in the $L_2$ norm are shown until roundoff values are reached (further increasing the number of gridpoints causes the error to grow 
with the number of points involved). Figure \ref{conv_high2}, in turn, shows the obtained convergence 
factors. 

\subsection{Group Speed}
\label{groupspeed}
We now turn our attention to the group speed that different discrete modes
have when the above considered operators are used. To simplify the
discussion, we actually restrict ourselves to the periodic case, which lends itself
for a clean analytical calculation. In this case the operators of order two, four,
 six and eight satisfying SBP are the standard, centered ones
 ($D_0,D_+,D_-$ denote the standard centered second order, and forward
 and backward first order operators, respectively):
\begin{eqnarray}
D^{(2)} &=& D_0 \label{2c}\\
D^{(4)} &=& D_0(I-h^2/6 D_+D_-) \label{4c}\\
D^{(6)} &=& D_0(I-h^2/6 D_+D_- + h^4/30 D^2_+D^2_-) \label{6c} \\
D^{(8)} &=& D_0(I-h^2/6 D_+D_- + h^4/30 D^2_+D^2_- - h^6/140 D^3_+ D^3_-) \label{8c}\\
\end{eqnarray} 
In discrete Fourier space, the eigenvalues for these operators are,
respectively, 
\begin{eqnarray}
\lambda_2 &=& \sin(\zeta)/\zeta \, ,\\
\lambda_4 &=& \sin(\zeta)/\zeta(1+2\sin(\zeta/2)/3) \, , \\
\lambda_6 &=& \sin(\zeta)/\zeta(1+2\sin(\zeta/2)/3+8\sin^4(\zeta/2)/15)\, ,\\
\lambda_8 &=& \sin(\zeta)/\zeta(1+2\sin(\zeta/2)/3+8\sin^4(\zeta/2)/15 + 64\sin^6(\zeta/2)/140) \, .
\end{eqnarray} 
where $\zeta = \omega h$, and $\omega$ is the associated wave number. 

The highest possible frequency is $\omega = N/2$ (with $N$ the number of
gridpoints). For that frequency,  $\zeta=\pi$ and the above eigenvalues are
all zero. Therefore the mode with highest possible frequency for a given
number of points does not 
propagate. Furthermore, if one examines the {\em group speed},
\begin{equation}
v_g = \frac{d (\lambda \omega)}{d\omega} \, ,
\end{equation} 
one finds that at $\zeta=\pi$ this speed is
\begin{eqnarray}
v_g^2 &=& -1 \, ,\\
v_g^4 &=& -5/3 \approx - 1.6 \, ,\\
v_g^6 &=& -33/15 \approx - 2.2 \, ,\\
v_g^8 &=& -341/135 \approx -2.6 \, . 
\end{eqnarray} 
Thus, for higher (than two) order operators the velocity of this mode is higher than
the continuum one (which is $1$). But, more importantly,  
in all cases the speed has the opposite sign. Of course, this effect goes away
with resolution, since the highest possible frequency moves to larger values as
resolution is increased. But, still, is an effect to be taken into account. 
For instance, if noise is produced at an interface, it propagates backwards, 
and with higher speed. Even though this effect is typically very small, it
might be noticeable in highly accurate simulations, or in simulations in which
the solution itself decays to very small values (see Section \ref{tails}). 
This could also be a source of difficulties in the presence of black holes
--or for this matter any system where some propagation speed changes sign-- 
since the event horizon traps these high frequency modes in a very narrow
region and then releases them as low frequency ones. 
We have observed this in some highly resolved one dimensional simulations,
and explains an observed convergence drop which goes away when numerical dissipation
is turned on. 

\section{Tests}
In this section we illustrate the behavior of the aforementioned penalty 
technique, together with the choice of different derivative operators. 
We present tests in one, two and three dimensions. In particular, we implement the linearized
Einstein equations (off a `gauge-wave' spacetime \cite{weakhyp,apples}) and
propagation of scalar fields in black hole backgrounds. The
former is cast in a way which yields a one-dimensional symmetric hyperbolic
system with coefficients depending both in space {\em and} time while the latter
provides  an hyperbolic system of equations with space varying
coefficients and sets a conforming grid for spherical black hole excision. 

Throughout this paper we employ a fourth order accurate Runge--Kutta time
integrator. In a number of tests aimed at examining the behavior of high
order operators we adopt a sufficiently small time step $\Delta t$
so that the time integrator does not play a role. Thus, we either 
choose 
a suitably small CFL factor or we scale the time step
quadratically with
the gridspacing $h$.

\subsection{One dimensional simulations: linearizations around a gauge wave} \label{1d}

As a first test we
 evolve Einstein's equations in one dimension, linearized around a background given by 
\begin{equation} 
ds^2 = e^{A \sin(\pi (x-t))} (-{dt}^2 + {dx}^2) + {dy}^2 + {dz}^2 . 
\label{gaugewavemetric} 
\end{equation} 
This background describes flat spacetime with a sinusoidal coordinate dependence,  
of amplitude $A$, along the $x$ direction. One of the interesting features of this
testbed is that while a linear problem, the coefficients in the equations to
solve are not only space but also time dependent. 

The non-trivial variables for this metric are 
\begin{eqnarray} 
\hat g_{xx} &=& e^{A \sin(\pi (x-t))}\, , \\ 
\hat K_{xx} &=& \frac{A}{2} \pi \cos\left(\pi \left(x-t \right)\right) 
e^{A/2 \sin\left(\pi \left(x-t\right)\right)} \;,  \\
\hat \alpha &=& e^{A/2 \sin\left(\pi \left(x-t \right)\right)} \, ,\\ 
\hat \beta^i &=& 0 \, . 
\end{eqnarray} 
We evolve the linearized Einstein equations using the symmetric hyperbolic formulation  
presented in Ref.~\cite{sarbach-tiglio} with a dynamical lapse given by 
the homogeneous time-harmonic condition (defined by requiring $\square t=0$). The formulation is cast in first order form  
by introducing the variables  
${\cal A}_x := \partial_x \alpha/\alpha$ and  $d_{xxx} := \partial_x g_{xx}$.  
The equations determining the dynamics of the (first order) perturbations,
which we assume to depend
solely on $(t,x)$, are obtained by considering 
linear deviations of a background metric given by Eq.~(\ref{gaugewavemetric}).  
That is, we consider 
\begin{eqnarray} 
g_{xx} &=& \hat g_{xx} + \delta g_{xx}  , \nonumber  \\  
K_{xx} &=& \hat K_{xx} + \delta K_{xx}  , \nonumber \\ 
d_{xxx} &=& \hat d_{xxx} + \delta d_{xxx}  , \nonumber \\ 
\alpha &=& \hat \alpha + \delta \alpha  , \nonumber \\ 
{\cal A}_{x} &=& \hat {\cal A}_{x} + \delta {\cal A}_{x}  \, ; \nonumber 
\end{eqnarray} 
replace these expressions in Einstein's equations and retain only first order
terms. The resulting equations are (henceforth dropping the $\delta$ notation) 
\begin{eqnarray} 
\dot{\alpha} &=& - A \pi \cos(\phi)  \alpha -  K_{xx}  
   + \frac{A \pi}{2 \hat{\alpha}} \cos(\phi)  g_{xx}\, , \label{gauge1}\\ 
\dot{\cal A}_x  &=&  -\frac{1}{\hat{\alpha}} \partial_x  K_{xx}  
       + \frac{A\pi}{2\hat{\alpha}} \cos(\phi)  K_{xx}\, , \nonumber \\ 
   & & -\frac{A\pi^2}{2\hat{\alpha}^2} \left ( A\cos(\phi)^2  
       + \sin(\phi) \right)  g_{xx}   \nonumber \\ 
   & & +\frac{A\pi}{2\hat{\alpha}^2} \cos(\phi)  d_{xxx}    \nonumber \\ 
   & &-\frac{A\pi}{2} \cos(\phi)  {\cal A}_{x} 
       + \frac{A\pi^2}{2\hat{\alpha}} \sin(\phi)  \alpha  \label{gauge2}\, ,\\ 
\dot{g}_{xx} &=& - A \hat{\alpha} \pi \cos(\phi)  \alpha  
       - 2 \hat{\alpha}  K_{xx} \label{gauge3}\, , \\ 
\dot{K}_{xx} &=& - \hat{\alpha} \partial_x  {\cal A}_{x}   
       - \frac{A \hat{\alpha} \pi}{2} \cos(\phi)  {\cal A}_{x} \nonumber \\ 
  &  & - \frac{A \pi^2}{4} \left ( -2 \sin(\phi)  
       + A \cos(\phi)^2 \right )  \alpha  \nonumber \\ 
  & &- A \pi \cos(\phi)  K_{xx} \nonumber \\ 
  & & + \frac{A\pi}{4\hat{\alpha}} \cos(\phi)  d_{xxx}\, , \label{gauge4} \\ 
\dot{d}_{xxx} &=& {A \hat{\alpha} \pi^2} \left ( \sin(\phi)  
      - A \cos(\phi)^2 \right ) \alpha  \nonumber \\ 
  & & - A \hat{\alpha} \pi \cos(\phi)  K_{xx} 
      - A \hat{\alpha}^2 \pi \cos(\phi)  {\cal A}_{x}   \nonumber \\ 
  & & - 2 \hat{\alpha} \partial_x  K_{xx} \, ,\label{gauge5} 
\end{eqnarray} 
where we have defined  $\phi := \pi (x-t)$. 
This system is symmetric hyperbolic and the symmetrizer used to define
the energy can be chosen so that the characteristic speeds which play a role
in the energy estimate are  
$0$ and $\pm 1$.  
 
We consider here a periodic initial boundary value problem on  
the domain $x\in[-1/2,3/2]$, where periodic boundary conditions at
$x=-1/2,3/2$ are implemented through an interface with penalty terms, as
described in Section  \ref{theory}. 
 
The system must satisfy two non-trivial constraint equations, corresponding to the  
definition of the variables $d_{xxx}$ and ${\cal A}_x$ (the linearized physical
constraints are automatically satisfied by the considered ansatz). When linearized, these 
 constraints are 
\begin{eqnarray} 
0 &=& C_x =  - \partial_x  g_{xx} +  d_{xxx} \;, \\ 
0 &=& C_{\cal A} =  {\cal A}_x - \frac{1}{\hat{\alpha }}  
\left ( \partial_x \alpha - \frac{A \pi}{2} \cos(\phi)  
\alpha \right ). 
\end{eqnarray} 

\begin{figure}[ht] 
\begin{center} 
\includegraphics*[height=7cm]{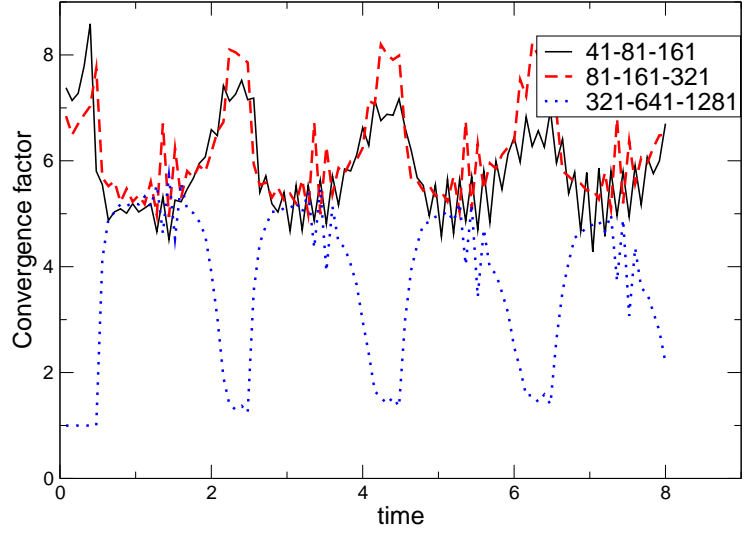} 
\caption{Evolutions of $1$d linearized Einstein's equations in a periodic
  domain, with periodicity enforced through an interface with penalty terms. Shown is the convergence
  factor for $K_{xx}$ when using the $D_{8-4}$ derivative, 
  CFL factor $\lambda=10^{-3}$, and dissipation
  $\sigma=5\times10^{-4}$. While the convergence factors obtained with
  $41$ to $321$ points oscillate between the expected order at the boundary
  and that one at the interior, the ones calculated with $321$ to $1281$ points
  are not meaningful when the pulse is located at interior points as round-off level
  is reached.}
\label{conv_k_48_001}  
\end{center} 
\end{figure}

In the first series of simulations we adopt a CFL factor $\lambda = 10^{-3}$\footnote{We do not claim
that this is the largest possible CFL factor for which the errors due to time
integration are negligible compared to the spatial ones, though. Actually, as
discussed later, in some cases one can use larger CFL factors.} and consider relatively short
evolutions corresponding to four crossing times. The $D_{8-4}$ derivative is used, and 
dissipation is added through the dissipative operator
constructed from $-\sigma h^7 D_+^4 D_-^4$, suitably modified at boundaries as
explained in Appendix \ref{dis} so as to make it non-positive definite with respect to the
appropriate scalar product. Thus, the use of this dissipative operator does {\em
  decrease} the order of the spatial discretization 
by one. The dissipation parameter used is $\sigma=5\times10^{-4}$. 
Figure \ref{conv_k_48_001} exemplifies the
behavior observed in the convergence of the field $K_{xx}$ (the other fields behave
similarly). As time progresses the convergence order
obtained oscillates in a way that is  consistent with the accuracy
obtained at interior and boundary points.

\begin{figure}[ht] 
\begin{center} 
\includegraphics*[height=7cm]{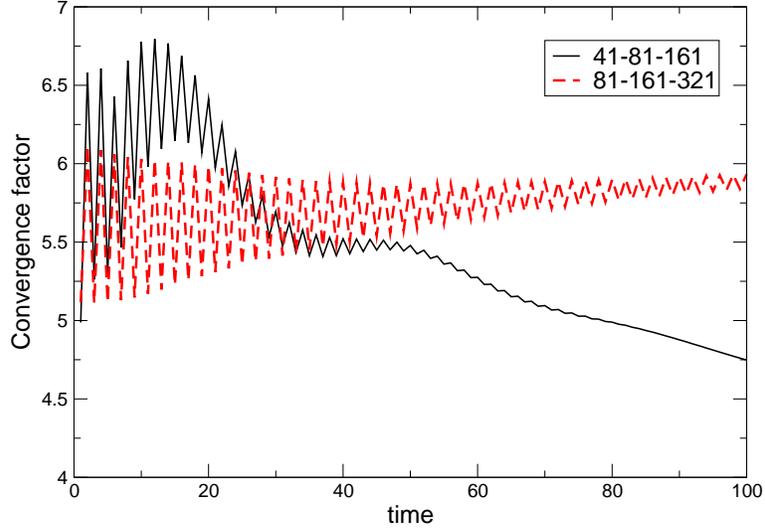} 
\caption{Same as previous figure, but with a decreasing CFL factor given
  by $\lambda=0.2/2^n$ with $(n=1..3)$.}. 
\label{conv_k_48_leco}  
\end{center} 
\end{figure}

Next, we adopt as a starting value for the CFL factor defined at the
coarsest grip to be $\lambda=0.2$ but in subsequent grids (refined by a factor of $2$)
we adopt $\lambda=0.2/2^n$ with $(n=1..3)$. Figure \ref{conv_k_48_leco} illustrates the
behavior observed; again, as time progresses the convergence order
obtained oscillates in between the order of accuracy of interior and boundary points,  with the additional
effect of accuracy loss due to the accumulation of error as time progresses. 

\begin{figure}[ht]
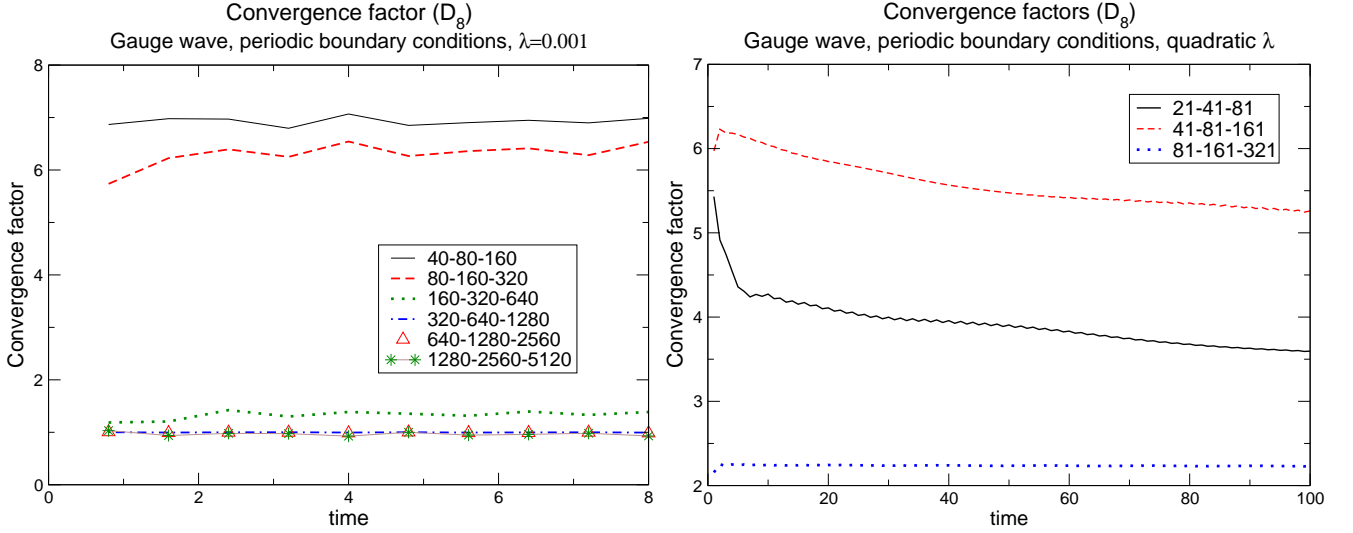

\begin{center}
\includegraphics*[height=7cm]{convg_per_48_cfl001.eps}
\includegraphics*[height=7cm]{convg_per_48_cflscaled.eps}
\caption{This figure shows evolutions similar to those of
  Figs.\ref{conv_k_48_001}, \ref{conv_k_48_leco}, with the only difference
  that periodicity is here enforced explicitly. The convergence factors for
  the metric component $g_{xx}$ are shown.}.
\label{ggauge1} 
\end{center}
\end{figure}

Finally, we compare the above results with those obtained in the ``truly periodic
case'', ie. when periodicity is used explicitly to employ the same
derivative operator at all points. We again consider cases where a sufficiently
small  CFL factor ($=10^{-3}$) is used or the time-step is scaled quadratically. 
Figures \ref{ggauge1} illustrates the observed convergence rate for the field
$g_{xx}$. As above, dissipation is added through a seventh order
dissipative operator (but now with no modification at boundaries needed) 
with same dissipative parameter: $\sigma=5\times10^{-4}$. 
While the errors remain above round-off level  
the observed convergence rate is consistent with the expected one of seven, as the orders
of the derivative and dissipative operators employed are eight and seven respectively.
Certainly, dissipation of higher order could have been introduced by simply employing
the KO style operator $h^9 (D_-D_+)^5$, but we have adopted this one to more
directly compare with the case with interface boundaries. For the highest
resolutions the errors reach round-off level and the obtained convergence
factors yield non-sensible values.

\begin{figure}[ht] 
\begin{center} 
\includegraphics*[height=7cm]{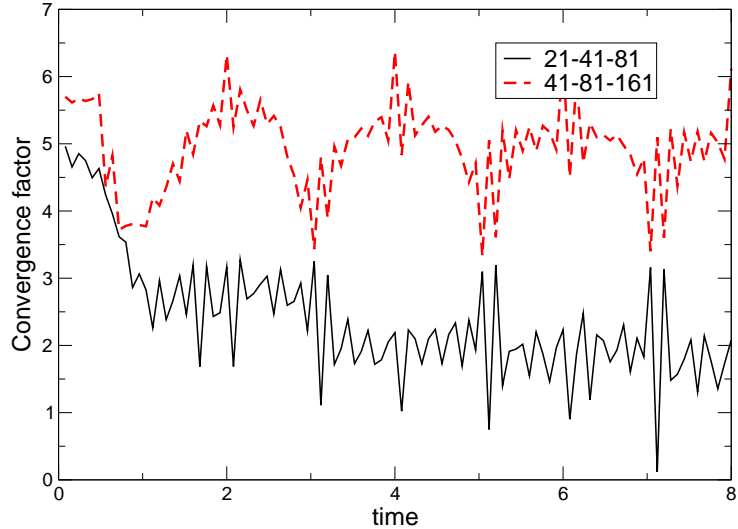} 
\caption{Similar to Fig. \ref{conv_k_48_001}, but using the $D_{6-3}$ derivative.
  The convergence factor for $K_{xx}$ is shown.} 
\label{conv_k_36_001}  
\end{center} 
\end{figure}

As an illustration of what is observed with other derivatives, we briefly discuss
some simulations using the $D_{6-3}$ operator and a fixed CFL factor
(given, as before, by $\lambda=10^{-3}$). 
Analogously as to was done above, dissipation is here added by
extending --as discussed in Appendix \ref{dis}-- the operator $\sigma h^6
D_-^3 D_+^3$ at and near boundaries in order to make it non-positive
definite with respect to the appropriate scalar product; a dissipation
parameter $\sigma=10^{-3}$ is used. The observed results are illustrated in 
Figure \ref{conv_k_36_001}, which shows the self-convergence factor for
$K_{xx}$. As before, it oscillates between the order of the scheme in the
interior and that one at boundary points. 

Summarizing, the results presented indicate that, in the case where boundaries 
are present, the worst case scenario --as far as the
expected convergence rate relates-- is determined by the accuracy order at and
close to boundary points.

\subsection*{Two and three-dimensional simulations. Problem set-up}

In this section we solve the wave equation for a scalar field $\phi$ propagating
on a fixed background, 
$$
\nabla^a \nabla_a \phi = 0 \, ,
$$
where $\nabla $ is the covariant derivative associated with the metric of the
background. We will consider two backgrounds: a $2$d one consisting of the unit
sphere with its standard metric and a $3$d one consisting of a rotating Kerr black hole background. 

We start by describing in more detail the equations solved and the multiple
coordinate system used, and then present the actual results of the
simulations. 

\subsection*{A strictly stable scheme for the wave equation in a time-independent, curved spacetime}
The wave equation in a time-independent background can be written, using any coordinates in which 
the metric is manifestly time independent as,
\begin{eqnarray}
\dot{\phi } &=& \alpha \Pi \label{phi_dot}\\
\dot{\Pi} &=& \beta^i \alpha^{-1} D_i(\alpha \Pi) +
h^{-1/2}D_i(h^{1/2}\beta^i\Pi + \alpha h^{1/2}H^{ij}d_j)\label{pi_dot}\\
\dot{d}_i &=& D_i (\alpha \pi) \label{d_dot}
\end{eqnarray}
where $H^{ij}:= h^{ij}-\alpha^{-2}\beta ^i \beta ^j$, $h^{ij}$ is the inverse
of the three-metric, $h={\rm det}(h_{ij})$, $\alpha$ is the lapse and $
\beta ^i$ the shift vector. 
The advantage of writing the equations in this way is that one can show that if $D$ is any
difference operator satisfying SBP, this form of the equations guarantees that
the semidiscrete version of the physical energy is a non-increasing function of
time. When the killing field is timelike this means that there is a norm in
which the solution is bounded for all times, 
thus suppressing artificial fast growing-modes without the need of
artificial dissipation (see \cite{strict} for
details). 

We now look at the characteristic variables and characteristic speeds with
respect to a ``coordinate'' observer. That is, the eigenfields and eigenvalues of the symbol
$A^in_i$, where $A^i$ denotes the principal part of the evolution equations
and $n_i$ the normal to the boundary \footnote{If the boundary is aligned with 
  the coordinates lines, as is indeed the case here, $n_i$ would be either $(1,0,0)$,
  $(0,1,0)$, or $(0,0,1)$.}. The characteristic variables with non-zero speeds
$$
\lambda^{\pm} =  (\pm \alpha + \beta^k\hat{n}_k)(h^{ij}n_i n_j)^{1/2}
$$ 
[where  $\hat{n}_k= n_k(h^{ij}n_i n_j)^{-1/2}$] are
$$
v^{\pm} = \lambda^{\pm }\Pi + \alpha H^{ij}  \hat{n}_id_j\;\;\; ; 
$$
while the zero speed modes are
$$
v^0_i = d_i - \hat{n}_i d_j\hat{n}^j \;.
$$

\subsection*{Cubed-sphere coordinates.}
The topology of the computational domain in our $2$d simulations is $S^2$, the
unit sphere, while in our $3$d ones it is $S^2 \times R^+$. Since it is not possible 
to cover the sphere with a single system of
coordinates which is regular everywhere, we employ multiple patches to
cover it. A convenient set of patches is defined by the
{\it cubed sphere coordinates}, defined as follows (for a related definition
see for instance \cite{cubed}). 

Each patch uses coordinates $a,b,c$, 
where $c=\sqrt{x^2+y^2+z^2}$, the standard radial coordinate, is the same 
for the six patches ($x,y,z$ are standard Cartesian coordinates). The other 
two coordinates, $a,b$
are defined as
\begin{itemize}
\item Patch 0 (neighborhood of $x=1$): $a=z/x,b=y/x$
\item Patch 1 (neighborhood of $y=1$): $a=z/y,b=-x/y$
\item Patch 2 (neighborhood of $x=-1$): $a=-z/x,b=y/x$
\item Patch 3 (neighborhood of $y=-1$): $a=-z/y,b=-x/y$
\item Patch 4 (neighborhood of $z=1$): $a=-x/z,b=y/z$
\item Patch 5 (neighborhood of $z=-1$): $a=-x/z,b=-y/z$
\end{itemize}

\begin{figure}[ht]
\begin{center}
\includegraphics*[height=7cm]{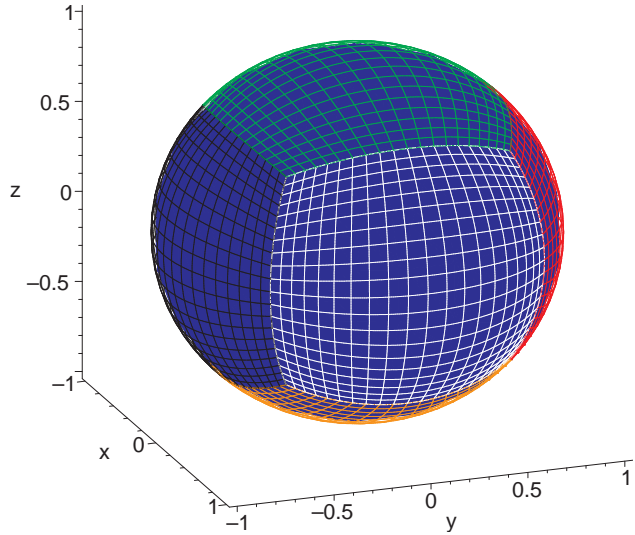}
\caption{Cubed-sphere coordinates for $S^2$.}.
\label{cubed} 
\end{center}
\end{figure}

Similarly, the inverse transformation is:
\begin{itemize}
\item Patch 0: $x=c/D, y=cb/D, z=ac/D$.
\item Patch 1: $x=-bc/D, y=c/D, z=ac/D$
\item Patch  2: $x=-c/D, y=-cb/D, z=ac/D$.
\item Patch 3: $x=bc/D, y=-c/D, z=ac/D$
\item Patch 4: $x=-ac/D, y= cb/D, z=c/D$
\item Patch 5: $x=ac/D, y= cb/D, z=-c/D$
\end{itemize}
with $D:=\sqrt{1+a^2+b^2}$. This provides a relatively simple multi-block 
structure for $S^2$ which can be exploited to implement the penalty technique
in a straightforward manner. Each patch is discretized with a uniform grid
in the coordinates $a$ and $b$, and the requirement of boundary points coinciding
in neighboring grids is indeed satisfied.  Figure \ref{cubed} shows this
gridstructure, for $20\times 20$ points on each patch. 

\subsection{$2$d Simulations}
We now discuss simulations of the wave equation on the unit sphere in
cubed-sphere coordinates, 
written in strictly stable
first order form [Eqs.(\ref{phi_dot},\ref{pi_dot},\ref{d_dot})]. The metric
used, therefore, is flat spacetime projected to the $r=1$ slice, which in local coordinates is
$$
ds^2 = -{dt}^2 + D^{-4}\left[ (1+b^2) \,{da}^2 + (1+a^2)\, {db}^2 - 2
  \, a \,b\, {da}\, {db} \right]\; , 
$$
where $D:=\sqrt{(1+a^2+b^2)}$.

\begin{figure}[ht]
\begin{center}
\includegraphics*[height=7cm]{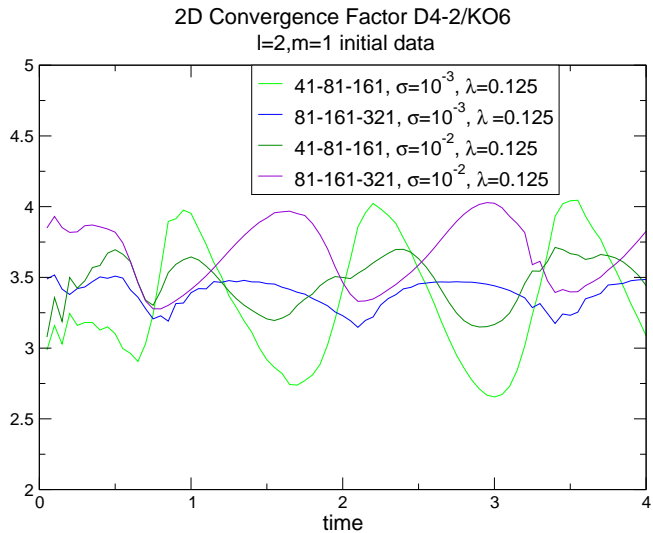}
\caption{Evolutions of the wave equation on the unit sphere, using
  cubed-sphere coordinates and a pure $l=2,m=1$ spherical harmonic as initial
  data. Shown is the convergence factor when the $D_{4-2}$
  derivative and the KO6 dissipation operators are used.}.
\label{fig1_2d} 
\end{center}
\end{figure}

Figure \ref{fig1_2d} shows simulations using the $D_{4-2}$ derivative and its associated 
dissipative operator constructed in Appendix \ref{dis}, which we call KO6, using $n\times n$ points on each of the six patches, where 
$n=41,81,161,321$. The initial data for $\Pi$ corresponds to a pure $l=2,m=1$ spherical
harmonic. The CFL factor used is $\lambda =0.125$ and for each set of runs two values 
of dissipation are used: $\sigma = 10^{-2}$ and $\sigma= 10^{-3}$. 
As can be seen from the Figure, the self convergence 
factor obtained with these resolutions 
is above the lower value (two)  expected from the order at the interfaces.  
The reason for the lower order at the interfaces not dominating is likely due to the
fact that the initial data is an 
eigenmode of the Laplacian operator, and the solution at the continuum is just an oscillation in 
time of this initial data, without propagation across the
interfaces. Indeed, the
oscillations in the convergence factors in Fig.\ref{fig1_2d} appear when the
numerical solution goes through
zero, and the frequency at
which this happens coincides approximately with the expected frequency at the
continuum for this mode.

The same initial data is now evolved with the $D_{6-3}$ derivative and KO8
dissipation (again, see Appendix \ref{dis}) and the results are shown in Figure \ref{fig2_2d}. 
As before, $\lambda=0.125$ and 
$n=41,81,161,321$ points are used, but 
the values of dissipation shown are now $\sigma=0$ (i.e., no dissipation) and
$\sigma=10^{-3}$. 
{\it At the same resolutions} there is a small 
difference in the obtained convergence factors, depending on the value of
$\sigma$, but with 
both values of this parameter the order of convergence is higher than the lower one expected from the time 
integrator if this one dominated. Figure 
\ref{fig2_2d} also presents a comparison made with a smaller CFL factor: $\lambda=0.0125$, 
keeping the dissipation at $\sigma=10^{-3}$. One more resolution is used ($641$ points) to look 
for differences between the solutions obtained with the two CFL factors, 
but they do not appear. This seems to suggest that 
at least in this case, and for these resolutions, 
 it is not necessary to use too small a CFL factor in order to avoid the 
time integrator's lowest order to dominate over the higher spatial discretization
(see Fig.\ref{conv_rad_3d} for another instance where this happens). It is also worth 
pointing out that the difference between the two highest resolutions is not quite at roundoff level, 
but it is rather small (of the order of $10^{-9}$ if scaled by the amplitude of
the initial data), as shown in Figure \ref{fig3_2d}. That figure shows the
$L_2$ norm of the differences between the solution at different resolutions,
for the simulations of Fig. \ref{fig2_2d} with $\lambda = 0.0125$ and $\sigma =
10^{-3}$. 

\begin{figure}[ht]
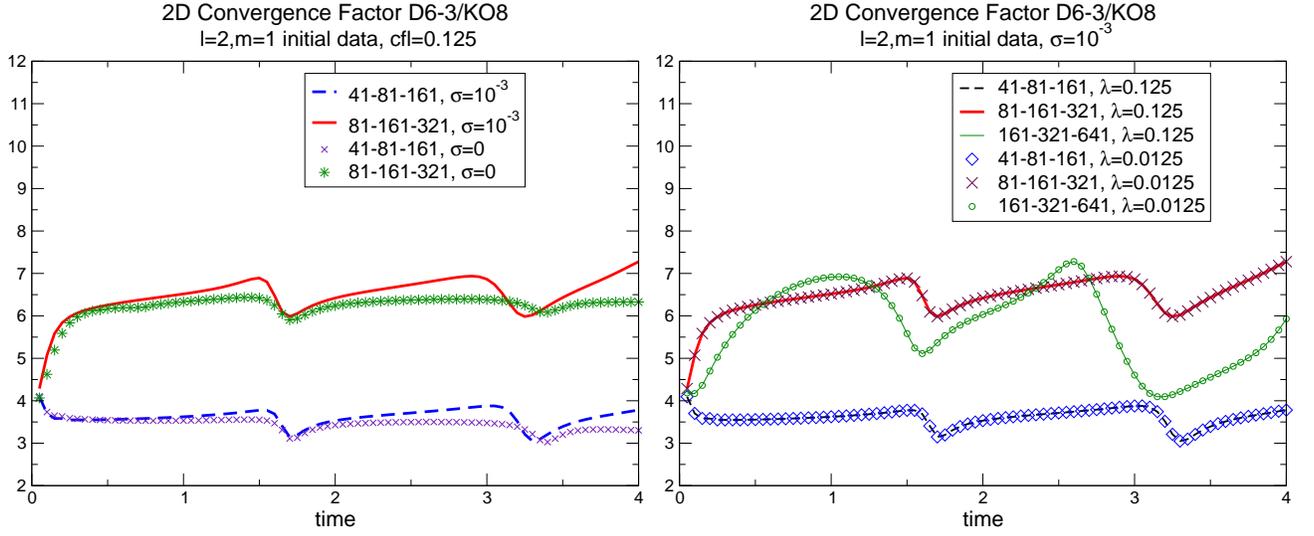

\begin{center}
\includegraphics*[height=7cm]{Q_factor_d6_3_KO8_l21a.eps}
\includegraphics*[height=7cm]{Q_factor_d6_3_KO8_l21.eps}
\caption{Same evolution equation and initial data as those used in Figure
  \ref{fig1_2d}, except that now the $D_{6-3}$ derivative and KO8 dissipation are used.}.
\label{fig2_2d} 
\end{center}
\end{figure}

\begin{figure}[ht]
\begin{center}
\includegraphics*[height=7cm]{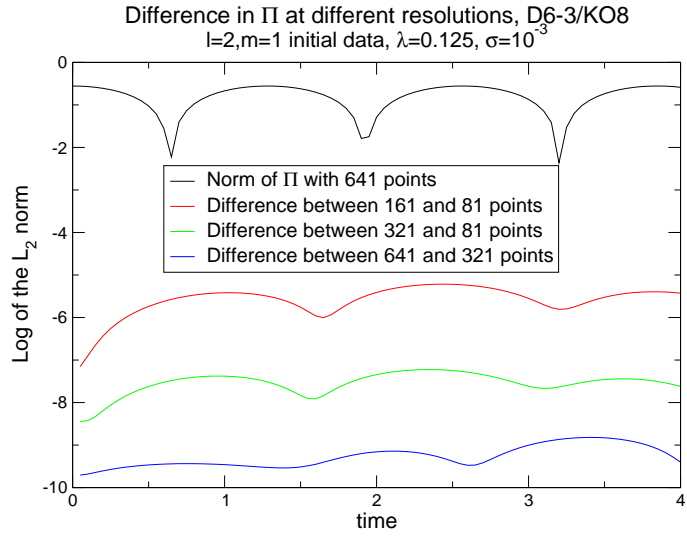}
\caption{$L_2$ norms of the differences between the solution at different
  resolutions, for the simulations of Figure \ref{fig2_2d} with $\lambda = 0.0125$ and $\sigma =
10^{-3}$.}.
\label{fig3_2d} 
\end{center}
\end{figure}

\begin{figure}[ht]
\begin{center}
\includegraphics*[height=7cm]{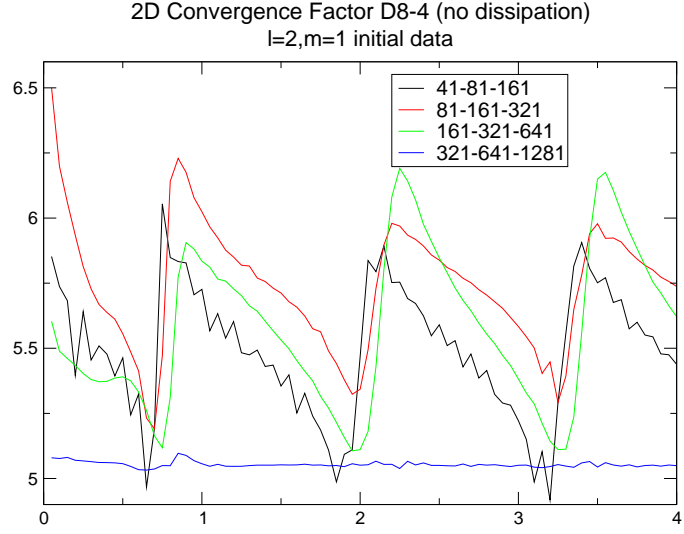}
\caption{Simulations of the same equation and initial data as those of the
previous figures, except that now the higher order, $D_{8-4}$ derivative is
used and no dissipation has been added.}.
\label{fig4_2d} 
\end{center}
\end{figure}

Finally, Figure \ref{fig4_2d} shows evolutions of the same initial data, with
the $D_{8-4}$ derivative, and 
no dissipation. The CFL factor is decreased when resolution is increased, much
as in Section  \ref{1d}, so that the order of the time integration does not
dominate over the higher one of the spatial discretization. That
is, for the resolutions shown we used
$\lambda=0.25,0.125,0.0625,0.03125,0.015625$.  
The convergence factor obtained is also higher than that one expected from the
lower order at the 
interfaces (presumably for the same reason as before, the initial data adopted) and higher 
than that one of the time integration. 

\subsection{Three dimensional simulations}
In this application we consider fields propagating on a Kerr black hole
background, as governed by equations 
(\ref{phi_dot},\ref{pi_dot},\ref{d_dot}), with the background metric written 
in Kerr-Schild form and cubed-sphere coordinates used for the angular directions. 
Homogeneous maximally dissipative boundary conditions are used at the outer
boundary, while no 
condition is needed at the inner one if it is {\em appropriately} placed inside
the black hole so that it constitutes a purely-outflow surface.

\subsubsection*{The Kerr-Schild metric in cubed-sphere coordinates}

The Kerr metric in Kerr-Schild form is 
$$
ds^2 = \eta_{\mu \nu} + 2H l_{\mu }l_{\nu }dx^{\mu }dx^{\nu }
$$
where $\eta_{\mu \nu }$ is the flat metric [with signature $(-,+,+,+)$], 
\begin{eqnarray}
H &=& \frac{mr}{r^2 + A^2\cos^2 {\theta }} \, ,\\
r^2 &=& \frac{1}{2}(\rho ^2-A^2) + \sqrt{\frac{1}{4}(\rho ^2-A^2)^2 + A^2z^2}  \, ,\\
\rho ^2 &=& x^2+y^2+z^2 \, ,
\end{eqnarray}
and $l^{\mu}$ is a null vector (both with respect to the flat metric and whole metric).

Therefore, in order to write the above metric in cubed-sphere coordinates one needs to 
write $\eta_{\mu \nu}$ and $l_{\mu}$ in these coordinates. A straightforward change of coordinates of 
the first one gives
$$
\eta_{\mu \nu}= -{dt}^2 +  {dc}^2 + c^2D^{-4}\left[
  (1+b^2){da}^2 + (1+a^2){db}^2 - 2a\,b\,{da}\,{db} \right]
$$
with $D:=\sqrt{(1+a^2+b^2)}$.  

In Cartesian coordinates the $l_{\mu}$ co-vector is, in turn, 
$$
l\equiv l_{\mu} dx^{\mu} = {dt} + \frac{rx + Ay}{r^2+A^2}\, {dx} +
\frac{ry-Ax}{r^2+A^2}\, {dy} + \frac{z}{r}\, {dz}
$$
with $x^{\mu} = (t,x,y,z)$ which, when changed to cubed-sphere coordinates $(t,a,b,c)$ gives
\begin{eqnarray}
l &=& {dt}+ \frac{-c^2aA^2(a^2-D^2)}{D^4r(r^2+A^2)} {da} +
 \frac{-c^2A(D^2r+Aba^2)}{D^4r(r^2+A^2)} {db} +
\frac{c(D^2r^2+a^2A^2)}{D^2r(r^2+A^2)} {dc}  \;\;\;\mbox{for patches 0-3} \\
&& \nonumber \\
l &=& {dt}+ \frac{-c^2A(D^2rb+aA)}{D^4r(r^2+A^2)}{da} +
\frac{c^2A(raD^2-Ab)}{D^4r(r^2+A^2)} {db} +
\frac{c(D^2r^2+A^2)}{D^2r(r^2+A^2)} {dc} \;\;\; \mbox{patch 4} \\
&& \nonumber \\
l &=& {dt}+ \frac{-c^2A(-D^2rb+aA)}{D^4r(r^2+A^2)}{da} + 
\frac{-c^2A(raD^2+Ab)}{D^4r(r^2+A^2)} {db} + 
\frac{c(D^2r^2+A^2)}{D^2r(r^2+A^2)}{dc} \;\;\; \mbox{patch 5}
\end{eqnarray}

To write the wave equation, one also needs the inverse metric, which is 
$$
g^{\mu \nu} = \eta^{\mu \nu} - 2H l^{\mu }l^{\nu } \, ,
$$
where all indices are raised with $\eta ^{\mu \nu}$ (the 
inverse of the flat metric). The non-zero components of the latter are:
\begin{eqnarray}
\eta^{aa} &=& \frac{D^2(1+a^2)}{c^2} \, , \\
\eta^{bb} &=& \frac{D^2(1+b^2)}{c^2} \, ,\\
\eta^{cc} &=& 1 \, ,\\
\eta^{tt} &=& -1 \, ,\\
\eta^{ab} &=& \frac{abD^2}{c^2} \, ,
\end{eqnarray}
and the vector $l^{\mu}$ in the cubed-sphere coordinates is,
\begin{eqnarray}
l^{\mu} &=& \left[-1, \frac{-aA(rb-A)}{r(r^2+A^2)},  
\frac{A(a^2-D^2)}{r^2+A^2},  \frac{c(D^2r^2+a^2A^2)}{D^2r(r^2+A^2)}  \right]^{\mu} 
\;\;\; \mbox{patches 0 to 3} \\
l^{\mu}&=& \left[-1,  -\frac{A(rb+aA)}{r(r^2+A^2)}, \frac{A(ar-bA)}{r(r^2+A^2)},
 \frac{c(D^2r^2+A^2)}{D^2r(r^2+A^2)}     \right]^{\mu} \;\;\; \mbox{ for patch 4} \\
l^{\mu}&=& \left[-1,  \frac{A(rb-aA)}{r(r^2+A^2)}, -\frac{A(ar+bA)}{r(r^2+A^2)},
 \frac{c(D^2r^2+A^2)}{D^2r(r^2+A^2)}       \right]^{\mu} \mbox{ for patch 5}\\
\end{eqnarray}

\subsubsection{Convergence tests}

\begin{figure}[ht]
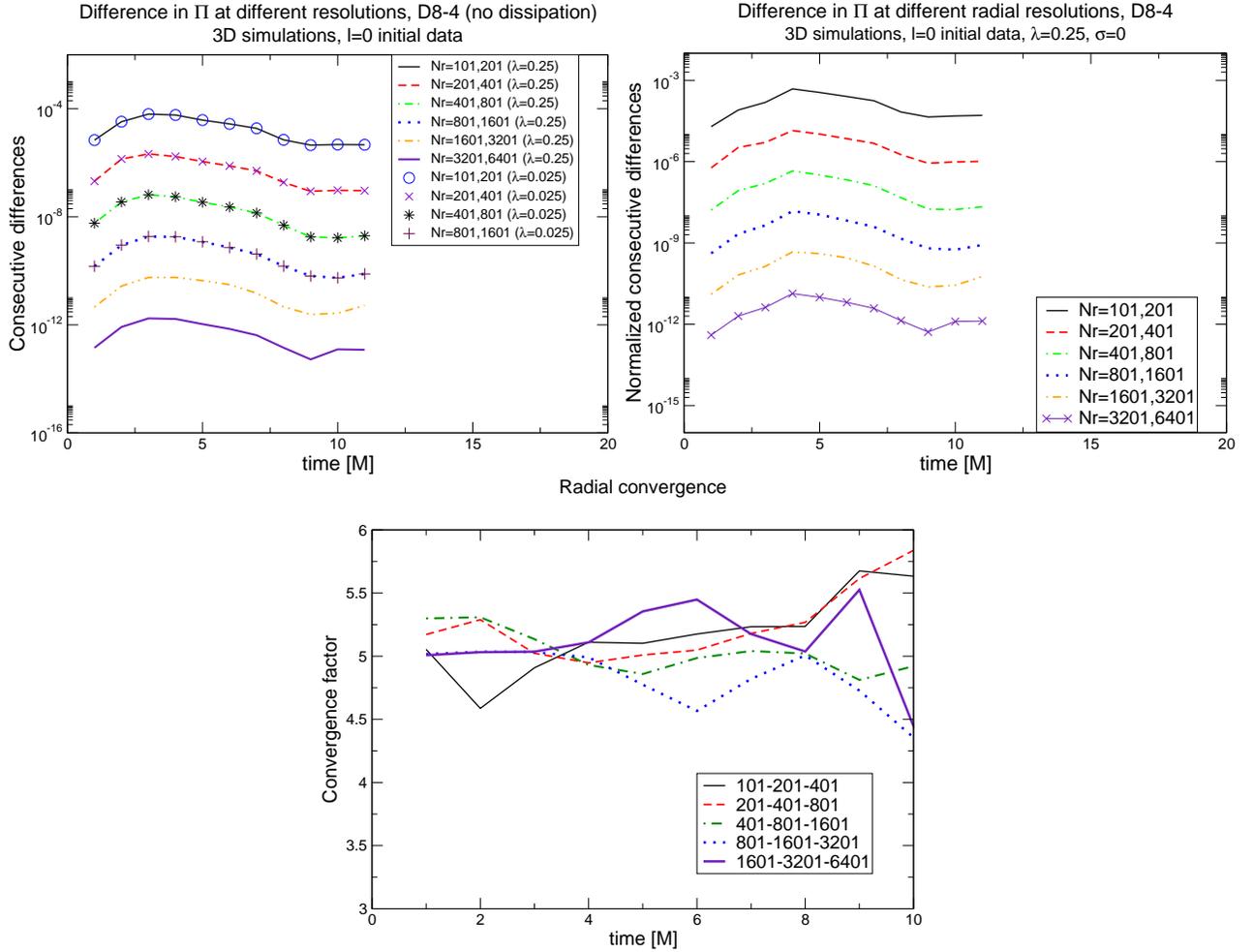

\begin{center}
\includegraphics*[height=6.5cm]{3d_small.eps}
\includegraphics*[height=6.5cm]{3d_small_normalized.eps}
\includegraphics*[height=6.5cm]{conv_3d_small.eps}
\caption{Convergence test in the radial direction, with CFL factors $\lambda
  =0.25,0.025$. Notice that no appreciable difference is found between
  $\lambda =0.25$ and a smaller value and even with $\lambda =0.25$ the convergence factor is
  not dominated by the time integrator.}
\label{conv_rad_3d} 
\end{center}
\end{figure}

\begin{figure}[ht]
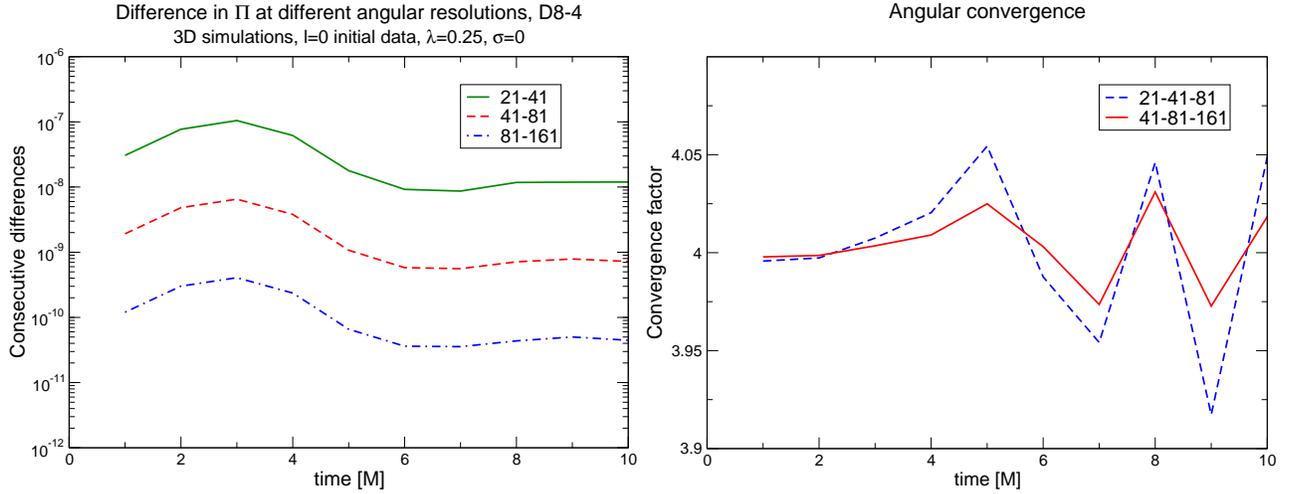

\begin{center}
\includegraphics*[height=6.5cm]{3d_small_ang.eps}
\includegraphics*[height=6.5cm]{conv_3d_small_ang.eps}
\caption{Convergence test in the angular direction, with a CFL factor $\lambda =1/4$}.
\label{conv_rad_3d_ang} 
\end{center}
\end{figure}

Figure \ref{conv_rad_3d} shows the differences, in the $L_2$ norm, between the numerical 
solutions at consecutive resolutions, using the $D_{8-4}$ scheme, with no
dissipation. The number of points in the 
angular directions is kept fixed to $16\times16$ points on each of the six patches, and the number 
of radial points ranges from $101$ to $6401$. The background is defined by a non-spinning black hole, and the 
inner and outer boundaries are at $1.9M$ and $11.9M$, respectively. Non-trivial initial data 
is given only to $\Pi$, in the form of a spherically symmetric Gaussian multipole:
\begin{equation}
\Pi(0,\vec{x}) = A\exp{(r-r_0)^2/\sigma_0^2} \label{gauss} \, ,
\end{equation}
with $r_0=5M,\sigma=M, A=1$. The resulting convergence factors, the normalized differences $||u_N -
  u_{2N}||/||u_N||$ and the non-normalized ones, $
||u_N - u_{2N}||$, are shown. The use of multiple patches not only allows for
non-trivial geometries, but additionally one is able to 
  define coordinates in a way such that resolution is adapted to the problem of interest. For example, in the
geometry being considered, $S^2\times R$, one employs a number of points in the angular direction limited
by the expected multipoles of interest and concentrates resources to increase the number of points 
in the radial direction. As
an example, the relative differences between the solution at different
resolutions shown in Figure \ref{conv_rad_3d} reaches values close to
roundoff, with modest computational resources. Even though the solution
here evolved is spherically symmetric at the continuum, as discussed below the number of points
used on each of the six patches that cover the sphere can 
reasonably resolve an $l=2$ multipole.
Next, Figure \ref{conv_rad_3d_ang} shows similar plots, but keeping the
 number of radial points fixed (to $101$), using $N_a\times N_a$
points on each of the six patches in the sphere, with $N_a=21,41,81$. 

\subsubsection{Tail runs} \label{tails}
To illustrate the 
behavior of the described techniques in  $3$d simulations we examine the
propagation of scalar fields on a Kerr black hole
background. The numerical undertaking of such problem has been previously
treated using pseudo-spectral methods \cite{tail_spectral}, which for smooth
solutions allows the construction of very efficient schemes. 
As explained next, the combination of multi-block evolutions with high order schemes
also lets one to treat the problem quite efficiently. A detailed study of this problem will
be presented elsewhere \cite{nils}; we here concentrate on two representative examples of
what is achievable.  

In the first case we examine the behavior of the scalar field propagating
on a background defined by a black hole with mass $M=1$ and spin parameter $a=0.5$.
Non trivial initial data is given only to $\Pi$, with a radial profile
given by a Gaussian pulse 
as in Eq.(\ref{gauss}) and angular dependence given by a pure $l=2$ multipole.
The inner and outer boundaries are placed at $r=1.8M$ and $r=1001.8M$ respectively. We adopt
a grid composed by six cubed-sphere patches, each of which is discretized with $20\times 20$ points in the
angular directions and $10001$ points in the radial one.
This translates into a relatively inexpensive calculation.

\begin{figure}[ht]
\begin{center}
\includegraphics*[height=7cm]{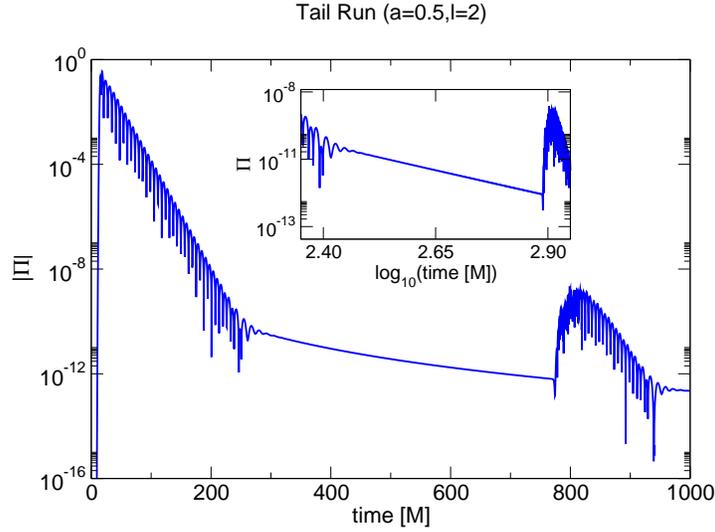}
\caption{Behavior of the time derivative of the scalar field
  for a pure multipole $l=2$ initial data, on a
  Kerr background, with $a=0.5$. Inner and outer boundaries are at $1.8M$ and
  $1,001.8M$, respectively and the six-patches grid is covered by
  $20\times20\times 10,001$ points 
  on each patch. The $D_{8-4}$ derivative
  is used, with no artificial dissipation. The noise at $t\approx 800M$ is due to the
  mode with group speed $-2.6$ discussed in Section \ref{HO}
  hitting the outer boundary and reaching the observer at the black hole
  horizon. This noise goes away by either pushing the outer boundary, increasing
  resolution and/or adding dissipation. 
The average slope for $\Pi$ in the interval $t\in [350M,750M]$ gives a decay for $\Pi$ of $t^{-4.07}$, 
in good agreement with the expected decay of $t^{-4}$. The inset shows a zoom in at the tail behavior.}
\label{tailkerr} 
\end{center}
\end{figure}

We adopt the  $D_{8-4}$ derivative operator, add no artificial dissipation and choose
a CFL factor $\lambda=0.25$. The salient features of the solution's  behavior observed 
are summarized in figure \ref{tailkerr} which shows the 
the time derivative of the scalar field, as a function of time,
at a point in the equatorial plane, on the even horizon. 
At earlier stages, the familiar quasi-normal ringing is observed. Next the late-time
behavior of the field reveals the expected tail-behavior as a fit 
in the interval $t\in [350M,750M]$ gives a decay for $\Pi$ of $t^{-4.07}$,
which agrees quite well the expected decay of $t^{-4}$. This can be
understood in terms of the generation of an $l=0$ mode in the solution due to
the spin of the black hole \cite{tail_spectral,burko-khanna}. Finally, noise can be observed appearing at
$t\approx 800M$  due to 
the outer boundary. This noise, however, is not related to physical information propagating to the outer boundary 
and coming back (for this one would have to wait till $t\approx 2,000M$) but,
rather, is related to spurious modes with 
high group velocities traveling in the wrong direction, as described in
Section \ref{HO}. As discussed there, 
for an eighth order centered derivative the speed of this spurious mode is around $-2.6$, which roughly matches 
with this noise appearing at $t\approx 800M$. We have checked that this boundary effect does 
go away with resolution, by introducing some amount of dissipation or 
by pushing the outer boundary farther out. Notice that while the first two options
allow one to observe the tail behavior for much longer, eventually physical
information would travel back from the outer boundary and ``cavity'' effects which
affect the decay would take place. Indeed, the behavior would no longer be determined by
a power law tail but by an exponential decay \cite{priceburko}.

\begin{figure}[ht]
\begin{center}
\includegraphics*[height=7cm]{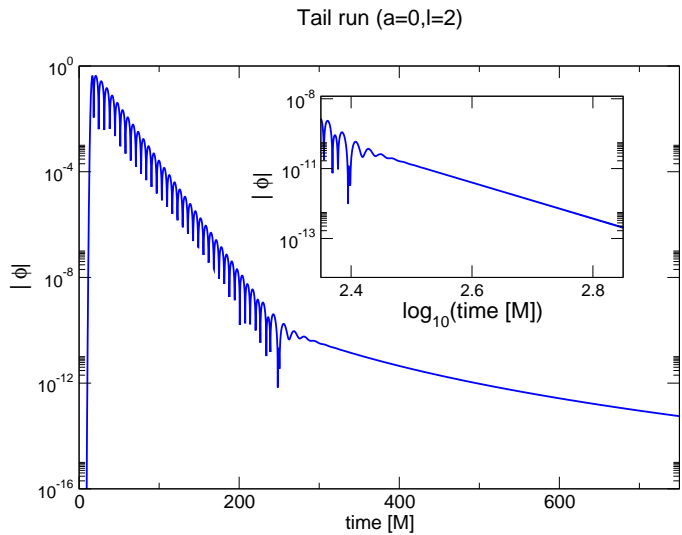}
\caption{A simulation with the same parameters as those of
  Fig. \ref{tailkerr}, but with a non-spinning black hole. 
The average slope for $\Phi$ in tail regime gives a decay of $t^{-6.96}$, 
again in good agreement with the expected decay of $t^{-7}$.}
\label{tailsch} 
\end{center}
\end{figure}

Figure \ref{tailsch} shows a similar run, in this case however the black 
is not spinning. A fit to the solution in the tail regime gives a decay for $\phi$ of $t^{-6.96}$, which again
matches quite well the expected decay of $t^{-7}$ \cite{price}. 

\section{Final Comments}

As illustrated in this work, the combination of the penalty technique
together with those guaranteeing a stable single grid implementation for
hyperbolic systems provides
a way to achieve stable implementations of
multi-block schemes of {\em arbitrary high order}. A similar penalty technique for
multi-block evolutions is also being pursued in conjuction with
pseudo-spectral methods \cite{kps}.

The flexibility provided by multiple grids can be exploited
to address a number of issues currently faced in simulations of Einstein's
equations,  among these
\begin{itemize}
\item The desire for a conforming inner boundary. This plays a central role
in ensuring a consistent implementation of the excision technique together
with a saving in the computational cost of the implementation.
\item The need for a smooth outer boundary. This removes the presence of corners and
edges which have proved difficult to dealt with even at the analytical level \cite{ibv,gioel}. Furthermore,
a smooth $S^2$ outer boundary simplifies tremendously the search for an efficient
matching strategy to an outside formulation aimed to cover a much larger region of the spacetime
with a formalism better suited to the asymptotic region (see, for example, \cite{ccm} and \cite{cpm1,cpm2,cpm3}).
\item The use of a grid that is better adapted to the description of wave phenomena as
they propagate in the region far from the sources.
\end{itemize}

\section{Acknowledgments}
 
This research was supported in part by the NSF under Grants No PHY0244335,
PHY0244699, PHY0326311, INT0204937, and PHY0505761 to Louisiana State University, 
No. PHY0354631 and PHY0312072 to Cornell University, No PHY9907949 to the 
University of California as Santa Barbara; the Horace Hearne Jr.
Institute for Theoretical Physics; CONICET and SECYT-UNC. This research
employed the resources of the Center for Computation and Technology at Louisiana
State University, which is supported by funding from the Louisiana
legislature's Information Technology Initiative. L. L. was partially supported
by the Alfred P. Sloan Foundation.

We thank Mark Carpenter, Peter Diener, Nils Dorband, Ian Hawke, Larry Kidder, Ken Mattsson, 
Jorge Pullin, Olivier Sarbach, Erik Schnetter, 
Magnus Svard, Saul Teukolsky, Jonathan Thornburg, and Burkhard Zink for
helpful discussions, suggestions and/or comments on the manuscript.

The authors thank Louisiana State University and the Horace Hearne Jr. Institute for
Theoretical Physics, the Universidad
Nacional de Cordoba and FaMAF, and
the University of California at Santa Barbara for hospitality 
at different stages of this work.

\appendix

\section{Coefficients for high order operators with diagonal metrics} \label{appendix_der}
For completeness we point out here some misprints in Ref.\cite{strand} in some 
of the expressions for the diagonal metric cases. 
\begin{itemize}
\item $D_{2-1}$: No typos. 
\item $D_{4-2}$: It says 
$\alpha_2=-1/2$, but it should be $\alpha_2 = -1/12$. The
expressions for $q_{2i}$ are also missing; they should be the following:
$$
q_{20}=\frac{8}{86}\;\;\; = \;\;\; q_{21}=-\frac{59}{86} \;\;\; =
\;\;\;q_{23}=-q_{21} \;\;\; , \;\;\; q_{24}=-q_{20}
$$ 
\item $D_{6-3}$: No typos in the scalar product or coefficients for the derivative, 
neither in the general case nor in the minimum bandwidth one. 

\item $D_{8-4}$: The operator that has the minimum bandwidth is correct, 
 but the three-parametric one has a typo in one of the coefficients 
(the scalar product is correct): it says
$$
q_{06} = 49(-1244160x_1 + 18661400x_3 - 13322233)/17977668
$$
when it should say
$$
q_{06} = 49(-1244160x_1 + 18662400x_3 - 13322233)/17977668
$$

\end{itemize}

\section{Dissipation for high order difference operators with diagonal norms} \label{dis}
The addition of artificial dissipation 
typically involves considering a  dissipative operator $Q_d$ which is
non-positive with respect to the scalar product with respect to which SBP holds, i.e.,
\be
<u,Q_du> \leq 0 \;\;\; \forall u \;. \label{neg}
\ee
We start the construction of such operators satisfying this
property by defining them as 
\begin{equation}
Q_d = (-1)^{m-1}\sigma h^n(D_+D_-)^m \label{dis_int}
\end{equation}
on gridpoints lying within the range $[r,s]$ contained in the
interval in which the weight used in the scalar product is
one. That is, the interval in which the difference operator is one of the centered
ones of Eqs.(\ref{2c},\ref{4c},\ref{6c},\ref{8c}); for example, if the gridpoints
range from $0$ to $N$, then $(r,s)$ must satisfy
\be
r\ge 1, s\le N-1 \;\;\; \mbox{ for } D_{2-1} \label{range21} \, ;
\ee
\be
r\ge 4, s\le N-4 \;\;\; \mbox{ for } D_{4-2} \label{range42}\, ;
\ee
\be
r\ge 6, s\le N-6 \;\;\; \mbox{ for } D_{6-3} \label{range63}\, ;
\ee
\be
r\ge 8, s\le N-8 \;\;\; \mbox{ for } D_{8-4} \label{range84}\, .
\ee
As we will see later, in some cases our construction of dissipative operators 
imposes stricter constraints on the range of allowed values for $r,s$ for each derivative.

If $n=2m-1$, the operator (\ref{dis_int}) is 
the standard Kreiss--Oliger dissipation (KO) \cite{ko}, which is negative definite
in the absence of boundaries (when the weight in the scalar product is identically
one).  The choice $n=2m-1$ ensures that the added dissipation
has the same `scale' as the principal part (that is, length$^{-1}$) and
that the resulting amplification factor is independent of resolution.

For each of the derivatives $D_{2-1},D_{4-2},D_{6-3},D_{8-4}$, 
we seek to extend $Q_d$ in Eq.(\ref{dis_int}) so that the resulting operator is negative
definite with respect to the corresponding SBP scalar product. As mentioned
above, we denote the points in which $Q_d$ is given  by
Eq.(\ref{dis_int}) as $i=r\ldots s$.

The identities of Appendix C are used in the calculations needed for the
construction of the operators below. 
These identities let one express the
norm of the dissipation in the interior, which is proportional to 
$(-1)^{n-1}(u,(D_+D_-)^nu)^{[r,s]}$, as 
\begin{equation}
(-1)^{n-1}(u,(D_+D_-)^nu)^{[r,s]} = \mbox{non-positive
  definite terms} + \sum_{0}^{r-1} u_i(\ldots ) + \sum_{s+1}^N u_i(\ldots ) \label{split} \, .
\end{equation}
Once this is done, the norm of the whole dissipative operator can be written as 
\begin{equation}
(u,Q_d u)^{[0,N]} = \mbox{non-positive
  definite terms} + \sum_{0}^{r-1} u_i[(\ldots )_i + h\sigma_iQ_du_i] +
\sum_{s+1}^N u_i[(\ldots)_i + h\sigma_iQ_du_i ] \label{zeroing}
\end{equation}
and it is straightforward to make it non-positive definite. For
example, by choosing 
\begin{equation}
Q_du_i = -\frac{(\ldots)_i}{\sigma_ih} \;, \label{easy}
\end{equation}
which is how we proceed below. Notice however this is not the only way to proceed. 
For instance, one could try to make the two sums of 
Eq.(\ref{zeroing}) cancel (as opposed to requiring the terms cancel at each
gridpoint).  

A more general approach for constructing dissipative operators that are negative
definite with respect to the appropriate (SBP) scalar product has been
recently presented by Mattson, Svard and Nordstrom \cite{mattsson}. 

\subsection{Fourth derivative dissipation, and KO type for $D_{2-1}$}
In this case the interior operator is 
$$
Q_d = -\sigma h^n(D_+D_-)^2 \, . 
$$
One can then split the norm of the dissipative operator over the whole grid
into terms involving the left and right ranges $[0,r], [s,N]$ (of the yet
undefined operator) and the interior terms:
$$
(u, Q_du)_{\Sigma}^{[0,N]} = h\sum_{0}^{r-1} \sigma_i u_iQ_du_i +
h\sum_{s+1}^N \sigma_i u_iQ_du_i 
- \sigma h^n(u, (D_+D_-)^2) ^{[r,s]}
$$
where $\sigma_i$ are the scalar product weights. Using the identities of
Appendix \ref{prop} One can see that
\begin{eqnarray}
(u, Q_du)_{\Sigma}^{[0,N]} &=&   -  \sigma h^n ||D_+D_-u||^2_{[r-1,s+1]}+  
 h\sum_{0}^{r-3} \sigma_i u_iQ_du_i +
 h\sum_{s+3}^N \sigma_i u_iQ_du_i \nonumber  \\
& & +  hu_{r-2}\left(\sigma_{r-2} Q_d + \sigma
h^{n-2}D_+^2 \right)u_{r-2} + 
  h u_{r-1}\left[ \sigma_{r-1} Q_d - \sigma h^{n-2}(2D_+D_- - D_+^2]
 \right)u_{r-1}  \nonumber \\
& & +  h u_{s+2}\left(\sigma_{s+2} Q_d + \sigma h^{n-2}
 D_-^2\right)u_{s+2} + hu_{s+1}\left[\sigma _{s+1}Q_d -
  \sigma h^{n-2} (2D_+D_- - D_-^2) \right]u_{s+1}  \nonumber
\end{eqnarray}

There are several options now to control the contribution of these terms, 
the simplest one is
\begin{eqnarray}
Q_du_{r-2} & =& -\frac{\sigma h^{n-2}}{\sigma_{r-2}} D_+^2 u_{r-2} \nonumber \\
Q_d u_{r-1} &=& \frac{\sigma h^{n-2}}{\sigma_{r-1}} (2D_+D_- - D_+^2)u_{r-1}
\nonumber \\
Q_du_{s+1} &=& \frac{\sigma h^{n-2}}{\sigma_{s+1}} (2D_+D_- - D_-^2)u_{s+1} \nonumber \\
Q_du_{s+2} &=& -\frac{2\sigma}{\sigma_{s+2}} h^{n-2} D_-^2 u_{s+2} \nonumber
\end{eqnarray}
and $Q_d=0$ everywhere else, which implies $r\ge
2,s\le N-2$. Notice that, as anticipated, this is a further constraint 
on the possible range of $r,s$ [c.f. Eq.(\ref{range21})].
The operator constructed not only satisfies the non-positivity requirement, but also
transforms according to $D_- \rightarrow -D_+$ under the symmetry $i
-\rightarrow N-i$ (as it should).

\textbf{Preferred choice:}

A KO type dissipation for $D_{2-1}$, which we call KO4 (because it involves
fourth derivatives in the interior) corresponds to 
$$
n=3 \; .
$$
With this choice the order of the scheme is one 
in at least two points (while in the absence of dissipation would be of order
one in just one point). In general, the best option for $r,s$ is 
$$
r=2, s=N-2\;, 
$$
as in that case 
 the dissipation operator is non-trivial everywhere
(while choosing $r> 2,s< N-2$ would imply that the dissipation is zero at
--and even possibly close to-- the boundaries). With this choice for $r,s$ the
order of the scheme is one at the last two boundary points.

\subsection{Sixth
  derivative dissipation, and KO type for $D_{4-2}$}
We now proceed to construct a higher derivative
dissipative operator. As before, we adopt at the interior
points the standard dissipation operator,
$$
Q_d =  \sigma h^n D_+^3D_-^3
$$
Then, defining $v = D_+D_- u$ and employing the identities 
of Appendix \ref{prop} one obtains 

\begin{eqnarray*}
(u,Q_du)_{\Sigma}^{[0,N]} &=& -\sigma h^n||D_- v||^2_{[r-1,s+2]} + h\sum_{0}^{r-1} \sigma_i u_iQ_du_i +
 h\sum_{s+1}^N \sigma_i u_iQ_du_i   \\ 
& & \\
& & + u_{r-3} h\left( - \sigma h^{n-3}D_+v_{r-2} + \sigma_{r-3} Q_d u_{r-3}\right)  \\
& & \\
& & +  u_{r-2} h \left( \sigma h^{n-3}(-hD_+^2 + 2D_+ )v_{r-2} + \sigma_{r-2} Q_d u_{r-2}
  \right) \\
& & \\
& & +  u_{r-1} h\left[ \sigma h^{n-3}\left( h(2D_+D_--D_+^2)v_{r-1}  - D_+v_{r-2} \right)
    + \sigma_{r-1}Q_du_{r-1}  \right] + \\
& & \\
& &  + u_{s+1}h\left[ \sigma h^{n-3}\left( h(2D_+D_--D_-^2) v_{s+1} +
 D_-v_{s+2}  \right) + \sigma_{s+1}Q_du_{s+1}  \right] \\
& & \\
& & + u_{s+2}h\left( \sigma h^{n-3}(-hD_-^2-2D_-)v_{s+2} +
  \sigma_{s+2}Q_du_{s+2}  \right)  \\
& & \\
& & + u_{s+3}h\left( \sigma h^{n-3}D_- v_{s+2} + \sigma_{s+3}Q_du_{s+3}
  \right)  \, .
\end{eqnarray*}

Here again one has several options, the simplest one is:
\begin{eqnarray*}
Q_du_{r-3} &=& \frac{\sigma h^{n-3}}{\sigma_{r-3}}D_+ v_{r-2} \\
Q_du_{r-2} &=& \frac{\sigma h^{n-2}}{\sigma_{r-2}}(hD_+^2  - 2D_+)v_{r-2} \\
Q_d u_{r-1} &=& -\frac{\sigma h^{n-2}}{\sigma_{r-1}}\left[ h(2D_+D_- - D_+^2)v_{r-1} -
  D_+v_{r-2} \right] \\
Q_d u_{s+1} &=& -\frac{\sigma h^{n-2}}{\sigma_{s+1}}\left[ h(2D_+D_- - D_-^2)v_{s+1} +
  D_-v_{s+2} \right] \\
Q_du_{s+2} &=& \frac{\sigma h^{n-2}}{\sigma_{s+2}}(hD_-^2  + 2D_-)v_{s+2} \\
Q_du_{s+3} &=& -\frac{\sigma h^{n-3}}{\sigma_{s+3}}D_- v_{s+3}
\end{eqnarray*}
and $Q_d=0$ everywhere else. Notice that, as in the $D_{2-1}$ case, $Q_d$ transforms under the symmetry  $i \rightarrow
N-i$ $Q_d$ as it should. Notice, however, that since we are modifying the
dissipative operator in three points near the boundary, no further constraint
occurs in the allowed range for $r,s$ [c.f. Eq.(\ref{range42})], unlike the
$D_{2-1}$ case. \\

\textbf{Preferred choice:}

A KO type dissipation for $D_{4-2}$, which we call KO6 corresponds to 
$$
n=5
$$
In general, the best choice for $r,s$ is 
$$
r=4,s=N-4 \;, 
$$
as in that case the dissipation does not reduce the order of the
overall scheme \footnote{Since the $D_{4-2}$ operator already has four points
close to each boundary where its order is two, while the above dissipative operator has only three
points where that happens.}. However, a drawback of our
simplified method for constructing dissipative operators (which cannot be
remedied by adopting different values for $r,s$) is that in this case the dissipation is
zero at the last gridpoint.

\subsection{Eighth derivative dissipation, and KO type for $D_{6-3}$}
As usual, we start with 
\be
Q_d = -\sigma h^n D_+^4 D_-^4 \;\;\;\; i=r\ldots s \label{diss63}
\ee
Then, using the identities of appendix \ref{prop} 
the norm of the dissipative operator results 
\begin{eqnarray*}
(u,Q_du)_{\Sigma}^{[0,N]} &=& h\sum_{0}^{r-1} \sigma_i u_iQ_du_i +
h\sum_{s+1}^N \sigma_i u_iQ_du_i - \sigma
h^n(u,(D_+D_-)^4u)_{[r,s]}\\
& & \\
&= & -\sigma h^n||D_-^3D_+u||^2_{[r,s+2]}  \\
& & \\
& & +  u_{r-1}\left[-\sigma h^{n-3}\left(h^2 (2D_+D_--D_+^2)w_{r-1} - h
  D_+w_{r-2}+D_+\alpha_{r-2} \right) + \sigma_{r-1}hQ_du_{r-1} \right]
 \\
& & \\
& & + u_{r-2}\left[-\sigma h^{n-3}\left( (-h^2D_+^2+2hD_+)w_{r-2} -
  3D_+\alpha_{r-2} \right) + \sigma_{r-2}hQ_du_{r-2}  \right] \\
& & \\
& & + u_{r-3}\left[ -\sigma h^{n-3} (-h D_+w_{r-2} + 3D_+\alpha_{r-2}) +
  \sigma_{r-3}h Q_du_{r-3} \right]  \\
& & \\
& & + u_{r-4}\left( \sigma h ^{n-3} D_+\alpha_{r-2} + \sigma_{r-4}h
Q_du_{r-4} \right) + u_{s+4}\left( \sigma h^{n-3}D_+\alpha _{s+2} +
\sigma_{s+4} h Q_d u_{s+4} \right)  \\
& & \\
& & + u_{s+3}\left[ -\sigma h^{n-3} (h D_-w_{s+2} + 3D_+\alpha_{s+2}) +
  \sigma_{s+3}h Q_du_{s+3} \right]  \\
& & \\
& & + u_{s+2}\left[-\sigma h^{n-3}\left( -(h^2D_-^2+2hD_-)w_{s+2} -
  3D_+\alpha_{s+2} \right) + \sigma_{s+2}hQ_du_{s+2}  \right] \\
& & \\
& & + u_{s+1}\left[-\sigma h^{n-3}\left(h^2 (2D_+D_--D_-^2)w_{s+1} + h
  D_-w_{s+2} + D_+\alpha_{s+2} \right) + \sigma_{s+1}hQ_du_{s+1} \right]
\end{eqnarray*}
where $w=(D_+D_-)^4u, \alpha=D_-D_+D_-u$. The simplest choice is given
by setting to zero all these coefficients, which, when expanding $w$ and
$\alpha$ gives 
\begin{eqnarray}
Q_du_{r-1} &=& \frac{\sigma h^{n-4}}{\sigma_{r-1}}\left[
  h^2(2D_+D_--D_+^2)(D_+D_-)^4u_{r-1} - hD_+(D_+D_-)^4u_{r-2} +
  (D_+D_-)^2u_{r-2} \right] \label{diss1} \\
& & \nonumber \\
Q_du_{r-2} &=& \frac{\sigma h^{n-4}}{\sigma_{r-2}}\left[
  (-h^2D_+^2 + 2hD_+)(D_+D_-)^4 - 3(D_+D_-)^2\right] u_{r-2} \label{diss2}\\
& & \nonumber \\
Q_du_{r-3} &=& \frac{\sigma h^{n-4}}{\sigma_{r-3}}\left[
  -hD_+ (D_+D_-)^4 + 3(D_+D_-)^2\right] u_{r-2} \label{diss3} \\
& & \nonumber \\
Q_d u_{r-4} &=& -\frac{\sigma
  h^{n-4}}{\sigma_{r-4}}(D_+D_-)^2u_{r-2}\label{diss4} \\
& & \nonumber \\
Q_d u_{s+4} &=& -\frac{\sigma  h^{n-4}}{\sigma_{s+4}}(D_+D_-)^2u_{s+2} \label{diss5}\\
& & \nonumber \\
Q_du_{s+3} &=& \frac{\sigma h^{n-4}}{\sigma_{s+3}}\left[
  hD_- (D_+D_-)^4 + 3(D_+D_-)^2\right] u_{s+2}  \label{diss6} \\
& & \nonumber \\
Q_du_{s+2} &=& \frac{\sigma h^{n-4}}{\sigma_{s+2}}\left[
  -(h^2D_-^2 + 2hD_-)(D_+D_-)^4 - 3(D_+D_-)^2\right] u_{s+2} \label{diss7} \\
& & \nonumber \\
aQ_du_{s+1} &=& \frac{\sigma h^{n-4}}{\sigma_{s+1}}\left[
  h^2(2D_+D_--D_-^2)(D_+D_-)^4u_{s+1} + hD_-(D_+D_-)^4u_{s+2} + (D_+D_-)^2u_{s+2} \right] \label{diss8}
\end{eqnarray}
and $Q_d=0$ everywhere else.

\textbf{Preferred choice:}

If we want a KO-type dissipation for the $D_{6-3}$ derivative, which we call
KO8, we have to choose 
$$
n=7 \; .
$$
There are six points near each
boundary where the difference operator has order three, while the
above dissipation has four points where that happens. Therefore, the
order of the whole scheme is not spoiled if 
$$
r=6,s=N-6
$$
As in the $D_{4-2}$ case, the drawback of our construction is that the
dissipation is zero near boundaries; in this case at the last two points.

\subsection{Dissipation for $D_{8-4}$}
In this case we have not been able to write an interior KO dissipation which 
does not spoil the eighth order accuracy of the derivative (that is, one as in
Eq.(\ref{dis_int}), with $m=5$) in the form of Eq.(\ref{zeroing}). 
Therefore our simplified approach does not work in this
case. However, since the expressions of the previous subsections
 are valid for a general scalar product, we can use, for example, expressions
 (\ref{diss63}) and (\ref{diss1}-\ref{diss8}) with the weights corresponding to the $D_{8-4}$
 derivative, and 
$$
n=7,r=8,s=N-8
$$
This results in a KO dissipation of the KO8 type which is zero at the last
four points, with the drawback that the order of the scheme is reduced to seven 
at the interior points, and three near
boundaries (as opposed to eight and four, respectively, in the derivative itself). 

\section{Useful properties in the construction of dissipative operators} \label{prop}

It is straightforward to show that with respect to the scalar product and norm
\[
(u,v)^{[r,s]} \equiv \sum_{j=r}^s u_jv_j h_x, 
\qquad \left(\|u\|^{[r,s]} \right)^2 = (u,u)^{[r,s]}
\]
the following properties hold:
\begin{eqnarray}
(u,D_+v)^{[r,s]} &=& - (D_-u,v)^{[r+1,s+1]} + u_j v_j |_r^{s+1}\\
(u,D_-v)^{[r,s]} &=& - (D_+u,v)_{[r-1,s-1]} + u_j v_j |_{r-1}^s\\
(u,D_0v)^{[r,s]} &=& -(D_0u,v)_{[r,s]} + \frac{1}{2}(u_jv_{j+1} +
u_{j+1}v_j)|_{r-1}^s
\end{eqnarray}
The proofs for the first and third
identity can be found in Ref. \cite{gko}, and the second one is trivially obtained
from the first. 

The following equalities are also straightforward to check, though obtaining
them becomes increasingly more cumbersome as more derivatives are involved.
\begin{eqnarray*}
(u,D_+D_-v)^{[r,s]} &=&  - (D_-u,D_-v)^{[r+1,s+1]} + (u_{j}D_-v_j)_{r}^{s+1} \\
 & = & - (D_-u,D_-v)^{[r,s+1]} - u_{r-1}D_+u_{r-1} + 
u_{s+1}D_-u_{s+1} \label{prop1}\\
& & \\
(u,(D_+D_-)^2v)^{[r,s]} &=& (D_+D_-u,D_+D_-v)^{[r-1,s+1]} + \\
& & \\
& &  \frac{u_{s+1}}{h}\left(2D_+D_- -
D_-^2\right)v_{s+1} 
 - \frac{u_{s+2}}{h}D_-^2 v_{s+2}+\frac{u_{r-1}}{h}\left(2D_+D_- - 
D_+^2\right)v_{r-1} - \frac{u_{r-2}}{h}D_+^2 v_{r-2} \nonumber \\
& & \\
(u,(D_+D_-)^3v)^{[r,s]} &=& - (D_- p, D_- w)_{[r-1,s+2]} \\
& & \\
& & +  \frac{u_{r-1}}{h}\left((2D_+D_- - D_+^2)w_{r-1} -
\frac{1}{h}D_+w_{r-2} \right) + 
 \frac{u_{r-2}}{h}\left(-D_+^2 + \frac{2}{h}D_+\right)w_{r-2} - 
 \frac{u_{r-3}}{h^2}D_+w_{r-2} \\
& & \\
& & \frac{u_{s+1}}{h}\left( (2D_+D_--D_-^2)w_{s+1} +
\frac{1}{h}D_-w_{s+2}  \right)+ 
 \frac{u_{s+2}}{h}\left(-D_-^2 - \frac{2}{h}D_- \right) w_{s+2} +
 \frac{u_{s+3}}{h^2}D_-w_{s+2} \\
& & \\
(u,(D_+D_-)^4v)^{[r,s]} &=&  (D_- q, D_- \alpha)_{[r,s+2]} \\
& & \\
& & +  \frac{u_{r-1}}{h}\left((2D_+D_--D_+^2)w_{r-1} - \frac{1}{h}
  D_+w_{r-2}+ \frac{1}{h^2}D_+\alpha_{r-2} \right) 
 \\
& & \\
& & +  \frac{u_{r-2}}{h}\left((-D_+^2 + \frac{2}{h}D_+)w_{r-2} -
  \frac{3}{h^2}D_+\alpha_{r-2} \right)  \\
& & \\
& & +  \frac{u_{r-3}}{h^2}\left( - D_+w_{r-2} +
  \frac{3}{h}D_+\alpha_{r-2} \right)  \\
& & \\
& &  -  \frac{u_{r-4}}{h^3} D_+\alpha_{r-2} \\
& & \\
& & - \frac{u_{s+4}}{h^3} D_+\alpha _{s+2}  \\
& & \\
& & + \frac{u_{s+3}}{h^2}\left( D_-w_{s+2} + \frac{3}{h}D_+\alpha_{s+2}
  \right)  \\
& & \\
& & + \frac{u_{s+2}}{h}\left( -(D_-^2 + \frac{2}{h}D_-)w_{s+2} -
  \frac{3}{h^2}D_+\alpha_{s+2} \right)   \\
& & \\
& & + \frac{u_{s+1}}{h}\left((2D_+D_--D_-^2)w_{s+1} + \frac{1}{h}
  D_-w_{s+2} + \frac{1}{h^2}D_+\alpha_{s+2} \right)
\end{eqnarray*}
where $w=D_+D_-v, p=D_+D_-u$,$q=D_-^2D_+u,\alpha=D_-^2D_+v$, and 
$\sigma_i$ denotes the value of the scalar product at gridpoint $i$.


\end{document}